\documentclass[fleqn,usenatbib]{mnras}

\pdfoutput=1

\usepackage{mathptmx}

\usepackage[T1]{fontenc}

\DeclareRobustCommand{\VAN}[3]{#2}
\let\VANthebibliography\thebibliography
\def\thebibliography{\DeclareRobustCommand{\VAN}[3]{##3}\VANthebibliography}

\usepackage{graphicx}	
\usepackage{amsmath}	
\usepackage{amssymb}	
\usepackage{subcaption}

\title[GC kinematics of simulated lenticular galaxies]{The present-day globular cluster kinematics of lenticular galaxies from the E-MOSAICS simulations and their relation to the galaxy assembly histories}

\author[Dolfi et al.]{Arianna Dolfi,$^{1}$\thanks{E-mail: adolfi@swin.edu.au}
Joel Pfeffer,$^{2}$
Duncan A. Forbes,$^{1}$
Warrick J. Couch,$^{1}$
Kenji Bekki,$^{2}$
Jean P. Brodie,$^{1,3}$
\and Aaron J. Romanowsky,$^{3,4}$
and J. M. Diederik Kruijssen$^{5}$
\\
\\
$^{1}$ Centre for Astrophysics \& Supercomputing, Swinburne University of Technology, Hawthorn VIC 3122, Australia\\
$^{2}$ ICRAR, M468, The University of Western Australia 35 Stirling Highway, Crawley Western Australia, 6009, Australia\\
$^{3}$ University of California Observatories, 1156 High St., Santa Cruz, CA 95064, USA\\
$^{4}$ Department of Physics $\&$ Astronomy, San Jos\'e State University, One Washington Square, San Jose, CA 95192, USA\\
$^{5}$ Astronomisches Rechen-Institut, Zentrum fur Astronomie der Universitat Heidelberg, Monchhofstraße 12-14, 69120 Heidelberg, Germany
}

\date{Accepted 2022 January 25. Received 2022 January 13; in original form 2021 November 14}

\pubyear{2021}

\begin{document}
\label{firstpage}
\pagerange{\pageref{firstpage}--\pageref{lastpage}}
\maketitle

\begin{abstract}
We study the present-day rotational velocity ($V_{\mathrm{rot}}$) and velocity dispersion ($\sigma$) profiles of the globular cluster (GC) systems in a sample of $50$ lenticular (S0) galaxies from the E-MOSAICS galaxy formation simulations. 
We find that $82\%$ of the galaxies have GCs that are rotating along the photometric major axis of the galaxy (\textit{aligned}), while the remaining $18\%$ of the galaxies do not (\textit{misaligned}). This is generally consistent with the observations from the SLUGGS survey. 
For the \textit{aligned} galaxies, classified as \textit{peaked and outwardly decreasing} ($49\%$), \textit{flat} ($24\%$) and \textit{increasing} ($27\%$) based on the $V_{\mathrm{rot}}/\sigma$ profiles out to large radii, we do not find any clear correlation between these present-day $V_{\mathrm{rot}}/\sigma$ profiles of the GCs and the past merger histories of the S0 galaxies, unlike in previous simulations of galaxy stars.
For just over half of the \textit{misaligned} galaxies, we find that the GC misalignment is the result of a major merger within the last $10\, \mathrm{Gyr}$ so that the \textit{ex-situ} GCs are misaligned by an angle between $0\degr$ (co-rotation) to $180\degr$ (counter-rotation) with respect to the \textit{in-situ} GCs, depending on the orbital configuration of the merging galaxies.  
For the remaining \textit{misaligned} galaxies, we suggest that the \textit{in-situ}
metal-poor GCs, formed at early times, have undergone more frequent kinematic perturbations than the \textit{in-situ} metal-rich GCs.  
We also find that the GCs accreted early and the \textit{in-situ} GCs are predominantly located within $0.2$ virial radii ($R_{200}$) from the centre of galaxies in 3D phase-space diagrams.
\end{abstract}




\section{Introduction}
Lenticular (S0) galaxies are characterized by a bulge surrounded by a smooth disk with very little or no ongoing star-formation. The fraction of lenticular galaxies is found to increase within the high density environments of rich galaxy clusters at the expense of spiral galaxies \citep{Dressler1980,Dressler1997,Fasano2000}. 
For this reason, understanding how S0 galaxies form is important for constraining which physical processes play a dominant role in the transformation and quenching of galaxies in different environments.

Focusing on the kinematic properties of galaxies, previous studies have found that S0 galaxies in the field and small galaxy groups are more pressure-supported than S0 galaxies in clusters \citep{Coccato2020,Deeley2020}. Additionally, S0 galaxies in the field and small galaxy groups also showed more frequently misaligned stellar and gas kinematics than cluster S0s \citep{Deeley2020}. 
These results suggest that cluster S0 galaxies are consistent with spiral progenitors, i.e. faded spirals, that had their star formation quenched and their spiral arm structure suppressed as a result of the interaction with the environment, e.g. ram-pressure stripping, tidal interactions, starvation \citep{Gunn1972,Larson1980,Bekki2009,Bekki2011,Merluzzi2016}. On the other hand, S0 galaxies in the field and small galaxy groups are likely the result of more complex formation histories that involved multiple mergers and accretion events (e.g. \citealt{Tapia2017,ElicheMoral2018,Dolfi2020,Dolfi2021}).

Previous works have used simulations to study the formation of S0 galaxies through mergers and investigate how mergers influence their kinematic properties.
Results showed that both major (mass-ratio $>$1:4) and minor (1:10$<$ mass-ratio $<$1:4) mergers can produce S0 galaxies \citep{Bournaud2005,Wu2014,Naab2014,Tapia2017,ElicheMoral2018,Schulze2020} and, specifically, major mergers are expected to spin up the rotation of the stars out to large radii, while minor mergers are expected to decrease it \citep{Naab2014,Wu2014,Schulze2020}. 

The commonly proposed two-phase formation scenario for massive early-type galaxies (ETGs) suggests that the galaxies had an early ($z\geq2$) \textit{in-situ} formation during which they formed most of their stellar mass, followed by the late ($z\simeq0$) accretion of dwarf galaxies in mini mergers (mass-ratio $<$ 1:10) onto their outskirts \citep{Oser2010,Damjanov2014,Zolotov2015,Rodriguez2016}. These late mini mergers are expected to decrease the rotation of the stars at large radii, where the galaxy transitions from a fast-rotating disk to a slowly rotating spheroidal component \citep{Arnold2011,Arnold2014,Guerou2016,Bellstedt2017,Dolfi2020}.

These results suggest that it is important to investigate the full kinematic behaviour of the galaxies from the inner regions out to the outskirts, in order to identify the presence of transitions that can help us to constrain the galaxy specific merger and accretion histories.

In a more recent work, \citet{Schulze2020} investigated the kinematic profiles of stars out to $\sim5$ effective radii ($R_{\mathrm{e}}$) of a sample of ETGs with $\log( \mathrm{M_{*}}/\mathrm{M_{\odot}} ) \geq 10.3$ from the \texttt{Magneticum} simulations. From the ratio of the rotation velocity to velocity dispersion profiles ($V_{\mathrm{rot}}/\sigma$), they identified three distinct kinematic profile shapes: a \textit{peaked and outwardly decreasing} (\textit{peaked}, hereafter), an \textit{increasing}, and a \textit{flat} $V_{\mathrm{rot}}/\sigma$ profile out to $\sim5\, R_{\mathrm{e}}$ that were likely the result of the different merger histories of the galaxies. Specifically, they suggested that galaxies with \textit{peaked} $V_{\mathrm{rot}}/\sigma$ profiles had an assembly history characterized by the late accretion of dwarf galaxies in minor and mini mergers that were disrupted beyond $\sim2\, R_{\mathrm{e}}$, enhancing the random motion of the galaxy outskirts without disrupting its central disk-like kinematics. The galaxies with \textit{flat} and \textit{increasing} $V_{\mathrm{rot}}/\sigma$ profiles were more likely to have experienced a late (i.e. $z\lesssim1$) major merger event. This event was likely more gas-rich for galaxies with \textit{increasing} $V_{\mathrm{rot}}/\sigma$ profiles, leading to the re-formation of a disk component and the rising of the rotational velocity of the stars out to large radii. On the other hand, the late major merger event did not contribute to a significant gas fraction in the galaxies with \textit{flat} $V_{\mathrm{rot}}/\sigma$ profiles and likely disrupted the central disk-like kinematics of the galaxy.
Therefore, we expect that the present-day stellar $V_{\mathrm{rot}}/\sigma$ profiles of ETGs can be used to constrain their past assembly history.

In \citet{Dolfi2021}, we studied the kinematic properties of a selected sample of $9$ S0 galaxies from the SAGES Legacy Unifying Globulars and GalaxieS (SLUGGS; \citealt{Brodie2014}) survey extending out to $\sim4$-$6\, R_{\mathrm{e}}$. 
To reach out to $\sim5\, R_{\mathrm{e}}$, we combined the stellar kinematics \citep{Arnold2014,Foster2016} with those of the globular clusters (GCs; \citealt{Forbes2017_GC}) and planetary nebulae (PNe; \citealt{Pulsoni2018}), which can be more easily probe out to large radii due to their brightness. 
Both the GCs and PNe are expected to trace the underlying stellar population of the ETGs. In fact, the PNe are stars with initial masses $1\, \mathrm{M_{\odot}} < \mathrm{M}_{*}<8\, \mathrm{M_{\odot}}$ that have evolved past the main sequence towards the red giant phase and should, thus, trace the kinematics of the stars out to large radii (e.g. \citealt{Romanowsky2006,Buzzoni2006}). 
On the other hand, the GCs typically display a colour bimodality that reflects the two distinct sub-populations: the red, metal-rich and blue, metal-poor GCs. These two GC sub-populations are not necessarily equal to the \textit{in-situ} and \textit{ex-situ} (accreted) GCs, respectively \citep{Forbes2018,Kruijssen2019}, however they are expected to trace the galaxy kinematics differently \citep{Gomez2021}. The red, metal-rich GCs are expected to trace the kinematics of the bulge and spheroid in ETGs, since they largely formed \textit{in-situ} at roughly the same time as the bulk of the host galaxy stars. The blue, metal-poor GCs are expected to trace the kinematics of the stellar halo of the galaxy, since they likely formed at earlier times than the red, metal-rich GCs or were later accreted onto the galaxy \citep{Forbes1997,Strader2005,Brodie2006}. In a recent work, using the E-MOSAICS simulations, \citet{Campos2021} found that the metal-poor GCs typically have more extended and shallower surface density profiles than the metal-rich GCs, suggesting an accreted origin for a significant fraction of the former. The surface density profile of the metal-poor GCs is also found to correlate with the slope of the dark matter halo density profile, therefore suggesting that the GC sub-populations are excellent probes for studying the outskirts of the galaxies. 
Additionally, simulations have also shown that major mergers are expected to spin up the rotation of the GC population at large radii, as opposed to minor mergers that are expected to decrease it \citep{Bekki2005}.

From the study in \citet{Dolfi2021}, we found that $6$ of the $9$ galaxies had consistent stellar, GCs and PNe kinematics with their rotation occurring along the photometric major-axis of the galaxy (\textit{aligned} galaxies). In the remaining $3$ galaxies, the PNe and GCs were rotating along a kinematic axis misaligned with respect to that of the stars, which was consistent with the photometric major-axis of the galaxy (\textit{misaligned} galaxies). These results seem to suggest distinct assembly histories for the \textit{aligned} and \textit{misaligned} galaxies, with the latter having likely undergone more recent and multiple accretion events. Among the $6$ \textit{aligned} galaxies, we also found that $4$ galaxies show a \textit{peaked} $V_{\mathrm{rot}}/\sigma$ profile, while the remaining $2$ galaxies show a \textit{flat} $V_{\mathrm{rot}}/\sigma$ profile. Therefore, the comparison with the simulations of \citet{Schulze2020} would suggest an assembly from late minor and mini mergers for the \textit{peaked} galaxies, while the \textit{flat} galaxies were likely involved in a merger with higher mass-ratio.     

However, these results assume that both GCs and PNe are tracing the kinematics of the underlying stellar population out to large radii. We find that this seems to be the case for the $6$ \textit{aligned} galaxies, but not for the remaining $3$ \textit{misaligned} galaxies that show twists in the kinematic position angle as a function of the galactocentric radius (see figure 3 in \citealt{Dolfi2021}).

In light of this, the aim of this paper is to use the E-MOSAICS\footnote{MOdelling Star cluster population Assembly In Cosmological Simulations within EAGLE (\url{https://www.astro.ljmu.ac.uk/~astjpfef/e-mosaics/})} simulations \citep{Pfeffer2018,Kruijssen2019} to study the present-day kinematic properties of the GC systems of the S0 galaxies to understand how they are related to the past formation histories of the S0 galaxies. In a recent work, \citet{Deeley2021} have found that more than half (i.e. $57\%$) of the S0 galaxies from the IllustrisTNG simulations have formed through significant merger events, while the remaining $37\%$ have formed through gas stripping events as a result of the infall onto a galaxy group or cluster. Therefore, as one of the most dominant S0 formation pathways, it is important to further understand how the different merger events affect the present-day kinematic properties of S0 galaxies out to large galactocentric radii and, specifically, the kinematic properties of the GC systems that have been studied. We also aim to further investigate the evolution of the \textit{aligned} and \textit{misaligned} galaxies to understand what physical processes are responsible for the present-day GC misalignment that we see in the observations from the SLUGGS survey \citep{Dolfi2021}.

The structure of the paper is as follows. In Sec. \ref{sec:emosaics_simulations}, we describe the E-MOSAICS simulations. In Sec. \ref{sec:data}, we describe the selection criteria of the simulated S0 galaxy sample as well as of the GC systems of each galaxy. In Sec. \ref{sec:kinematic_analysis_results}, we describe the kinematic analysis and results of the GC systems of our selected S0 galaxies and we also investigate the kinematics of the metal-rich and metal-poor GCs separately for a sub-sample of the galaxies. In Sec. \ref{sec:phase_space_diagrams}, we study whether there exists a correlation between the location of the GC systems of the galaxies on the 3D phase-space diagrams and their accretion time onto the galaxy. In Sec. \ref{sec:formation_history_S0s}, we discuss the assembly histories of our simulated galaxies by investigating the relation between the $V_{\mathrm{rot}}/\sigma$ profiles of the GCs with the merger histories of the galaxies derived from the E-MOSAICS simulations. In Sec. \ref{sec:summary}, we summarize our results and conclusions.

\section{The E-MOSAICS simulations}
\label{sec:emosaics_simulations}
E-MOSAICS is a suite of models designed to study the formation and evolution of star clusters and of their host galaxies \citep{Pfeffer2018}. E-MOSAICS is obtained from a combination of the MOSAICS \citep{Kruijssen2008,Kruijssen2009,Kruijssen2011,Pfeffer2018} semi-analytic model for the star cluster formation and evolution, and the suite of cosmological, hydrodynamical simulations from EAGLE \citep{Schaye2015,Crain2015} for the formation and evolution of galaxies in a standard $\Lambda$ cold dark matter ($\Lambda$CDM) Universe. 

EAGLE is run using a modified version of the N-Body Tree-PM smoothed particle hydrodynamics (SPH) code, \texttt{GADGET3} \citep{Springel2005}, which includes modifications to the SPH code, time stepping scheme and sub-grid physics models implemented in the simulations. Some of the main relevant physical processes that are implemented in the simulations within these sub-grid models include radiative cooling, reionization, star formation, stellar mass loss and metal enrichment, feedback from star formation and active galactic nuclei, black hole growth from gas accretion and mergers.
The efficiency of the sub-grid models for feedback are calibrated in order to reproduce certain observables. Specifically, the efficiency from the AGN feedback is calibrated to ensure that the simulations are reproducing the observed present-day relation between the central black hole mass and galaxy stellar mass. On the other hand, the efficiency from stellar feedback is calibrated to reproduce the observed present-day galaxy stellar mass to halo mass relation \citep{Schaye2015}.

MOSAICS is a semi-analytic model for the formation and evolution of star clusters that is incorporated into the EAGLE simulations of galaxy formation as a sub-grid model \citep{Kruijssen2011}. Within this model star clusters are not simulated as individual particles, but they are formed as a sub-grid component of a star particle that is produced as a result of a gravitationally unstable gas particle in the simulations. Therefore, a fraction of the mass of the newborn star particle (which is set by the cluster formation efficiency; \citealt{Bastian2008}) is used to generate a sub-grid population of star clusters, of which the initial properties are derived from the properties of the local gas and star particle at the moment of their formation \citep{Kruijssen2012,Campos2017}.
Subsequently to their formation, star clusters evolve and undergo mass-loss as a result of stellar evolution, modelled following the implementation for the star particles in EAGLE \citep{Wiersma2009}, two-body relaxation \citep{Lamers2005} and tidal shocks from the interstellar medium.
Dynamical friction is also another important process which may be responsible for the mass-loss and disruption of star clusters and it is included in post-processing in E-MOSAICS \citep{Pfeffer2018}. The timescales of dynamical friction are calculated for each star cluster according to the definition by \citet{Lacey1993} and star clusters are assumed to be completely disrupted by dynamical friction if their ages exceed the timescale for dynamical friction.
Dynamical friction is most effective in the disruption of the most massive star clusters and at small galactocentric radii.

The present-day (i.e. $z=0$) stellar masses of the star clusters formed within the simulations can range between $10^{2} < \mathrm{M_{*}}/\mathrm{M_{\odot}} < 10^{8}$ \citep{Pfeffer2018}. However, the lowest mass star clusters with $\mathrm{M_{*}} < 5 \times 10^{3}\, \mathrm{M_{\odot}}$ are immediately removed after their formation. On the other hand, the star clusters with $\mathrm{M_{*}} \sim 5\times 10^{4}\, \mathrm{M_{\odot}}$ are typically disrupted due to tidal shocks over the course of a Hubble time, while the high-mass star clusters (i.e. $\mathrm{M_{*}} > 10^{5}\, \mathrm{M_{\odot}}$) are removed by dynamical friction.
The globular cluster mass function in E-MOSAICS is found to agree well with that of observed galaxies at the high-mass end (i.e. $\mathrm{M_{*}} > 10^{5}\, \mathrm{M_{\odot}}$). However, the simulations overproduce the number of star clusters with $\mathrm{M_{*}} \leq 10^{5}\, \mathrm{M_{\odot}}$. Indeed, star clusters with $\mathrm{M_{*}} \sim 10^{4}\, \mathrm{M_{\odot}}$ are overestimated by a factor of $\sim10$-$100$, while star clusters with $\mathrm{M_{*}} \sim 10^{5}\, \mathrm{M_{\odot}}$ are overestimated by a factor of $2$-$10$ in E-MOSAICS \citep{Pfeffer2018}.
This issue may be traced to the fact that EAGLE does not incorporate an explicit model of the cold interstellar medium (ISM) and, as a result, low-mass star clusters are not efficiently disrupted through tidal shocks from the interactions with the ISM.

The inefficiency of the simulations at disrupting star clusters, due to the lack of an explicit cold ISM model in EAGLE, is also responsible for the excess of metal-rich GCs in E-MOSAICS. Specifically, the simulations overproduce the number of metal-rich GCs by a factor of $5$ in the metallicity range $-0.5 < \mathrm{[Fe/H]} < 0.0$ and by a factor of $2.5$ in the metallicity range $-1.0 < \mathrm{[Fe/H]} < -0.5$ \citep{Kruijssen2019}. 
However, \citet{Kruijssen2019} found that the overproduction of the metal-rich GCs in the metallicity range $-1.0 < \mathrm{[Fe/H]} < -0.5$ in the simulations does not have a statistically significant impact on the derived properties and correlations describing the formation and evolution of the galaxies.

In Sec. \ref{sec:GC_systems_simulations}, we will discuss the adopted selection criteria to minimize the impact of the overproduction of the low-mass star clusters in our kinematic results. 

Overall, the E-MOSAICS simulations are found to be able to reproduce most of the key observed properties of both young and old star cluster populations in their host galaxies. For this reason, they have been used in previous works to reconstruct the merger and accretion history of the Milky Way by comparing the age, metallicity and orbital properties of the observed GCs with those from the simulations (e.g. \citealt{Kruijssen2019b,Pfeffer2020,Kruijssen2020,Gomez2021}).

In addition to the initial suite of $25$ zoom-in simulations of Milky Way-mass galaxies, the cosmological volume of the E-MOSAICS simulations spans $34.4\, \mathrm{cMpc}$ (Crain et al., in prep.; first reported in section 2.2 of \citealt{Bastian2020}) and the simulations have an initial gas particle mass of $2.26\times10^{5}\, \mathrm{M}_{\odot}$, thus they can resolve galaxies with $\mathrm{M}_{*} > 10^{7}\, \mathrm{M_{\odot}}$ with more than $100$ stellar particles. 
However, E-MOSAICS includes very few massive galaxies with $\mathrm{M}_{*} \geq 10^{11}\, \mathrm{M_{\odot}}$ that are mainly ellipticals \citep{Correa2017}. At the same time, the largest galaxy cluster included in the simulations has halo mass $\mathrm{M_{200}} \sim 6 \times 10^{13}\, \mathrm{M_{\odot}}$, i.e. Fornax cluster-like, meaning that E-MOSAICS does not probe the dense environments of rich galaxy clusters, e.g. Coma cluster-like. Therefore, E-MOSAICS is limited to the low-density environments of the field and galaxy groups.

\begin{figure*}
\centering
\large
\textbf{\textbf{GalaxyID = 4315753}}\par\medskip
    \includegraphics[width=0.4\textwidth]{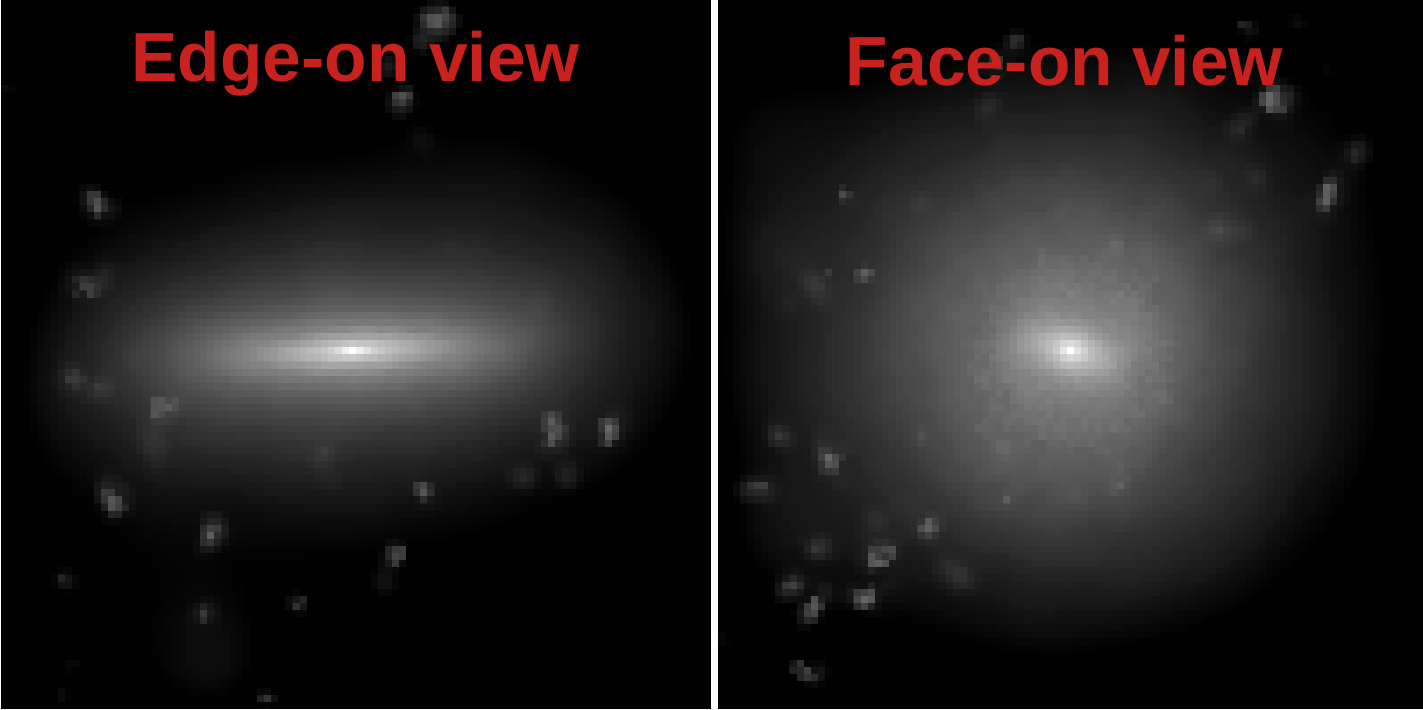}
    \includegraphics[width=0.41\textwidth]{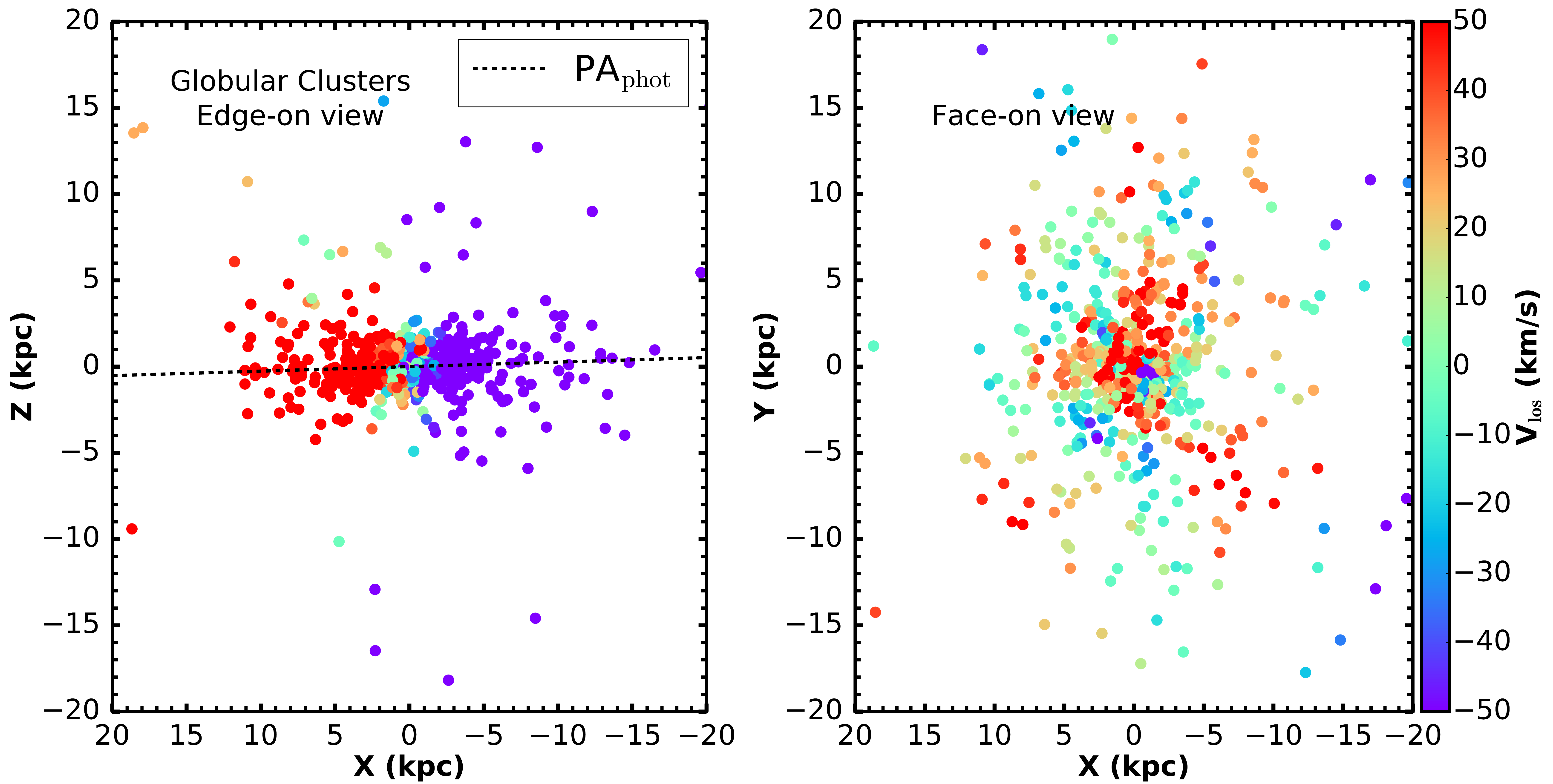}
\caption{\textit{Left:} mock SDSS $g$-band images of the edge-on and face-on view of one example S0 galaxy from the E-MOSAICS simulations, which is included in our final sample. The image size is $50\times50\, \mathrm{kpc}$ and the surface brightness reaches down to a minimum of $27\, \mathrm{mag\,  arcsec^{-2}}$. \textit{Right:} the edge-on and face-on spatial distributions of the GC system of the galaxy colour-coded by the line-of-sight velocity after applying the selection criteria described in Sec. \ref{sec:GC_stellar_mass_cut}-\ref{sec:GC_metallicity_cut}. The photometric major axis ($\mathrm{PA}_{\mathrm{phot}}$) of the galaxy is represented by the dashed line in the edge-on projection.}
\label{fig:mock_S0_images}
\end{figure*}

\begin{figure}
    \centering
    \includegraphics[width=0.5\textwidth]{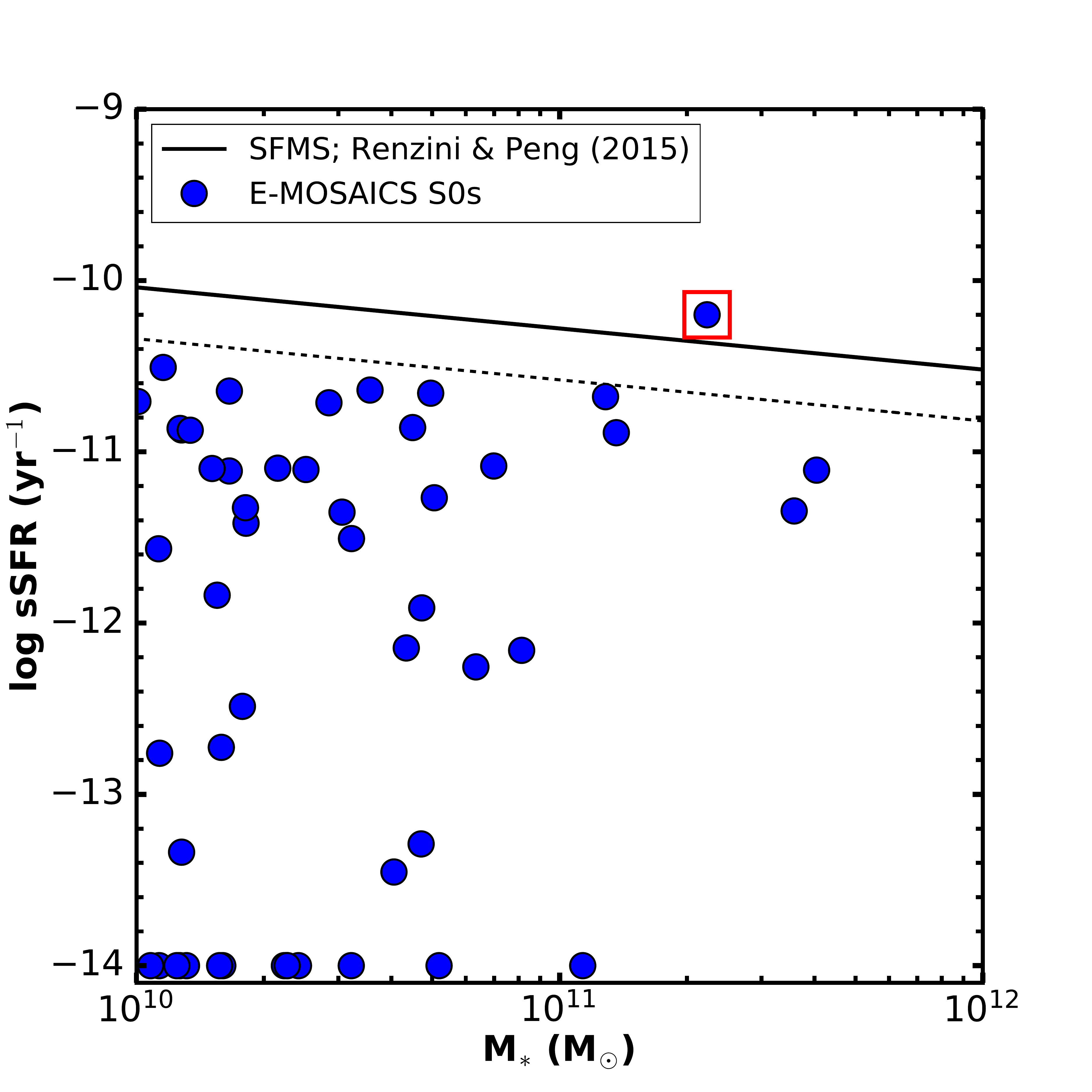}
    \caption{Distribution of our E-MOSAICS S0s on the specific star formation rate (sSFR)-stellar mass ($\mathrm{M_{*}}$) plane (blue dots). The red square represents the excluded galaxy with high sSFR. There are $11$ galaxies in the simulations with very low (almost equal to zero) star-formation rates, which we show here with a fixed $\mathrm{sSFR}=10^{-14}\, \mathrm{yr}^{-1}$. The solid black line represents the location of the star-forming main sequence (SFMS) from \citet{Renzini2015} and the dashed line represents the $0.3\, \mathrm{dex}$ scatter below the SFMS.}
    \label{fig:sSFR_StellarMass}
\end{figure}

\begin{figure}
    \centering
    \includegraphics[width=0.5\textwidth]{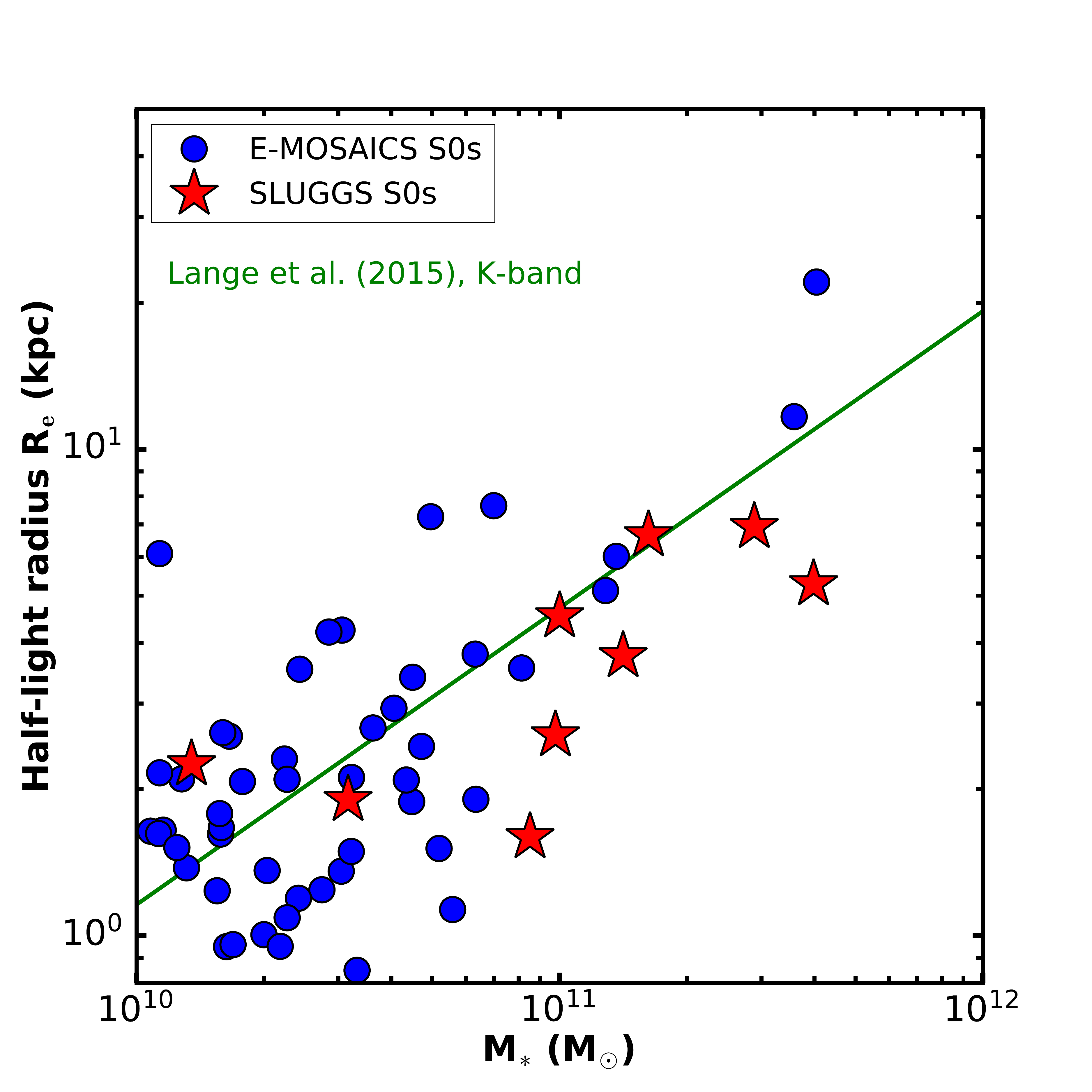}
    \caption{Distribution of our final sample of $50$ selected S0 galaxies from the E-MOSAICS simulations in the half-light radius ($R_{\mathrm{e}}$)-stellar mass ($\mathrm{M_{*}}$) plane (blue dots). For comparison, we also show the sample of nine S0 galaxies from the SLUGGS survey (red stars), studied in \citet{Dolfi2020,Dolfi2021}, with the circularized radius ($R_{\mathrm{e}}$) and total stellar mass ($\mathrm{M_{*}}$) taken from \citet{Forbes2017}. The solid line represents the $R_{\mathrm{e}}$-$\mathrm{M_{*}}$ single power-law relation of ETGs from \citet{Lange2015}, which is obtained using the best-fitting parameters in the $K$-band for the ETGs with $\mathrm{M_{*}} > 2 \times 10^{10}\, \mathrm{M_{\odot}}$ (see table B2 in \citealt{Lange2015}).} 
    \label{fig:StellarMass_Re}
\end{figure}

\section{The Data}
\label{sec:data}

\subsection{The S0 galaxy sample}
\label{sec:S0_galaxy_sample_selection}
From the simulations, we initially select $170$ galaxies with total stellar masses $\mathrm{M_{*}}>10^{10}\, \mathrm{M_{\odot}}$. This lower limit on the total galaxy stellar mass is chosen to match the stellar mass distribution of the observed early-type galaxies (ETGs) from the SLUGGS survey \citep{Brodie2014,Forbes2017}.

We produce mock SDSS $g$-band images of the edge-on and face-on view of each galaxy, colour-coded by the surface brightness calculated for each stellar particle using the \textsc{fsps} stellar population model \citep{Conroy2009,Conroy2010}. 
The images are modified to enhance the spiral arms so that we could more easily identify the S0 galaxies for our final sample. Therefore, following \citet{Trayford2017}, young stars (i.e. $<100\, \mathrm{Myr}$ old) are re-sampled from the star-forming gas particles at higher resolution (i.e. $1000\, \mathrm{M_{\odot}}$), assuming a constant star-formation rate, in order to better trace the spiral structure of galaxies. The synthetic images were generated using \textsc{SPHviewer} \citep{Llambay2015}. 

From the visual inspection of the edge-on and face-on images, which was performed by one of the authors of this paper (WJC), we identify $51$ S0 galaxies. The S0 galaxies are selected based on the identification of a disk component and absence of the spiral arms from the edge-on and face-on view images. However, despite the best effort to identify a representative sample of S0 galaxies, we note that there may still be few mis-classified S0s in our sample that may be prolate ellipticals. 
Fig. \ref{fig:mock_S0_images} shows the edge-on and face-on view of one example S0 galaxy from the E-MOSAICS simulations, which is included in our final sample. We also show the corresponding edge-on and face-on spatial distributions of the GC system of the galaxy colour-coded by the line-of-sight velocity after applying the selection criteria described in Sec. \ref{sec:GC_stellar_mass_cut}-\ref{sec:GC_metallicity_cut}. 
We note the presence of luminous clumps around the galaxy in Fig. \ref{fig:mock_S0_images}. However, these very disperse clumps are not necessarily a sign of spiral arm structure, but they are a result of the smoothing adopted to enhance the youngest stars with respect to the old stars.

Next, we consider the specific star formation rate (sSFR) of the $51$ visually selected S0 galaxies and we include in our final sample only those galaxies with sSFR more than $0.3\, \mathrm{dex}$ below the star-forming main sequence (SFMS). 
This constraint should ensure that we are reliably selecting a population of quenched galaxies, if the scatter of the SFMS is $\sim0.3\, \mathrm{dex}$, as found in previous studies (e.g. \citealt{Speagle2014}). 
Fig. \ref{fig:sSFR_StellarMass} shows the distribution of our S0 galaxies from the E-MOSAICS simulations on the sSFR-$\mathrm{M_{*}}$ plane (blue dots), with the best-fit line of the location of the SFMS (solid black line) from \citet{Renzini2015}. We note that in the EAGLE simulations the $\mathrm{sSFR}\simeq10^{-10}\, \mathrm{yr}^{-1}$ for the SFMS at $\mathrm{M_{*}}\simeq10^{10}\, \mathrm{M_{\odot}}$ (see figure 5 in \citealt{Furlong2015}), which is consistent with the value obtained from the SFMS relation of \citet{Renzini2015} for the same stellar mass, as shown in Fig. \ref{fig:sSFR_StellarMass}.
We find that only one galaxy fails to satisfy our selection criteria based on the sSFR, as it lies above the SFMS (red square in Fig. \ref{fig:sSFR_StellarMass}). Therefore, we exclude this galaxy from our final sample. We note that there are $11$ galaxies in the simulations with very low star-formation rates (SFR) that are practically equal to zero. In Fig. \ref{fig:sSFR_StellarMass}, we plot these galaxies with a fixed $\mathrm{sSFR}=10^{-14}\, \mathrm{yr}^{-1}$. After the sSFR cut, we obtain a final sample of $50$ S0 galaxies, with quenched star-formation, from the E-MOSAICS simulations.

In Fig. \ref{fig:StellarMass_Re}, we show the distribution of our final sample of $50$ simulated S0 galaxies in the half-light radius ($R_{\mathrm{e}}$)-stellar mass ($\mathrm{M_{*}}$) plane (blue dots). Here, we assume that mass follows light in the simulations and, therefore, that the 2D half-mass radius of our E-MOSAICS S0 galaxies is equal to the 2D half-light radius, $R_{\mathrm{e}}$. For comparison, we also show the nine S0 galaxies (red stars), observed from the SLUGGS survey \citep{Brodie2014}, that we have studied in our previous works \citep{Dolfi2020,Dolfi2021}. For the observed galaxies, the circularized radius ($R_{\mathrm{e}}$) and the total stellar mass ($\mathrm{M_{*}}$) are taken from \citet{Forbes2017}.
The solid line represents the $R_{\mathrm{e}}$-$\mathrm{M_{*}}$ single power-law relation of ETGs from \citet{Lange2015}, which is obtained using the best-fitting parameters in the $K$-band for the ETGs with $\mathrm{M_{*}} > 2 \times 10^{10}\, \mathrm{M_{\odot}}$ (see table B2 in \citealt{Lange2015}).

In Fig. \ref{fig:StellarMass_Re}, we note that there is quite a bit of scatter of both our simulated and observed S0 galaxies around the $K$-band single power-relation of ETGs with $\mathrm{M_{*}} > 2 \times 10^{10}\, \mathrm{M_{\odot}}$ from \citet{Lange2015}. Additionally, there is also one simulated S0 with $\mathrm{M_{*}} > 2 \times 10^{11}\, \mathrm{M_{\odot}}$ characterized by very large $R_{\mathrm{e}}\sim20\, \mathrm{kpc}$. In \citet{Lange2015}, the authors do not report the $1\sigma$ scatter of the observed data around the $R_{\mathrm{e}}$-$\mathrm{M_{*}}$ single power-relation of ETGs. However, we note that the ETGs from \citet{Lange2015} show an overall similar scatter around the corresponding $K$-band single power-law relation, with ETGs having $R_{\mathrm{e}}$ ranging between $\sim1$-$3\, \mathrm{kpc}$ at $\mathrm{M_{*}} = 10^{10}\, \mathrm{M_{\odot}}$ and between $\sim2$-$10\, \mathrm{kpc}$ at $\mathrm{M_{*}} = 10^{11}\, \mathrm{M_{\odot}}$. 
At $\mathrm{M_{*}} > 10^{11}\, \mathrm{M_{\odot}}$, some ETGs also show half-light radii as large as $R_{\mathrm{e}}\sim20\, \mathrm{kpc}$. This scatter in the $R_{\mathrm{e}}$-$\mathrm{M_{*}}$ plane of ETGs slightly depends on the imaging band from which the $R_{\mathrm{e}}$ of the galaxies was calculated, being the largest in the $u$-band \citep{Lange2015}. 
In a recent work, \citet{Graaff2021} produced mock $r$-band images of galaxies in the EAGLE simulations to compare the mass-size relation of simulated and observed galaxies. They found that using the $r$-band sizes measured for the simulated galaxies improves the agreement between the simulated and observed mass-size relation of quiescent and star-forming galaxies.
Therefore, we conclude that our simulated and observed S0 galaxies are, overall, consistent with the $R_{\mathrm{e}}$-$\mathrm{M_{*}}$ relation from \citet{Lange2015} in Fig. \ref{fig:StellarMass_Re}. 

Fig. \ref{fig:StellarMass_Re} also shows that our simulated S0 galaxies cover overall a similar stellar mass range as our observed S0 galaxies from the SLUGGS survey. However, we note a bias of the S0 galaxies from the SLUGGS survey to higher stellar masses (i.e. $\gtrsim10^{11}\, \mathrm{M_{\odot}}$) than the S0 galaxies from the E-MOSAICS simulations. 
This is due to the fact that the EAGLE simulations do not contain many high-mass galaxies with $\mathrm{M_{*}}\gtrsim10^{11}\, \mathrm{M_{\odot}}$ and that the majority of these massive galaxies in EAGLE are ellipticals \citep{Correa2017}.
Finally, due to the limited volume of the E-MOSAICS simulations, the majority of the simulated galaxies are located in low-density environments (i.e. field and galaxy groups; see Sec.\ref{sec:emosaics_simulations}), similarly to the observed S0 galaxies from the SLUGGS survey. 
Therefore, the consistency between the environment of the S0 galaxies from the SLUGGS survey and E-MOSAICS simulations allow us to make comparisons between our simulated and observed S0 galaxy sample, when investigating how their kinematic properties relate to the merger histories of the galaxies. Additionally, in the next Sec. \ref{sec:1D_kinematic_profiles_GC}, we will consider the high-mass (i.e. $10.5 < \log( \mathrm{M}_{*}/\mathrm{M}_{\odot} ) < 11.6$) S0 galaxy sub-sample in the simulations for the comparison with the SLUGGS survey.

\subsection{The GC systems of the S0 galaxies in the E-MOSAICS simulations}
\label{sec:GC_systems_simulations}

\subsubsection{GC selection: stellar mass cut}
\label{sec:GC_stellar_mass_cut}
For each of the $50$ S0 galaxies from the E-MOSAICS simulations in the previous Sec. \ref{sec:S0_galaxy_sample_selection}, we select all of their star clusters that are gravitationally bound to the host galaxy and have stellar masses $\mathrm{M_{*}}>10^{5}\, \mathrm{M_{\odot}}$ at $z=0$. This stellar mass cut ensures that our kinematic results will not be influenced by the overabundant population of the low-mass (i.e. $\mathrm{M_{*}}\lesssim10^{5}\, \mathrm{M_{\odot}}$) star clusters that were not destroyed from the interaction with the cold ISM (see Sec. \ref{sec:emosaics_simulations}).
The same stellar mass cut (i.e. $\mathrm{M_{*}}>10^{5}\, \mathrm{M_{\odot}}$) was also used by \citet{Kruijssen2019}, as they found that the GC mass function of their sample of simulated galaxies was consistent with that of the Milky Way after removing the lower mass GCs.

\subsubsection{GC selection: magnitude cut}
\label{sec:GC_magnitude_cut}
Fig. \ref{fig:luminosity_function_GCs} shows the $V$-band GC luminosity function of one S0 galaxy from our selected sample, as an example. 
The grey histogram shows the $V$-band GC luminosity function prior to applying any stellar mass cut to the GC system, while the black histogram is obtained after removing the low-mass GCs with $\mathrm{M_{*}}\lesssim10^{5}\, \mathrm{M_{\odot}}$. 
We notice that the $V$-band GC luminosity function resembles more the shape of a Gaussian with a luminosity peak at $\mathrm{M}_{V}\simeq-7$ when selecting only the GCs with $\mathrm{M_{*}}>10^{5}\, \mathrm{M_{\odot}}$ for typically all the GC systems of the S0 galaxies in our sample.   
This is more consistent with the $V$-band GC luminosity function of observed galaxies, which is characterized by a Gaussian-like shape with luminosity peak typically located at $\mathrm{M}_{V}\simeq-7.5$ (e.g. \citealt{Forbes2018}). 
The non Gaussian-like shape with prominent peak below $M_{\mathrm{V}}\simeq-7$ of the GC systems of our S0 galaxies seems likely to be the result of the overabundant population of the low-mass ($\mathrm{M_{*}}\lesssim10^{5}\, \mathrm{M_{\odot}}$) star clusters that are not being efficiently disrupted in the simulations (see Sec. \ref{sec:emosaics_simulations}), as shown in Fig. \ref{fig:luminosity_function_GCs} prior to applying any GC stellar mass cut. For this reason, we exclude these low-mass star clusters in order to avoid significant contamination from these objects.
The red vertical dashed line shows the $V$-band magnitude of the brightest GC that is included in the system (i.e. $\mathrm{M}_{V}^{\mathrm{\omega Cen}}=-10.5$). This luminosity limit is chosen to include up to the $V$-band magnitude of $\omega$-Centauri ($\mathrm{M}_{V}^{\mathrm{\omega Cen}}\simeq-10.3$), which is the most luminous GC of the Milky Way, consistent with the observations of the GC systems from the SLUGGS survey \citep{Pota2013}.

\begin{figure}\centering
    \includegraphics[width=0.5\textwidth]{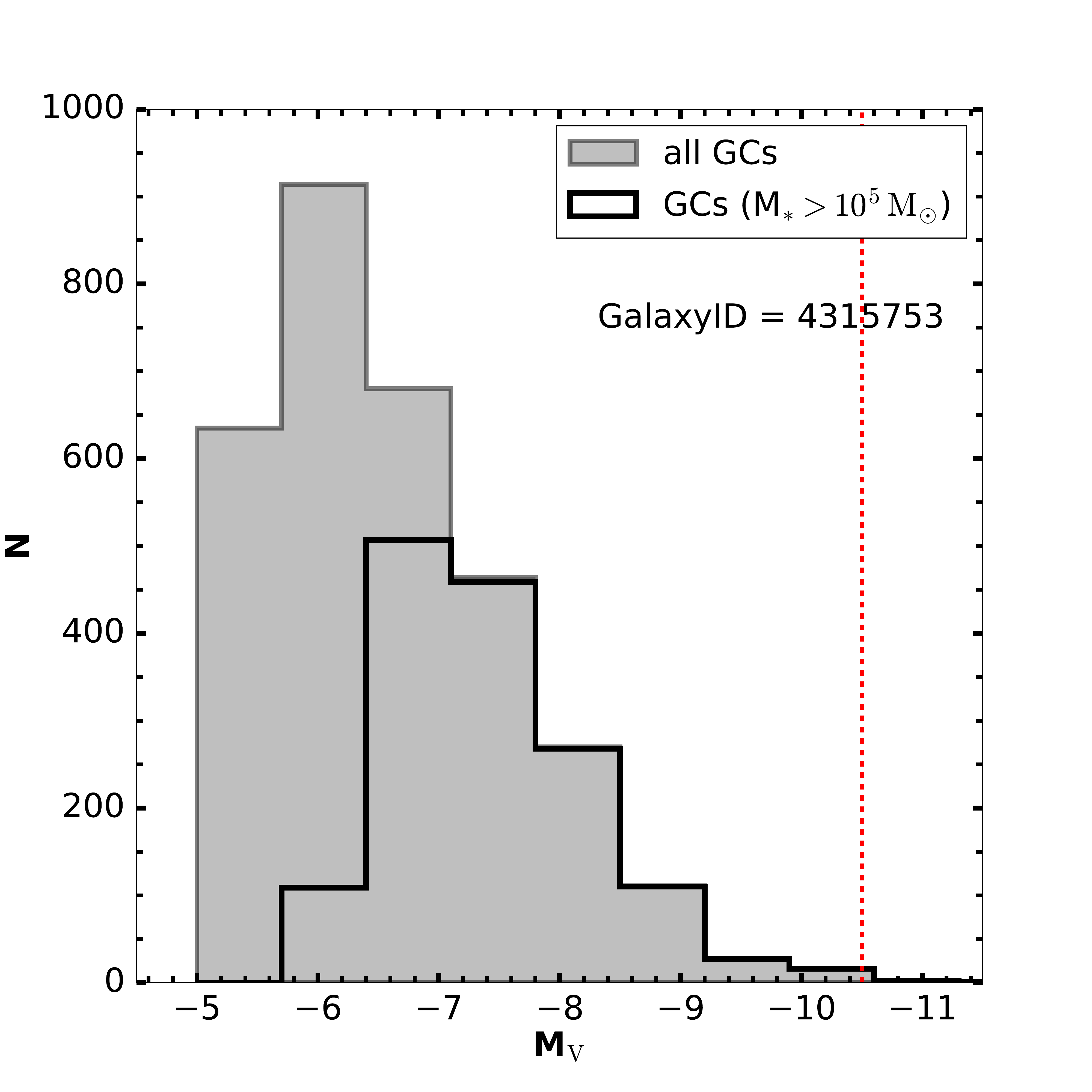}
    \caption{The $V$-band GC luminosity function of one S0 galaxy from our selected sample. The grey histogram is obtained without adopting any stellar mass cut. The black histogram is obtained when considering only the GCs with $\mathrm{M_{*}}>10^{5}\, \mathrm{M_{\odot}}$. The vertical red dashed line represents the $V$-band magnitude of the most luminous GCs included in our GC systems and is chosen to include up to $\omega$-Cen-like systems (i.e. $\mathrm{M}_{V}^{\mathrm{\omega Cen}}\simeq-10.3$), which are the most luminous in the Milky Way.}
    \label{fig:luminosity_function_GCs}
\end{figure}

\subsubsection{GC selection: age cut}
\label{sec:GC_age_cut}
Since, in this work, we are mainly interested in studying the kinematic properties of the GCs (i.e. old star clusters), we then select only those star clusters with ages above $8\, \mathrm{Gyr}$ in each galaxy. 
In the Milky Way, an age cut above $8\, \mathrm{Gyr}$ would only exclude three of its GCs, as the bulk of its GCs are older than $\sim10\, \mathrm{Gyr}$ (see figure 1 in \citealt{Forbes2020}).
While galaxies may have a larger spread in the ages of their GCs with some being $<5\, \mathrm{Gyr}$ old (see NGC 3377 in \citealt{Usher2019}), the bulk of the GCs has typically old ages ($>8\, \mathrm{Gyr}$) in the remaining galaxies \citep{Usher2019}.
Overall, we find that the fraction of the star clusters with ages below $8\, \mathrm{Gyr}$ accounts for $<30\%$ of the total GC population in $62\%$ of our simulated S0 galaxies. Therefore, the majority of our galaxies are dominated by an old GC population which is the main focus of this work.

\subsubsection{GC selection: metallicity cut}
\label{sec:GC_metallicity_cut}
Following \citet{Kruijssen2019}, we remove the most metal-poor GCs with $\mathrm{[Fe/H]} < -2.5$. Specifically, \citet{Kruijssen2019} restricted the analysis to the GC metallicity range $-2.5 < \mathrm{[Fe/H]} < -0.5$ to mimic the metallicity distributions spanned by the observed GCs of the Milky Way and M31, while reducing the contamination from the overproduced metal-rich GCs (see Sec. \ref{sec:emosaics_simulations}). In this work, the adopted low metallicity cut is also observationally motivated. In fact, previous spectroscopy analyses have measured the metallicity of the GCs of the observed ETGs from the SLUGGS survey and found that all the GCs had metallicity $\mathrm{[Fe/H]} > -2.5$ \citep{Pastorello2015,Usher2019}. Additionally, we find that the metal-poor GCs with $\mathrm{[Fe/H]} < -2.5$ are very low in numbers accounting for less than $10\%$ of the total GC population in each one of our simulated S0 galaxies, in accordance with the results discussed in the previous works of \citet{Pfeffer2018,Kruijssen2019}. In fact, previous works also predict the existence of a metallicity floor at $\mathrm{[Fe/H]}=-2.5$ as a result of the galaxy mass-metallicity relation, below which high-redshift low-mass galaxies could not form GCs that were massive enough (i.e. $\mathrm{M_{*}}>10^{5}\, \mathrm{M_{\odot}}$) to survive until the present-day \citep{Beasley2019,Kruijssen2019b}. Additionally, these low-mass galaxies are also not resolved in the E-MOSAICS simulations, since they are expected to have stellar mass $\mathrm{M_{*}}\sim10^{5}\, \mathrm{M_{\odot}}$ at $\mathrm{[Fe/H]}=-3$ \citep{Kruijssen2019b}, which is the particle resolution of the simulations. For this reason, metal-poor stars with $\mathrm{[Fe/H]}<-2.5$ are not reliable as they are not modelled in the simulations. However, these results do not rule out the existence of GCs with metallicity $\mathrm{[Fe/H]}<-2.5$, as suggested by the discovery of a very metal-poor GC with $\mathrm{[Fe/H]}=-2.9$ in M31 \citep{Larsen2020}. 
Therefore, while GCs with $\mathrm{[Fe/H]}<-2.5$ may likely exist in the local Universe, they are expected to be rare and low in numbers as seen in the simulations. For this reason, their removal should not significantly influence the results in this work. 

On the other hand, we do not apply any cut to the overproduced GC population in the metallicity range $-1.0 < \mathrm{[Fe/H]} < 0.0$, since our tests show that the inclusion of this GC population does not significantly impact the kinematic properties of the GC systems of our simulated S0 galaxies. Additionally, recent results have found that the high metallicity GCs are not overproduced in massive galaxies (i.e. $\mathrm{M_{*}}\gtrsim10^{11}\mathrm{M_{\odot}}$), but mainly in galaxies with stellar masses between $10^{9.5}\, \mathrm{M_\odot} < \mathrm{M_{*}} < 10^{10.5}\mathrm{M_{\odot}}$ (Pfeffer et al., in prep.). 
Overall, the metallicity range $-2.5 < \mathrm{[Fe/H]} < 0.5$ of the GC systems of our sample of selected S0 galaxies from the E-MOSAICS simulations is consistent with the metallicity measured for the GC systems of the observed ETGs from the SLUGGS survey (e.g. \citealt{Pastorello2015,Usher2019}), which we have studied in our previous work \citep{Dolfi2021}.

\subsubsection{GC selection: surface density cut}
\label{sec:GC_density_cut}
We find that a very small fraction of our S0 galaxies are characterized by very extended GC systems that reach out to galactocentric radii as large as $\sim0.5$-$1\, \mathrm{Mpc}$. 
Most of these galaxies also show evidence of shells or tidal tails from the visual inspection of their images with Mpc-scale field-of-views (FoV), suggesting that they may have recently experienced interactions. Therefore, we apply an observationally motivated cut to the surface density profile of the GC systems of our S0 galaxies, in order to avoid contamination from potential outliers GCs that may belong to the outer regions of the satellite galaxies orbiting around the central galaxy. \citet{Pota2013} found that the surface density profiles of the red and blue GCs of the ETGs from the SLUGGS survey reach levels as low as $\log(\Sigma_{\mathrm{GCs}}/\mathrm{arcmin^{2}}) \simeq -2$, corresponding to $\log(\Sigma_{\mathrm{GCs}}/\mathrm{kpc^{2}}) \simeq -2$. 
After we apply the same density cut to the GC surface density profiles of all our simulated S0 galaxies, the GC systems extend out to galactocentric radii ranging between $\sim10$-$200\, \mathrm{kpc}$. This is overall consistent with the radial extension of the observed S0 galaxies from the SLUGGS survey, with the most extended galaxy reaching out to $\sim100\, \mathrm{kpc}$


\begin{figure*}
\centering
\Large 
\textbf{{\it \textbf{Aligned}} galaxies \\ $10 < \log_{10} (\mathrm{M}/\mathrm{M}_{\odot}) < 10.5$}\par\medskip
    \includegraphics[width=0.29\textwidth]{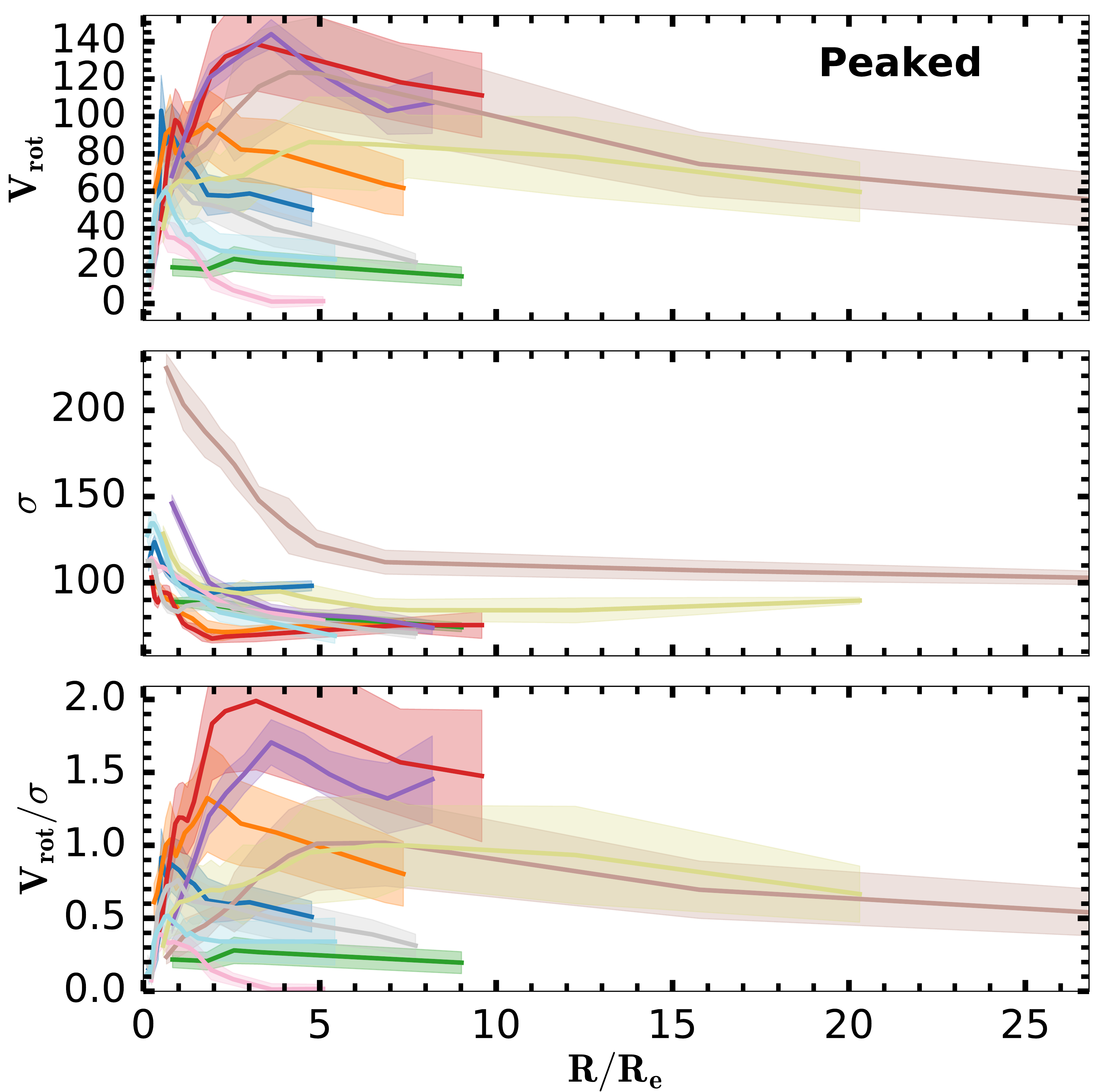}
    \includegraphics[width=0.29\textwidth]{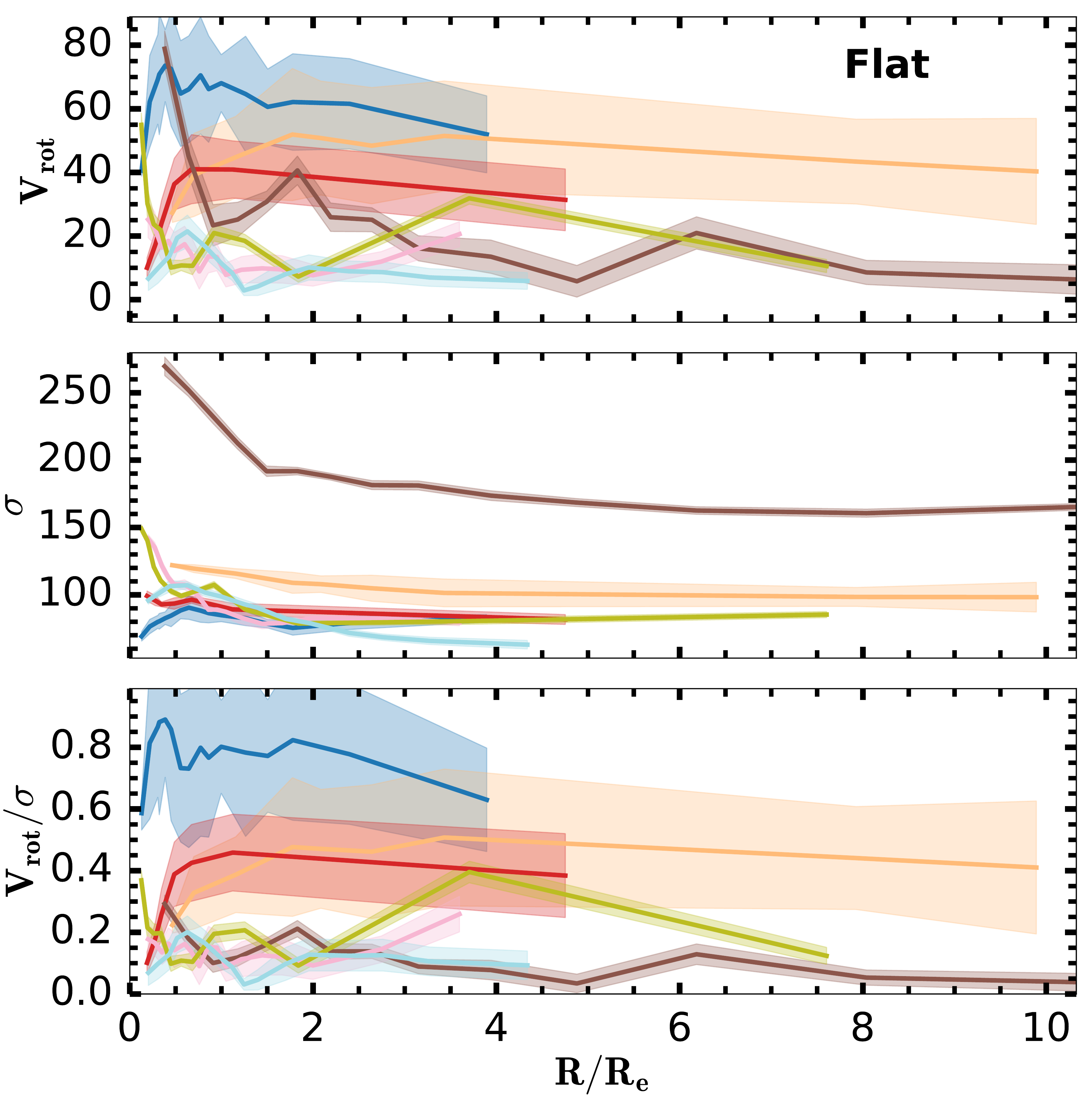}
    \includegraphics[width=0.29\textwidth]{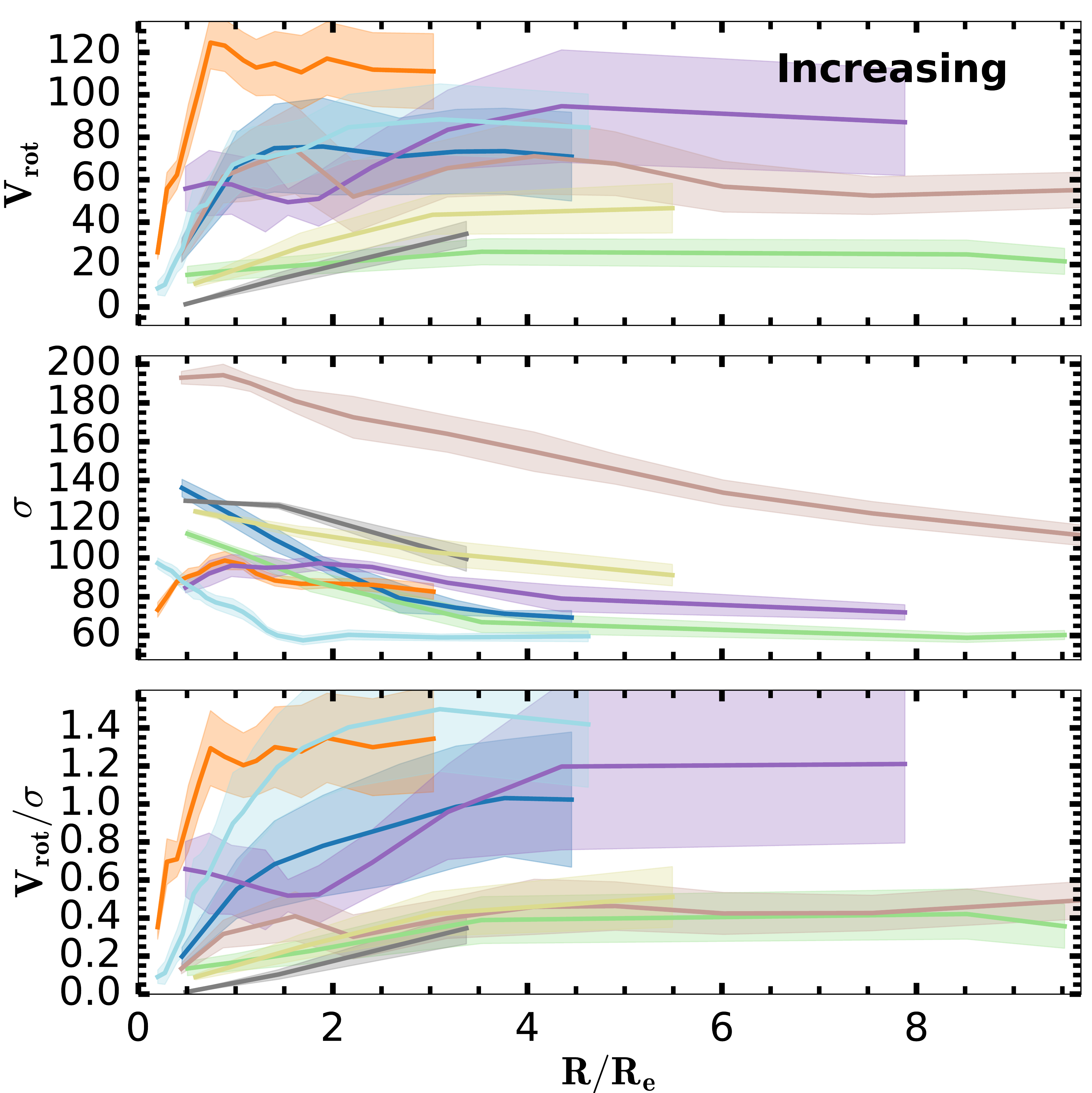}
\caption{\textit{Continued}}
\end{figure*}
\begin{figure*}
\ContinuedFloat
\centering
\Large 
\textbf{$10.5 < \log_{10} (\mathrm{M}/\mathrm{M}_{\odot}) < 11.5$}\par\medskip
    \includegraphics[width=0.29\textwidth]{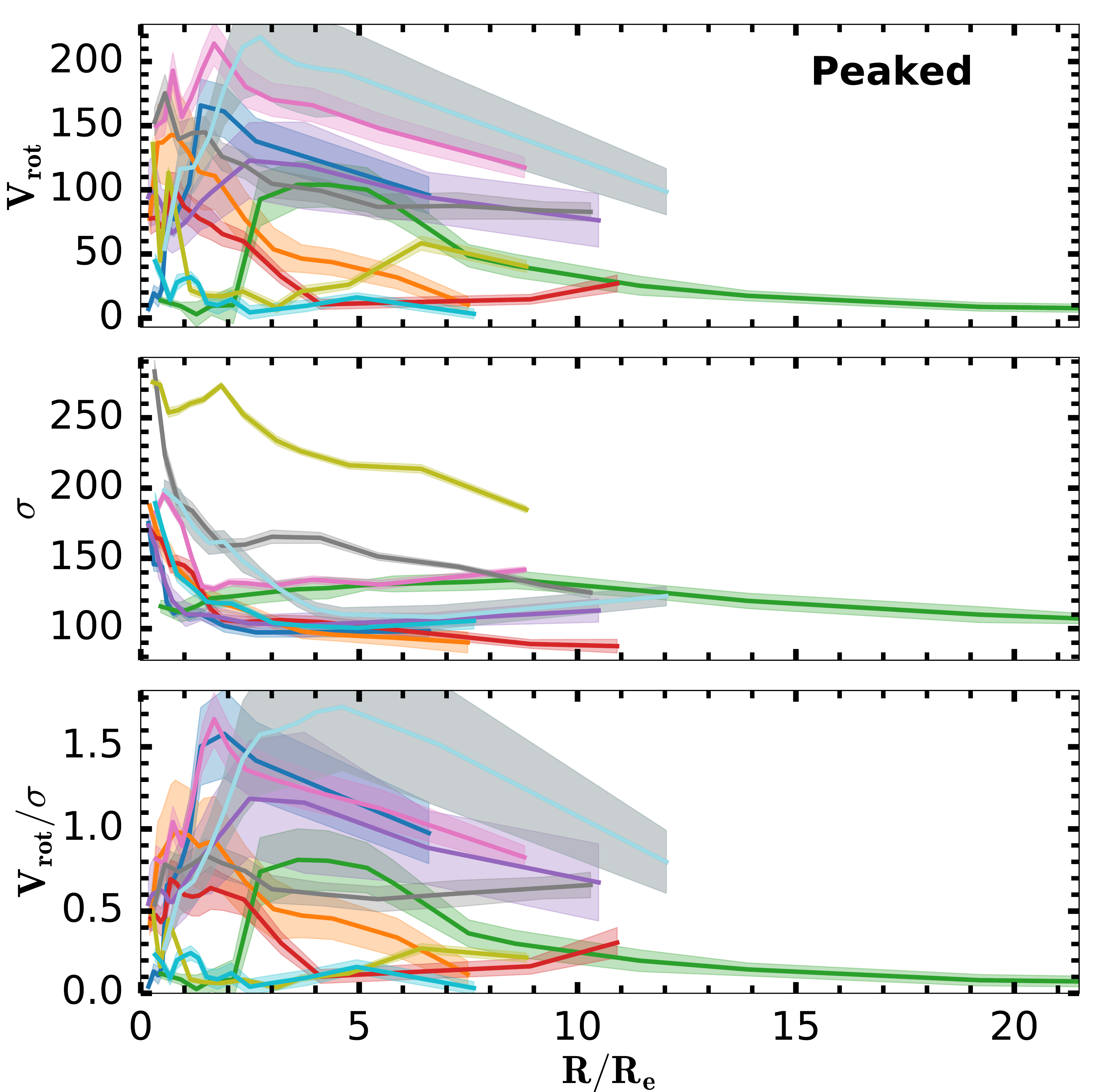}
    \includegraphics[width=0.29\textwidth]{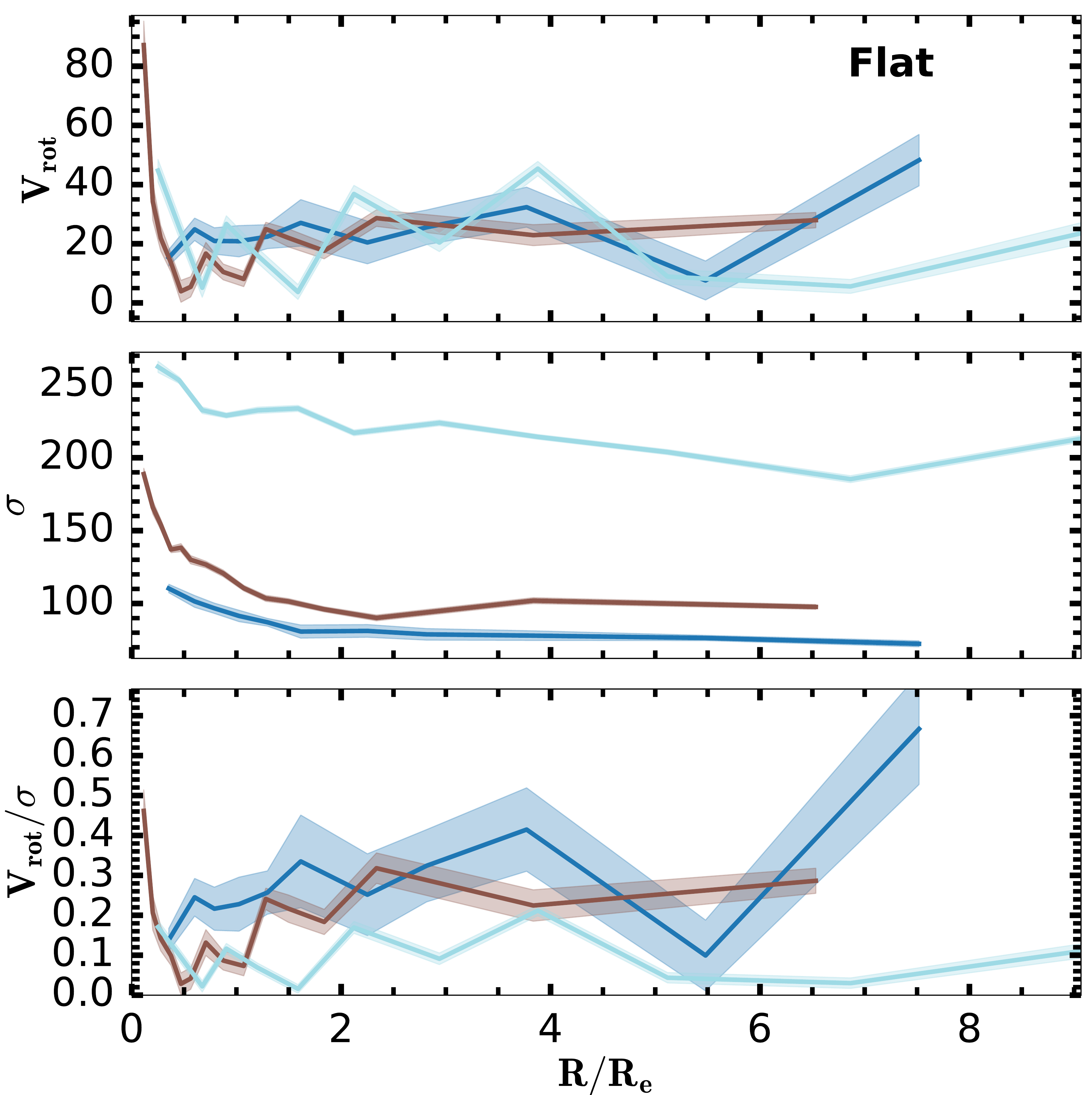} 
    \includegraphics[width=0.29\textwidth]{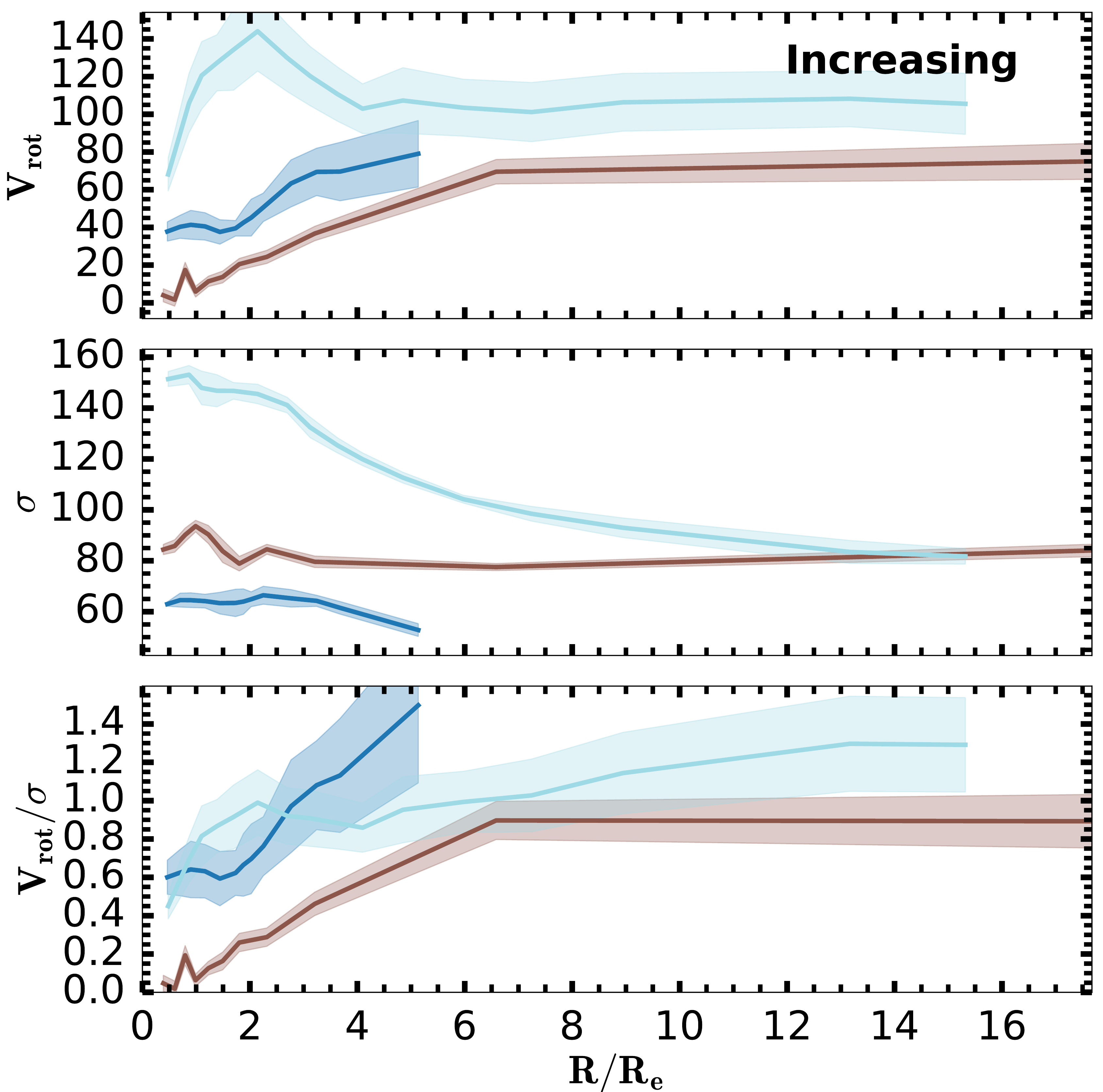}
\end{figure*}
\begin{figure*}
\ContinuedFloat
\centering
\Large 
\textbf{{\it \textbf{Misaligned}} galaxies}\par\medskip
    \includegraphics[width=0.29\textwidth]{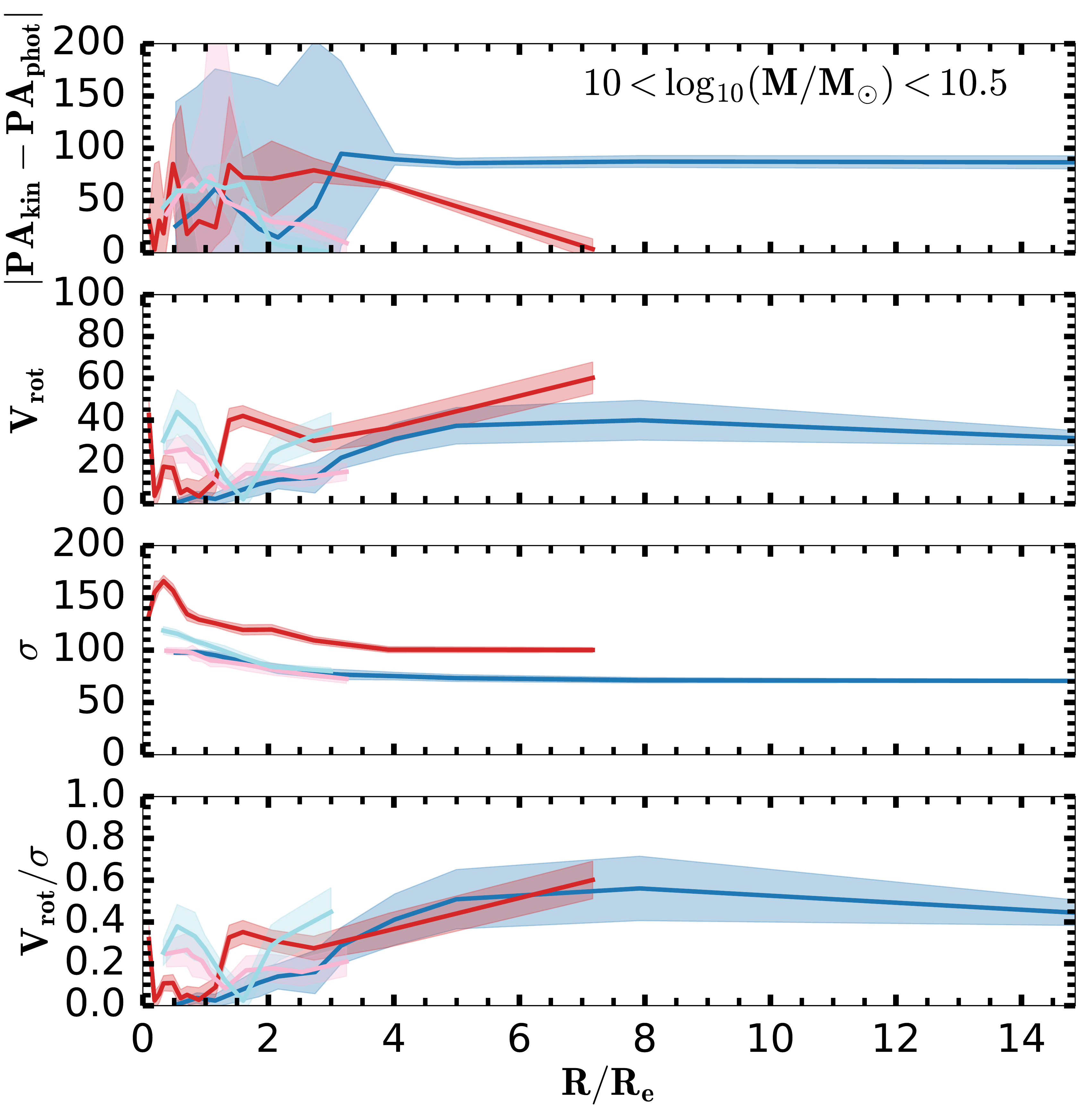}
    \includegraphics[width=0.29\textwidth]{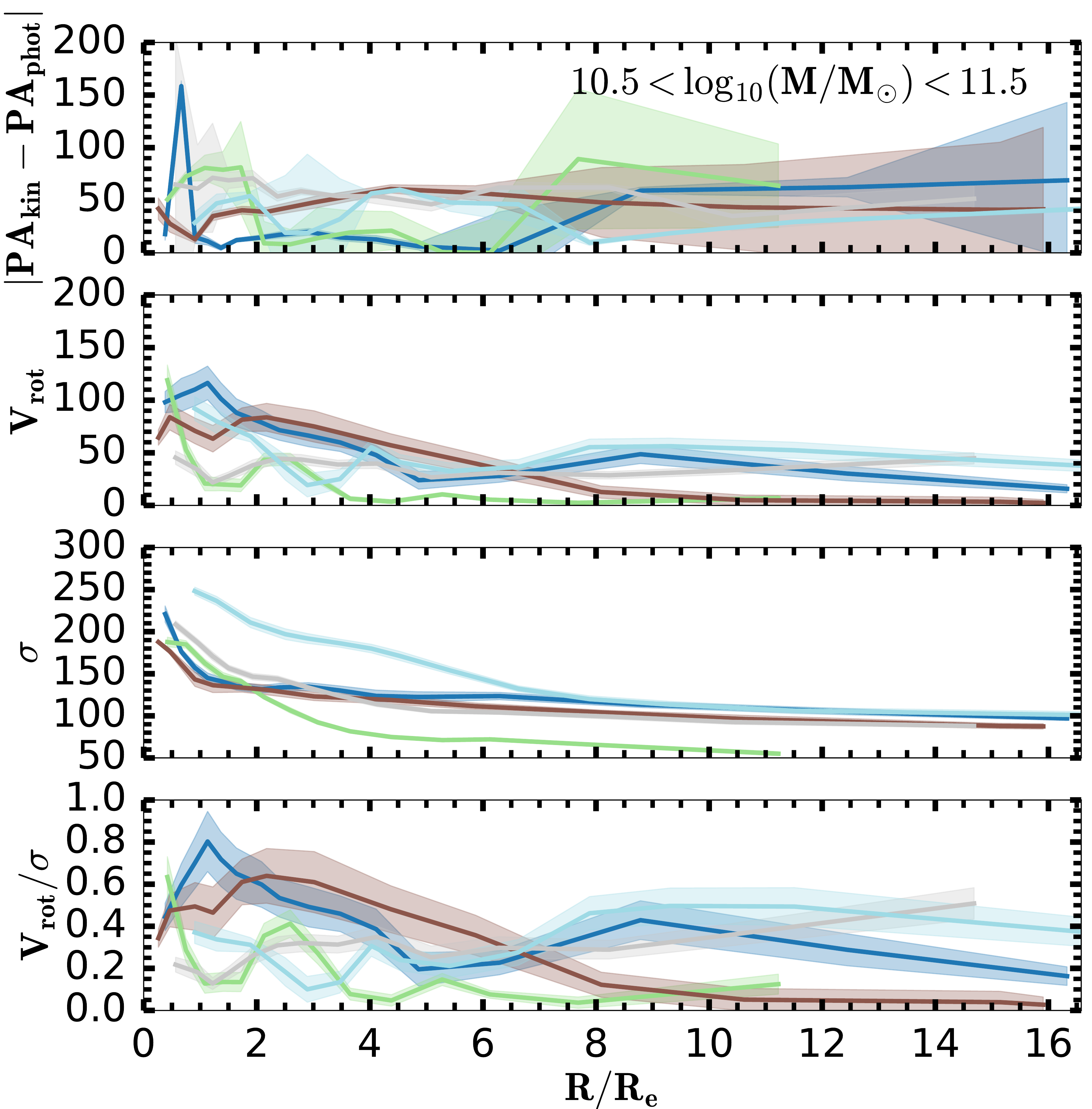}
    \caption{1D kinematic profiles of the GC systems of the $41$ \textit{aligned} (top) and $9$ \textit{misaligned} (bottom) galaxies. For the \textit{aligned} galaxies, we show the 1D $V_{\mathrm{rot}}$, $\sigma$ and $V_{\mathrm{rot}}/\sigma$ kinematic profiles of the GCs from the top to the bottom panel of each sub-figure, respectively. For the \textit{misaligned} galaxies, we also show the absolute difference between the 1D $\mathrm{PA}_{\mathrm{kin}}$ profile of the GCs and the $\mathrm{PA}_{\mathrm{phot}}$ of the galaxy in the top panel of each sub-figure. The shaded areas represent the $1\sigma$ errors calculated as described in Sec. \ref{sec:kinematic_analysis_results}. We separate both the \textit{aligned} and \textit{misaligned} galaxies into the low ($10.0 < \log( \mathrm{M}_{*}/\mathrm{M}_{\odot} ) < 10.5$) and high ($10.5 < \log( \mathrm{M}_{*}/\mathrm{M}_{\odot} ) < 11.6$) stellar mass bin, where the high-mass bin covers the same range of stellar masses as the observed S0 galaxies from the SLUGGS survey studied in \citet{Dolfi2021}. From the $V_{\mathrm{rot}}/\sigma$ profiles of the GCs, we classify our \textit{aligned} S0 galaxies as \textit{peaked} ($49\%$; left panel), \textit{flat} ($24\%$; middle panel) and \textit{increasing} ($27\%$; right panel). We find that the three different $V_{\mathrm{rot}}/\sigma$ profile shapes equally dominate in the low stellar mass bin with similar numbers, i.e. $\sim7$-$10$ galaxies in each category. On the other hand, the galaxies with \textit{peaked} $V_{\mathrm{rot}}/\sigma$ profiles dominate in the high stellar mass bin with a total of $10$, as opposed to the $3$ galaxies in each category of \textit{flat} or \textit{increasing} $V_{\mathrm{rot}}/\sigma$ profiles.}
\label{fig:GC_kinematic_profiles}
\end{figure*}

\section{Kinematic Analysis and Results}
\label{sec:kinematic_analysis_results}
In this section, we calculate the 1D kinematic profiles of the GC systems of our selected sample of $50$ S0 galaxies from the E-MOSAICS simulations after applying the GC selection criteria described in Sec. \ref{sec:GC_systems_simulations}.
Here, all the galaxies are rotated such that the angular momentum vector of the stars aligns with the $z$-axis (i.e. the disk lies on the $xy$-plane). 
We carry out the kinematic analysis focusing on the edge-on projection of the galaxy. We note that the viewing angle may have an influence for the comparison with the observations from the SLUGGS survey that are characterized by random viewing angles as well as for the comparison with other simulations.

To calculate the 1D kinematic profiles of the GC systems of our simulated S0 galaxies, we use the \texttt{kinemetry}\footnote{\url{http://davor.krajnovic.org/idl/}} method developed by \citet{Krajnovic2011}. Within this method, we bin our data in elliptical annuli centered on the galaxy and we recover the best-fitting moments of the line-of-sight-velocity-distribution (LOSVD) along each annulus. 
The best-fitting LOSVD moments are calculated by performing a least-square minimization between the disk model, described by equations 3 and 4 shown in \citet{Dolfi2020}, and our kinematic data. 
The details of the application of this method to sparse and non-homogeneous kinematic datasets are given in \citet{Proctor2009,Bellstedt2017,Dolfi2020}, while the details describing the $\chi^{2}$-minimization are given in \citet{Foster2011,Foster2016}.  

As previously done in \citet{Dolfi2020,Dolfi2021}, we calculate the first two moments of the LOSVD of the GC systems of our S0 galaxies, i.e. rotation velocity ($V_{\mathrm{rot}}$) and velocity dispersion ($\sigma$) profiles, from which we derive the corresponding $V_{\mathrm{rot}}/\sigma$ profiles. The velocity dispersion of the GCs is calculated by performing a nearest neighbour binning, as described in our previous work \citep{Dolfi2020}. We also fit for both the kinematic position angle ($\mathrm{PA}_{\mathrm{kin}}$) and kinematic axial-ratio ($\mathrm{q}_{\mathrm{kin}}$), where they are well constrained and do not excessively vary as a function of the galactocentric radius. If this is not the case, then we fix the $\mathrm{PA}_{\mathrm{kin}}$ to the photometric position angle ($\mathrm{PA}_{\mathrm{phot}}$)\footnote{$\mathrm{PA}_{\mathrm{phot}}$ is estimated from the edge-on view of the images of the S0 galaxies shown in Fig. \ref{fig:mock_S0_images}. It is measured from North towards East for consistency with the $\mathrm{PA}_{\mathrm{kin}}$.} of the galaxy and we fix the $\mathrm{q}_{\mathrm{kin}}$ to the mean value obtained from the unconstrained fit.
The $1\sigma$ errors on the $\mathrm{PA}_{\mathrm{kin}}$ and $\mathrm{q}_{\mathrm{kin}}$ profiles are calculated by running \texttt{kinemetry} on $100$ bootstrapped samples obtained by sampling with replacement the original dataset. On the other hand, the $1\sigma$ errors on the $V_{\mathrm{rot}}$ and $1\sigma$ profiles are the standard errors of the mean, calculated from the standard deviation of the velocity and velocity dispersion measurements of the data points in each elliptical annulus divided by the square root of the number of points in each bin.

\subsection{1D kinematic profiles of the GC systems of the S0 galaxies in the E-MOSAICS simulations}
\label{sec:1D_kinematic_profiles_GC}
Fig. \ref{fig:GC_kinematic_profiles} shows the 1D kinematic profiles of the GC systems of our selected sample of $50$ S0 galaxies from the E-MOSAICS simulations. 
Since the colour-magnitude diagrams of the GC systems do not show evidence of any colour bimodality, as it was the case for the S0 galaxies from the SLUGGS survey \citep{Pota2013}, we do not split here between the red and blue GC sub-populations. 
However, in the next Sec. \ref{sec:GC_subpopulations}, we will investigate whether the metal-rich and metal-poor GC sub-populations show different kinematic behaviours by adopting the same fixed metallicity cut at $\mathrm{[Fe/H]}=-1$ for all the GC systems of our S0 galaxies.

We classify the galaxies as \textit{aligned} (Sec. \ref{sec:aligned_galaxies}) if the GCs are rotating along the photometric major axis ($\mathrm{PA}_{\mathrm{phot}}$) of the galaxy within the $1\sigma$ errors. Otherwise, we classify the galaxies as \textit{misaligned} (Sec. \ref{sec:misaligned_galaxies}).

Finally, we divide both \textit{aligned} and \textit{misaligned} galaxies into two stellar mass bins: low-mass $10 < \log( \mathrm{M}_{*}/\mathrm{M}_{\odot} ) < 10.5$ and high-mass $10.5 < \log( \mathrm{M}_{*}/\mathrm{M}_{\odot} ) < 11.6$. The high stellar mass bin is consistent with the stellar mass range covered by all the observed S0 galaxies from the SLUGGS survey that we have studied in \citet{Dolfi2021}, with the exception of one very low-mass S0 (i.e. NGC 7457). 

\subsubsection{\textit{Aligned} galaxies}
\label{sec:aligned_galaxies}
Following the simulations from \citet{Schulze2020}, we classify the $V_{\mathrm{rot}}/\sigma$ profiles of the GCs of the \textit{aligned} galaxies (see Fig. \ref{fig:GC_kinematic_profiles}, top) as \textit{peaked}, \textit{flat} or \textit{increasing} based on the $V_{\mathrm{rot}}/\sigma$ gradient calculated between the outer and inner regions of the GC $V_{\mathrm{rot}}/\sigma$ profiles. 
Specifically, we calculate the $V_{\mathrm{rot}}/\sigma$ gradient (i.e. $\Delta V_{\mathrm{rot}}/\sigma$) as the difference between the mean value of the $V_{\mathrm{rot}}/\sigma$ in the outer (i.e. $2.0 < R/R_{\mathrm{e}} < 3.5$) and inner (i.e $0.5 < R/R_{\mathrm{e}} < 2.0$) radial bins. 
We classify the galaxy as \textit{peaked}, \textit{flat} or \textit{increasing} if the $V_{\mathrm{rot}}/\sigma$ gradient is $\Delta V_{\mathrm{rot}}/\sigma < -0.04$, $-0.04 < \Delta V_{\mathrm{rot}}/\sigma < 0.04$ or $\Delta V_{\mathrm{rot}}/\sigma > 0.04$, respectively, as defined by \citet{Schulze2020}.

However, we note here that the stellar $V_{\mathrm{rot}}/\sigma$ profiles of \citet{Schulze2020} do not extend beyond $\sim5\, R_{\mathrm{e}}$. 
On the other hand, the $V_{\mathrm{rot}}/\sigma$ profiles of the GC systems of our simulated S0s also typically extend out to $\sim10\, R_{\mathrm{e}}$, with $10$ galaxies also reaching beyond $10\, R_{\mathrm{e}}$. Additionally, the \textit{peaked} $V_{\mathrm{rot}}/\sigma$ profiles of the GC systems of our S0 galaxies can have rotational velocities that reach their peak value at larger radii (i.e. beyond $2\, R_{\mathrm{e}}$) than found by \citet{Schulze2020} for their \textit{peaked} galaxies. Indeed, some of our S0 galaxies show $V_{\mathrm{rot}}/\sigma$ profiles peaking at $\sim5\, \mathrm{R_{\mathrm{e}}}$ (or beyond) and decreasing outwardly, so they would be classified as \textit{increasing} according to the definition by \citet{Schulze2020} out to $\sim5\, R_{\mathrm{e}}$. Therefore, for these galaxies whose $V_{\mathrm{rot}}/\sigma$ profiles extend beyond $\sim5\, R_{\mathrm{e}}$, we also look at their kinematic behaviour at larger radii in order to properly classify them into one of the three categories from \citet{Schulze2020}.

Fig. \ref{fig:GC_kinematic_profiles} (top) shows the $41$ \textit{aligned} galaxies ($82\%$ of total) split into the \textit{peaked}, \textit{flat} and \textit{increasing} $V_{\mathrm{rot}}/\sigma$ profile shapes from the top to the bottom panels, respectively. Each sub-figure of Fig. \ref{fig:GC_kinematic_profiles} (top) shows the 1D $V_{\mathrm{rot}}$, $\sigma$ and $V_{\mathrm{rot}}/\sigma$ profiles of GC systems of the S0 galaxies in the low ($10 < \log( \mathrm{M}_{*}/\mathrm{M}_{\odot} ) < 10.5$; left column) and high ($10.5 < \log( \mathrm{M}_{*}/\mathrm{M}_{\odot} ) < 11.6$; right column) stellar mass bins.

We find that that the \textit{peaked}, \textit{flat} and \textit{increasing} $V_{\mathrm{rot}}/\sigma$ profile shapes occur in similar numbers, i.e. $10$, $7$ and $8$, respectively, in the low stellar mass bin, while the \textit{peaked} $V_{\mathrm{rot}}/\sigma$ profile shape is the dominant one in the high stellar mass bin. In fact, $10$ galaxies show a \textit{peaked} $V_{\mathrm{rot}}/\sigma$ profile shape as opposed to the \textit{flat} and \textit{increasing} $V_{\mathrm{rot}}/\sigma$ profile shapes that contain only $3$ galaxies each at the high stellar masses.
We also investigated the kinematics of the GC systems of the central and satellite galaxies. However, we did not see any clear differences, unlike previous works that found that S0 galaxies in clusters are typically more rotationally supported than S0 galaxies in the field \citep{Deeley2020,Coccato2020}. We suggest that the lack of clear differences between the central and satellite galaxies in this work may be due to the fact that the E-MOSAICS simulations do not include the highest density environments of clusters, but they are limited to the field and group environments (i.e. Fornax cluster-like), as mentioned in Sec. \ref{sec:emosaics_simulations}.

The comparison with \citet{Schulze2020} simulations suggests that low-mass S0 galaxies (Fig. \ref{fig:GC_kinematic_profiles}, left-hand side) can form through a range of different merger and accretion histories. Specifically, galaxies with \textit{peaked} $V_{\mathrm{rot}}/\sigma$ profiles are expected to have typically formed through late (i.e. $z<1$) mini mergers that mainly influence the kinematic properties of the outer regions of the galaxies without destroying their central disk-like kinematics. On the other hand, galaxies with \textit{flat} $V_{\mathrm{rot}}/\sigma$ profiles are more likely to have experienced a late major merger that destroyed the central disk-like kinematics of the galaxies. This major merger may have also occurred in the form of multiple minor mergers, as a single major merger event is typically expected to spin up the rotation of the GCs at large radii \citep{Bekki2005}.
Finally, galaxies with \textit{increasing} $V_{\mathrm{rot}}/\sigma$ profiles are suspected to have likely formed through a late gas-rich major merger that may have re-formed the central kinematically cold disk component of the galaxies, thus enhancing the rotation of the GCs out large radii \citep{Bekki2005}. Using the E-MOSAICS simulations, \citet{Gomez2021} also found that a late major merger is likely to produce more eccentric spatial distributions of the metal-rich GCs compared to those of the metal-poor GCs (i.e. \textit{increasing} $V_{\mathrm{rot}}/\sigma$ profiles). On the other hand, galaxies characterized by a larger spread in GC eccentricities (low $V_{\mathrm{rot}}/\sigma$ profiles) are likely to have assembled $50\%$ of their mass at earlier times \citep{Gomez2021}, consistent with the assembly history predicted for the \textit{peaked} galaxies (i.e. two-phase formation scenario)

On the other hand, the high-mass S0 galaxies (Fig. \ref{fig:GC_kinematic_profiles}, right-hand side) seem to be mostly dominated by a formation history from late mini mergers, as shown by the larger number of the \textit{peaked} galaxies than the \textit{flat} and \textit{increasing} galaxies in the high stellar mass bin.

In \citet{Dolfi2021}, we have studied the kinematic profiles of an observed sample of $9$ selected S0 galaxies from the SLUGGS survey extending out to $\sim5\, R_{\mathrm{e}}$ by combining the kinematic of the stars, GCs and planetary nebulae (PNe). As mentioned above, the $8$ out of the $9$ S0s from the SLUGGS survey cover similar stellar masses to the high-mass S0s from the simulations, i.e. $10.5 < \log( \mathrm{M}_{*}/\mathrm{M}_{\odot} ) < 11.6$.  
From the observations, we find that $6$ out of $9$ S0 galaxies ($67\%$) have GCs and PNe with consistent kinematics with respect to the underlying stars that rotate along the $\mathrm{PA}_{\mathrm{phot}}$ of the galaxies (and, thus, are referred to as \textit{aligned} galaxies)

Overall, the kinematic results of the GCs of these \textit{aligned} galaxies from the SLUGGS survey seem to be consistent with those of the simulated S0 galaxy sample from E-MOSAICS. Specifically, in the range of stellar masses in common between the observations and simulations (i.e. $10.5 < \log( \mathrm{M}_{*}/\mathrm{M}_{\odot} ) < 11.6$), we find that galaxies with \textit{peaked} $V_{\mathrm{rot}}/\sigma$ profiles dominate with a $\sim62\%$ fraction in the simulations and $\sim60\%$ fraction in the observations. These are followed by the galaxies with \textit{flat} $V_{\mathrm{rot}}/\sigma$ profiles that occur with a $\sim19\%$ fraction in the simulations and $\sim40\%$ fraction in the observations. 
We do not find evidence of any \textit{increasing} $V_{\mathrm{rot}}/\sigma$ profile shapes in the observations, but this could simply be a result of sample selection effects or low number statistics. However, according to the simulations, galaxies with \textit{increasing} $V_{\mathrm{rot}}/\sigma$ profile shapes should occur with similar frequencies as the galaxies with \textit{flat} $V_{\mathrm{rot}}/\sigma$ profile shapes, i.e. $\sim19\%$.

\begin{figure*}
\centering
\huge
\textbf{{\it \textbf{Peaked}} galaxies}\par\medskip
\begin{subfigure}[b]{0.24\textwidth}
\centering 
    \includegraphics[width=\textwidth]{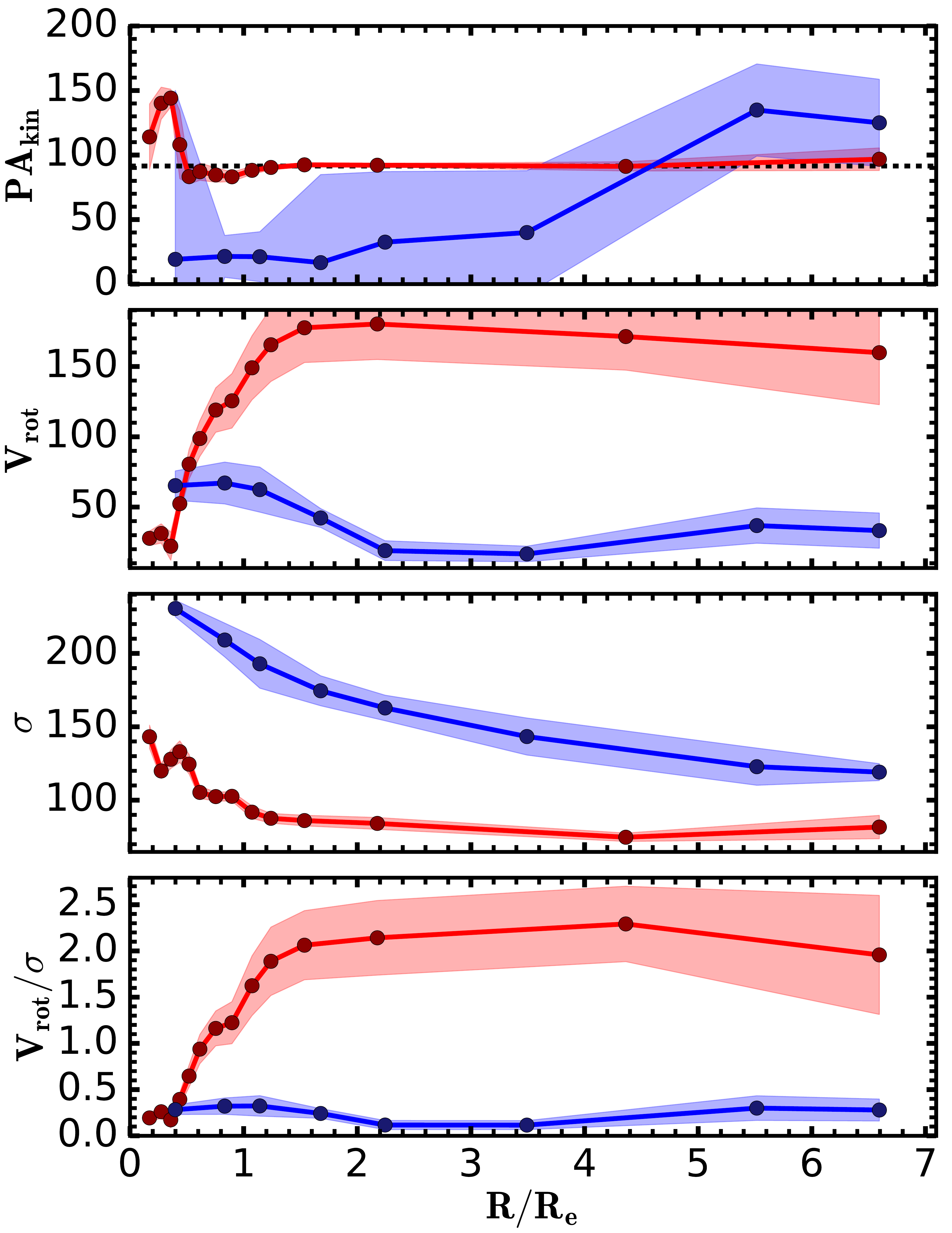}
\caption{GalaxyID = 4315753}
\label{fig:a}
\end{subfigure}
\begin{subfigure}[b]{0.24\textwidth}
\centering 
    \includegraphics[width=\textwidth]{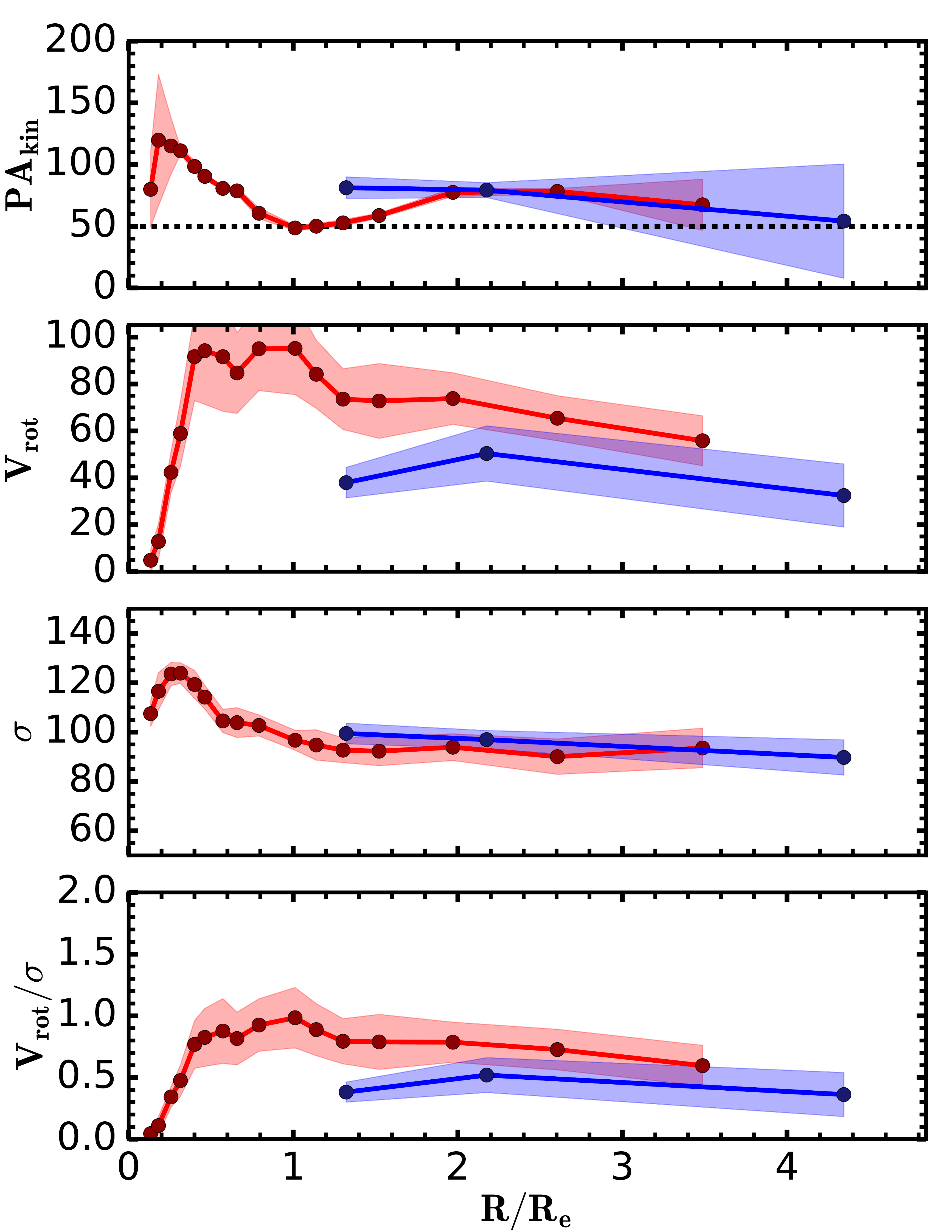}
\caption{GalaxyID = 8050835}
\label{fig:b}
\end{subfigure}
\begin{subfigure}[b]{0.24\textwidth}
\centering 
    \includegraphics[width=\textwidth]{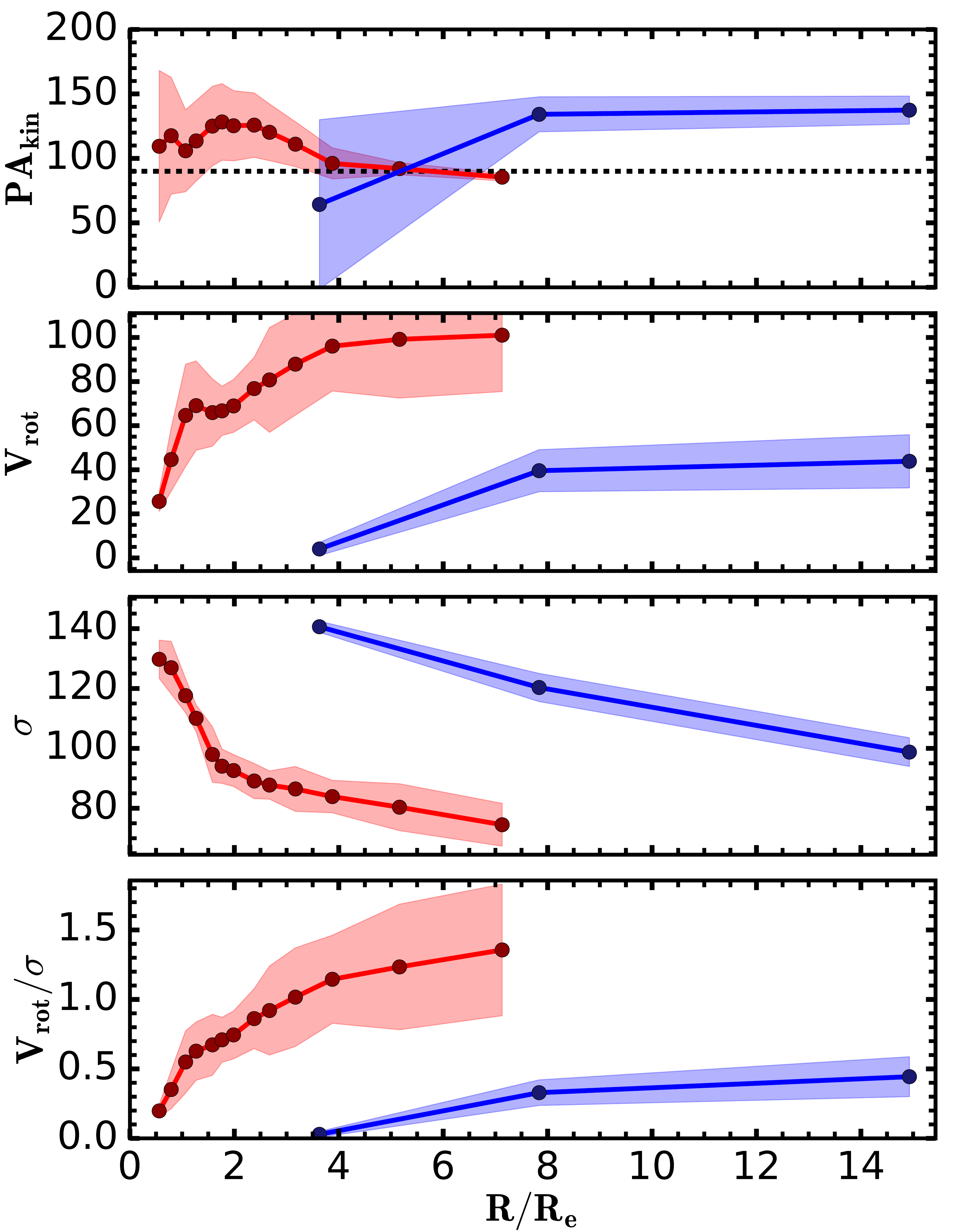}
\caption{GalaxyID = 8057006}
\label{fig:c}
\end{subfigure}
\begin{subfigure}[b]{0.24\textwidth}
\centering 
    \includegraphics[width=\textwidth]{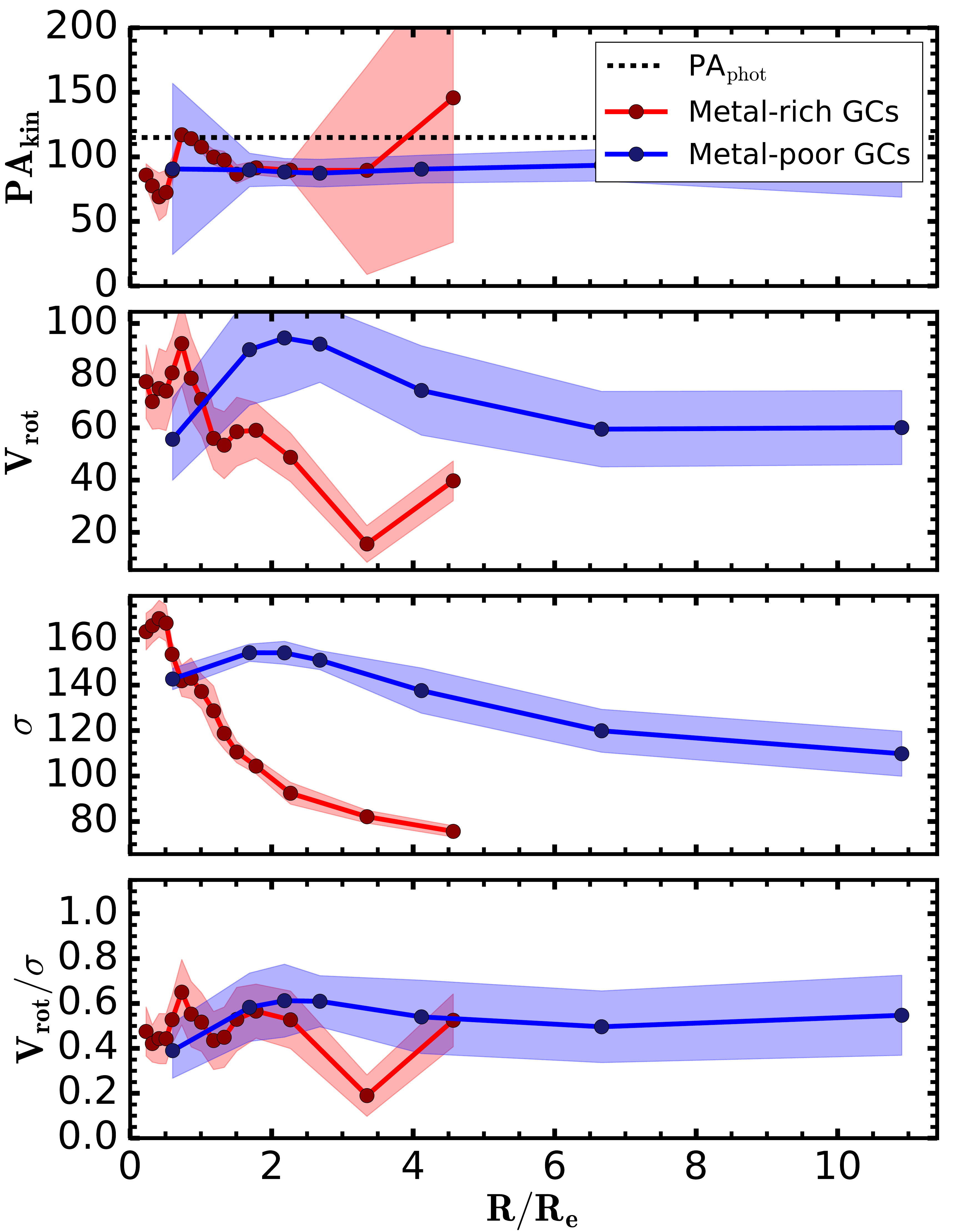}
\caption{GalaxyID = 13450192}
\label{fig:d}
\end{subfigure}
\begin{subfigure}[b]{0.24\textwidth}
\centering 
    \includegraphics[width=\textwidth]{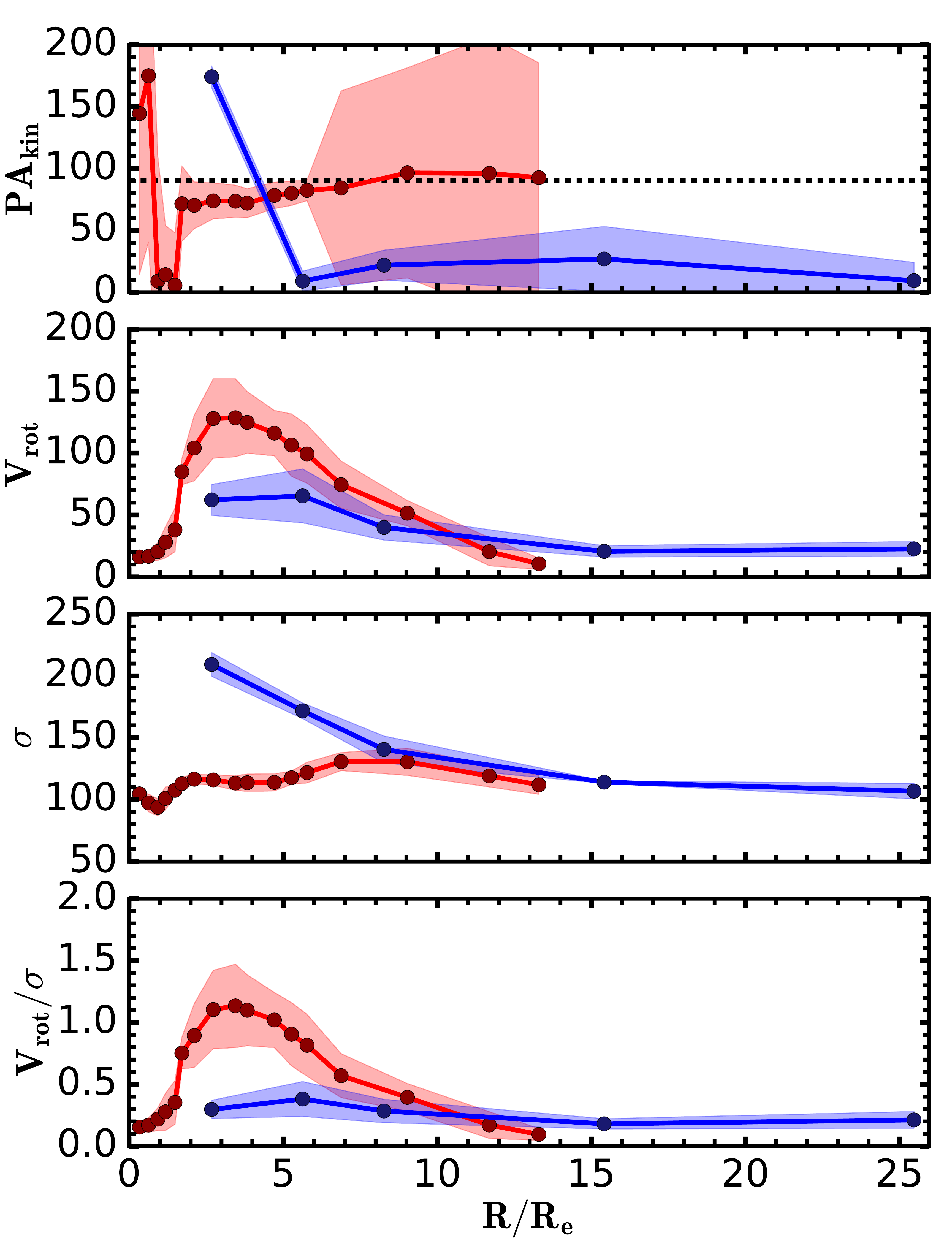}
\caption{GalaxyID = 13394165}
\label{fig:e}
\end{subfigure}
\begin{subfigure}[b]{0.24\textwidth}
\centering 
    \includegraphics[width=\textwidth]{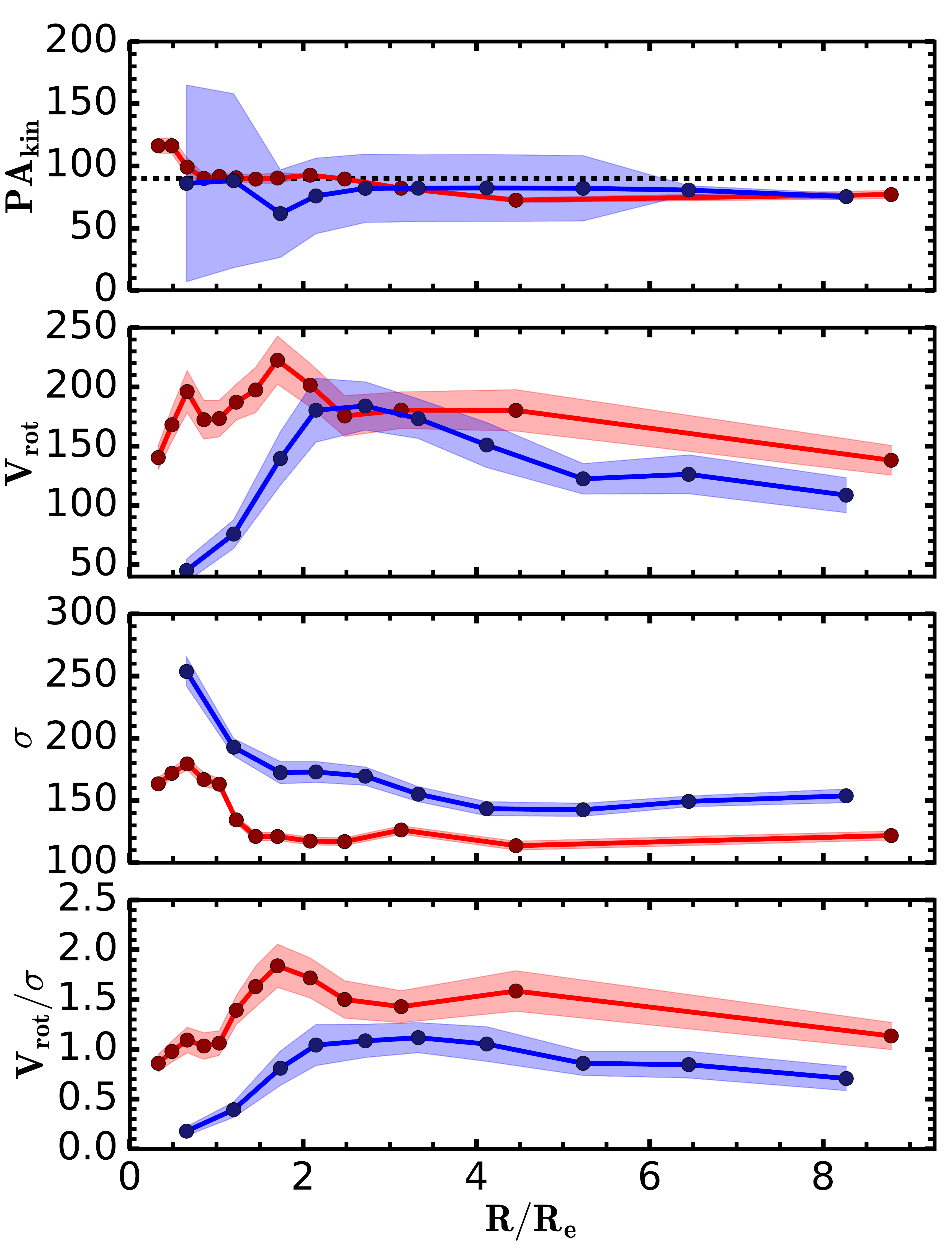}
\caption{GalaxyID = 15613877}
\label{fig:f}
\end{subfigure}
\begin{subfigure}[b]{0.24\textwidth}
\centering 
    \includegraphics[width=\textwidth]{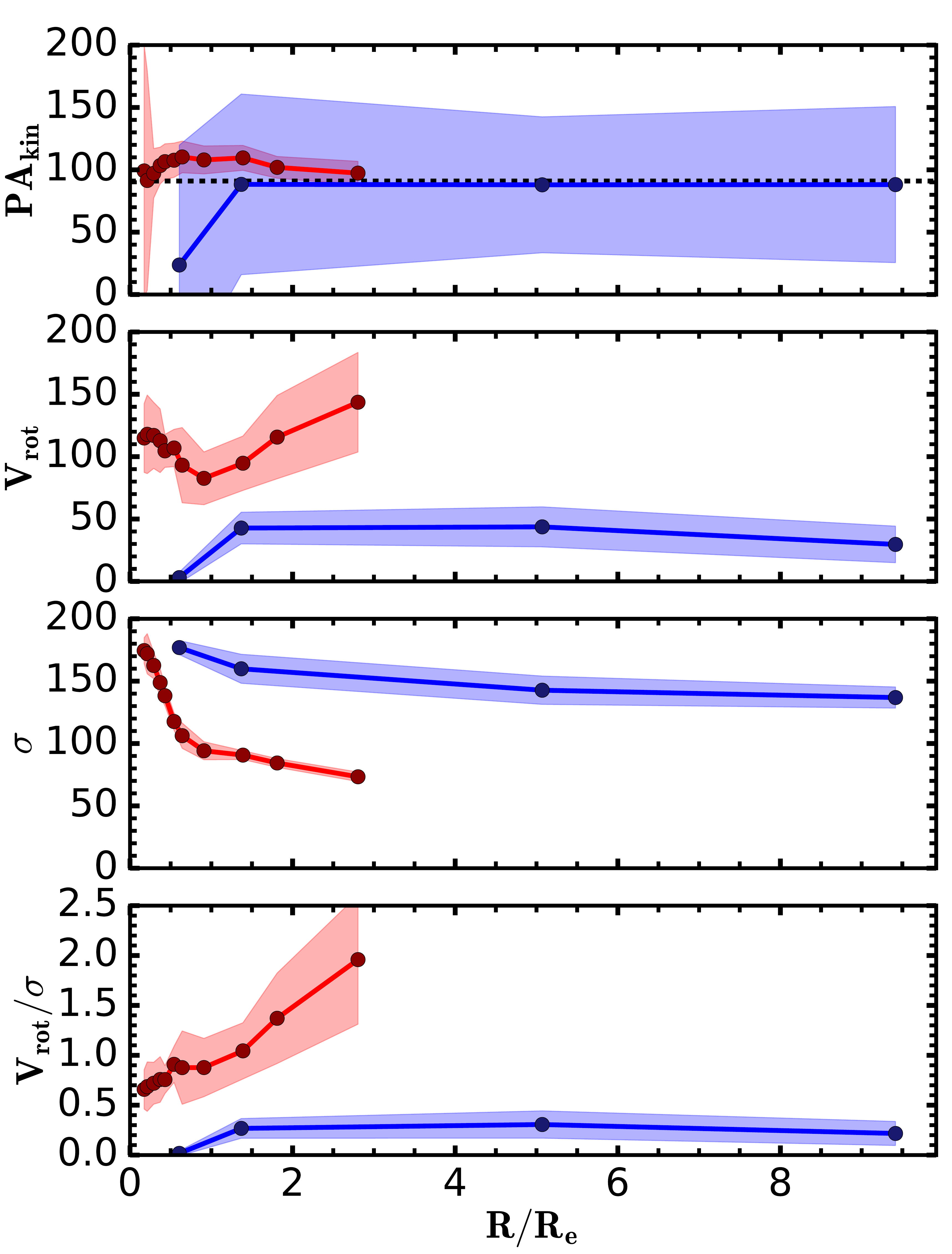}
\caption{GalaxyID = 15854963}
\label{fig:g}
\end{subfigure}
\begin{subfigure}[b]{0.24\textwidth}
\centering 
    \includegraphics[width=\textwidth]{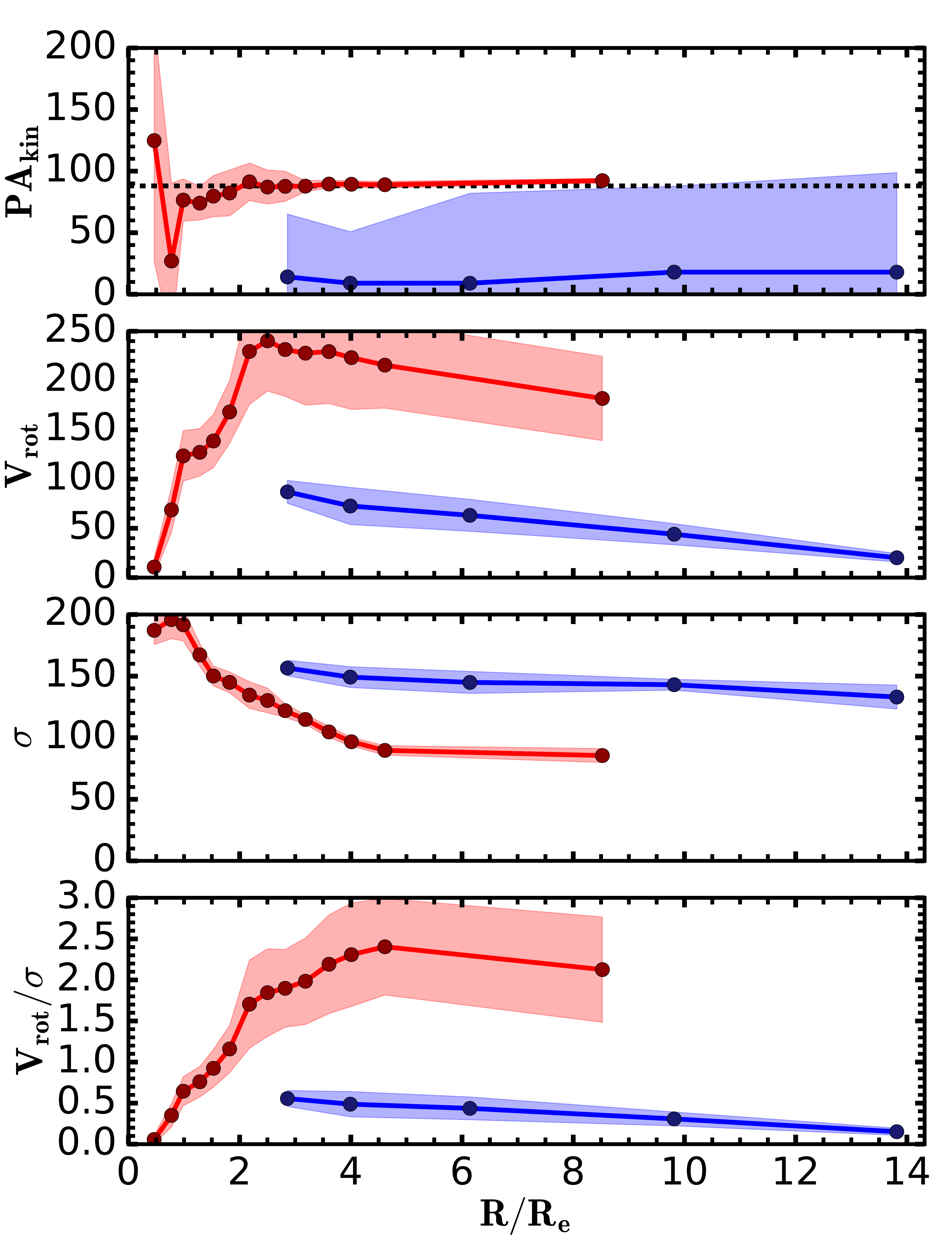}
\caption{GalaxyID = 17827697}
\label{fig:h}
\end{subfigure}
\begin{subfigure}[b]{0.24\textwidth}
\centering 
    \includegraphics[width=\textwidth]{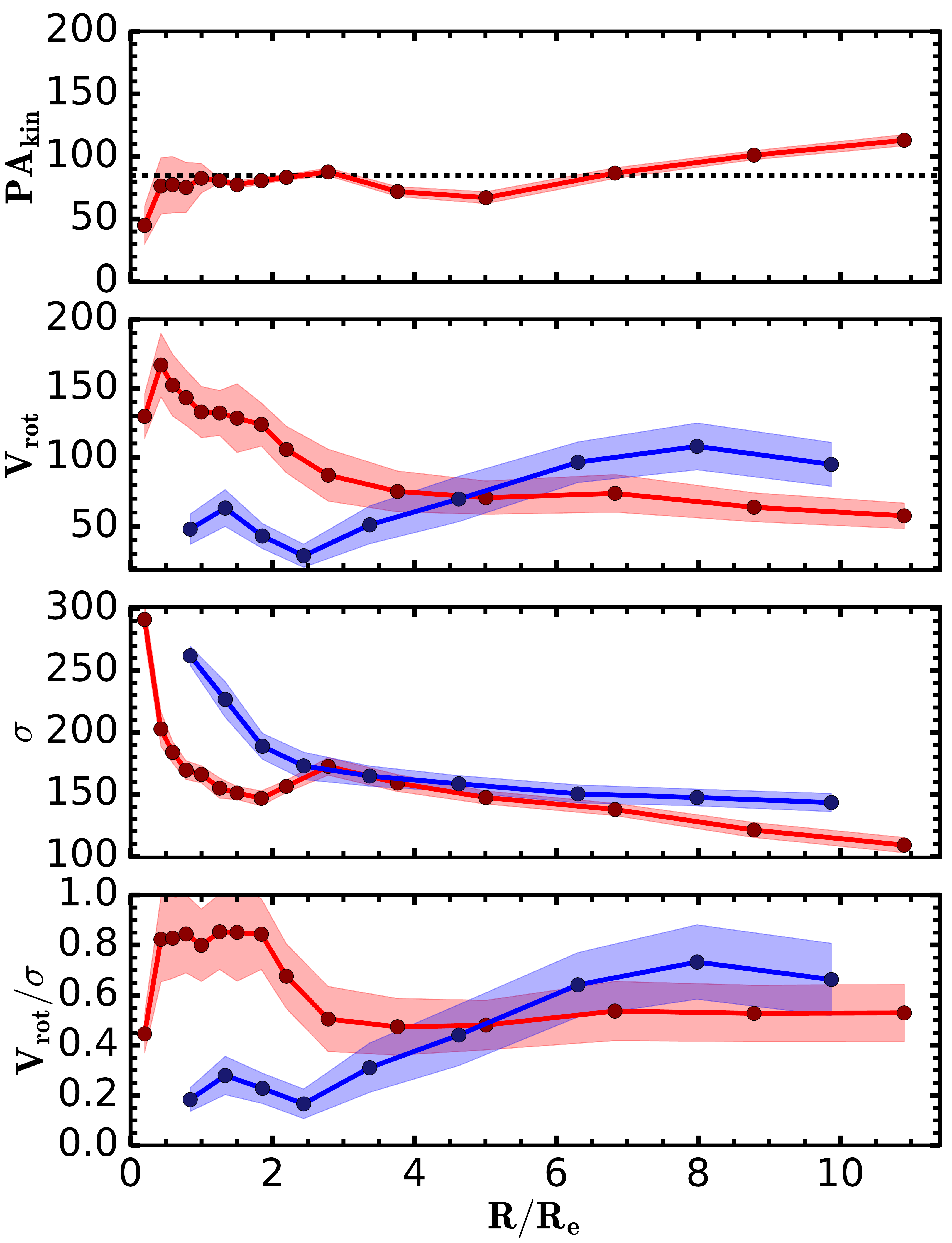}
\caption{GalaxyID = 14455505}
\label{fig:i}
\end{subfigure}
\begin{subfigure}[b]{0.24\textwidth}
\centering 
    \includegraphics[width=\textwidth]{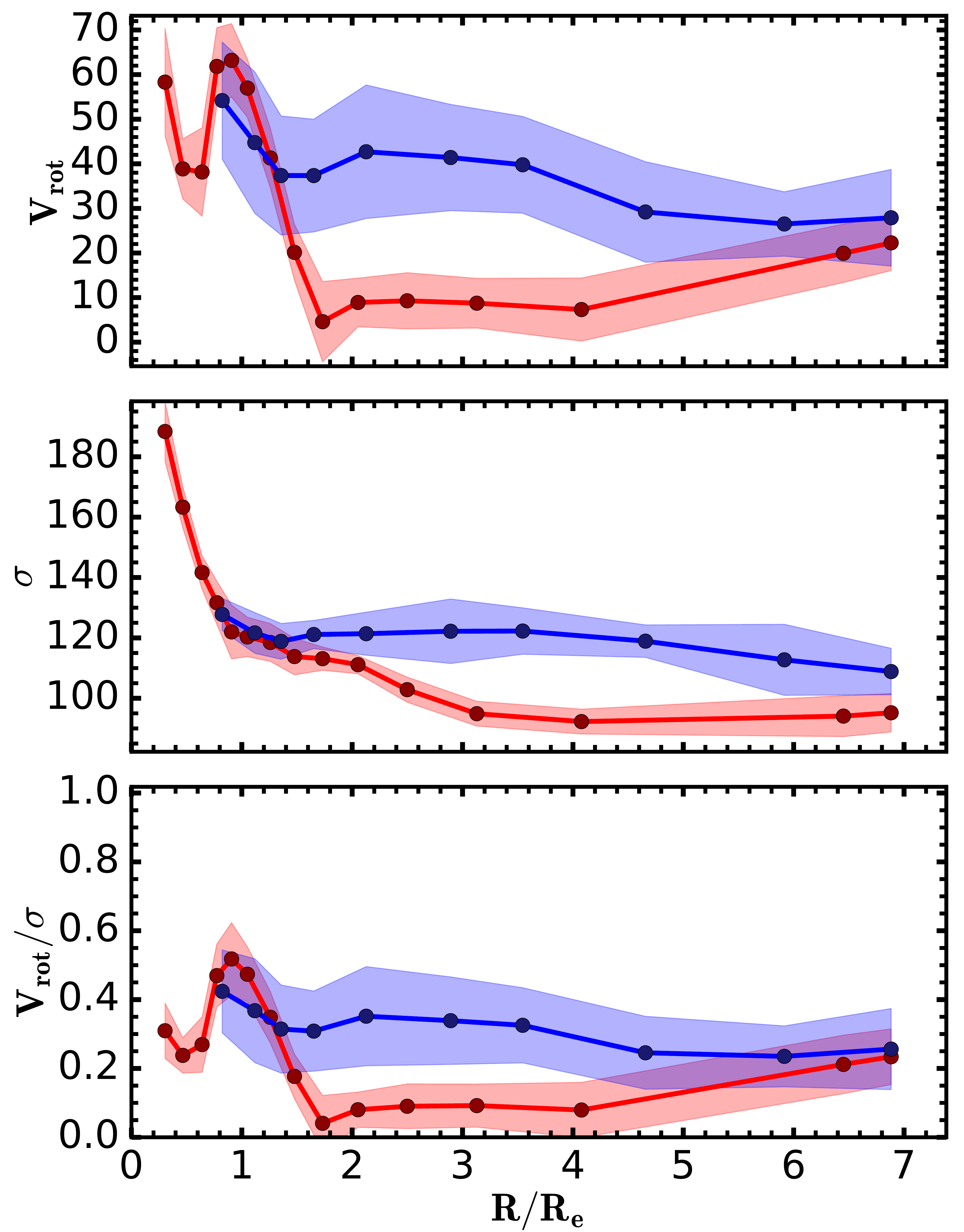}
\caption{GalaxyID = 16787302}
\label{fig:j}
\end{subfigure}
\begin{subfigure}[b]{0.24\textwidth}
\centering 
    \includegraphics[width=\textwidth]{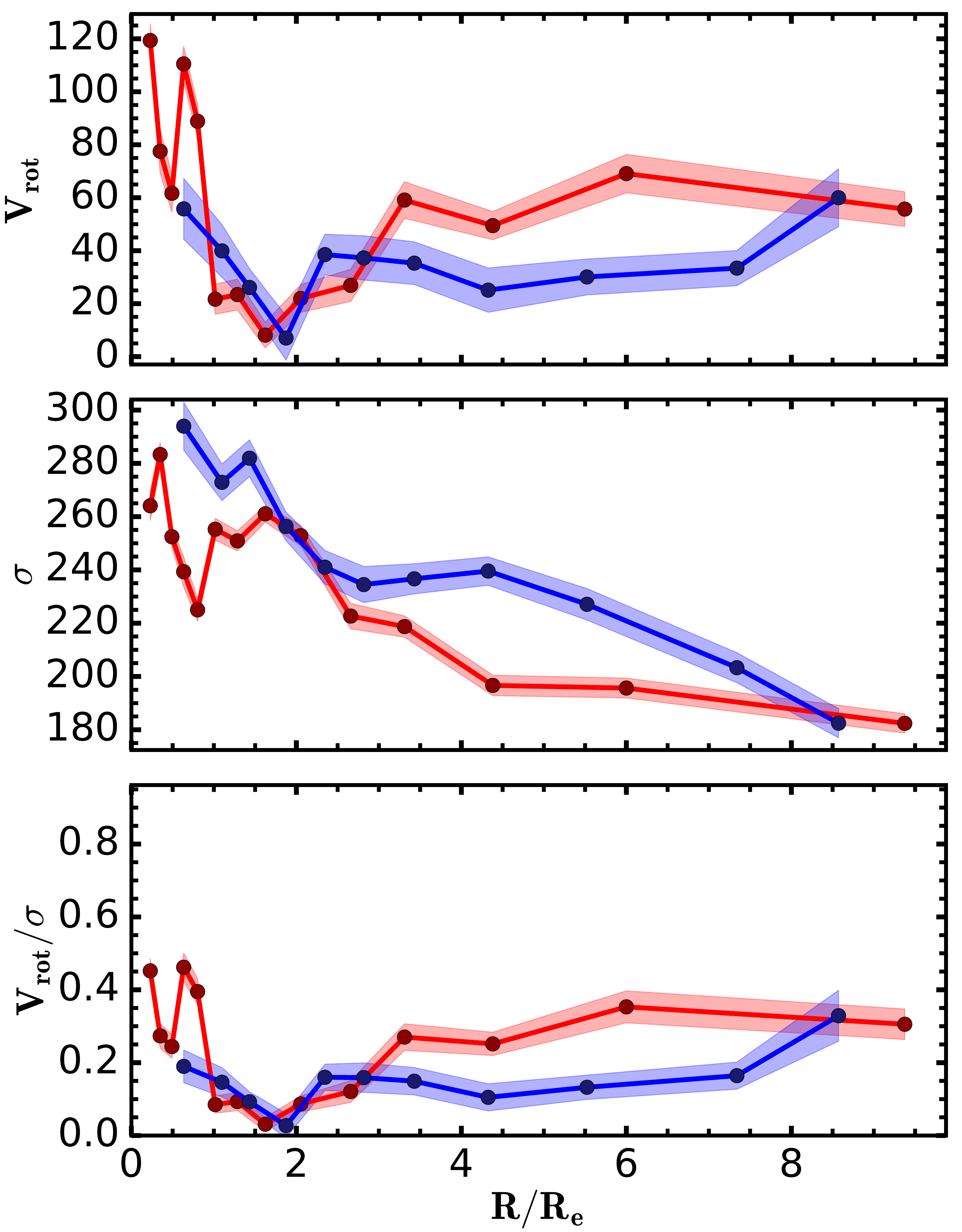}
\caption{GalaxyID = 6276428}
\label{fig:k}
\end{subfigure}
\caption{1D kinematic profiles of the metal-rich (red line) and metal-poor (blue line) GCs obtained by adopting a fixed metallicity split at $\mathrm{[Fe/H]}=-1$ for the $11$ \textit{aligned} S0 galaxies with \textit{peaked} $V_{\mathrm{rot}}/\sigma$ profiles characterized by a large enough number of GCs in each sub-population (i.e. more than $40$). Each sub-figure shows the $\mathrm{PA}_{\mathrm{kin}}$ (where fitted), $V_{\mathrm{rot}}$, $\sigma$ and $V_{\mathrm{rot}}/\sigma$ profiles from the top to the bottom panel for the metal-rich and metal-poor GCs of each galaxy. In all of the galaxies where the $\mathrm{PA}_{\mathrm{kin}}$ is fitted, we see that the metal-rich GCs are rotating along the photometric major axis of the galaxy. On the other hand, the metal-poor GCs are characterized by lower rotational velocity than the metal-rich GCs and they show kinematic misalignment with respect to the metal-rich GCs in some of the galaxies.}
\label{fig:GC_subpopulations_peaked}
\end{figure*}

\begin{figure*}
\centering
\huge
\textbf{{\it \textbf{Increasing}} galaxies}\par\medskip
\begin{subfigure}[b]{0.24\textwidth}
\centering 
    \includegraphics[width=\textwidth]{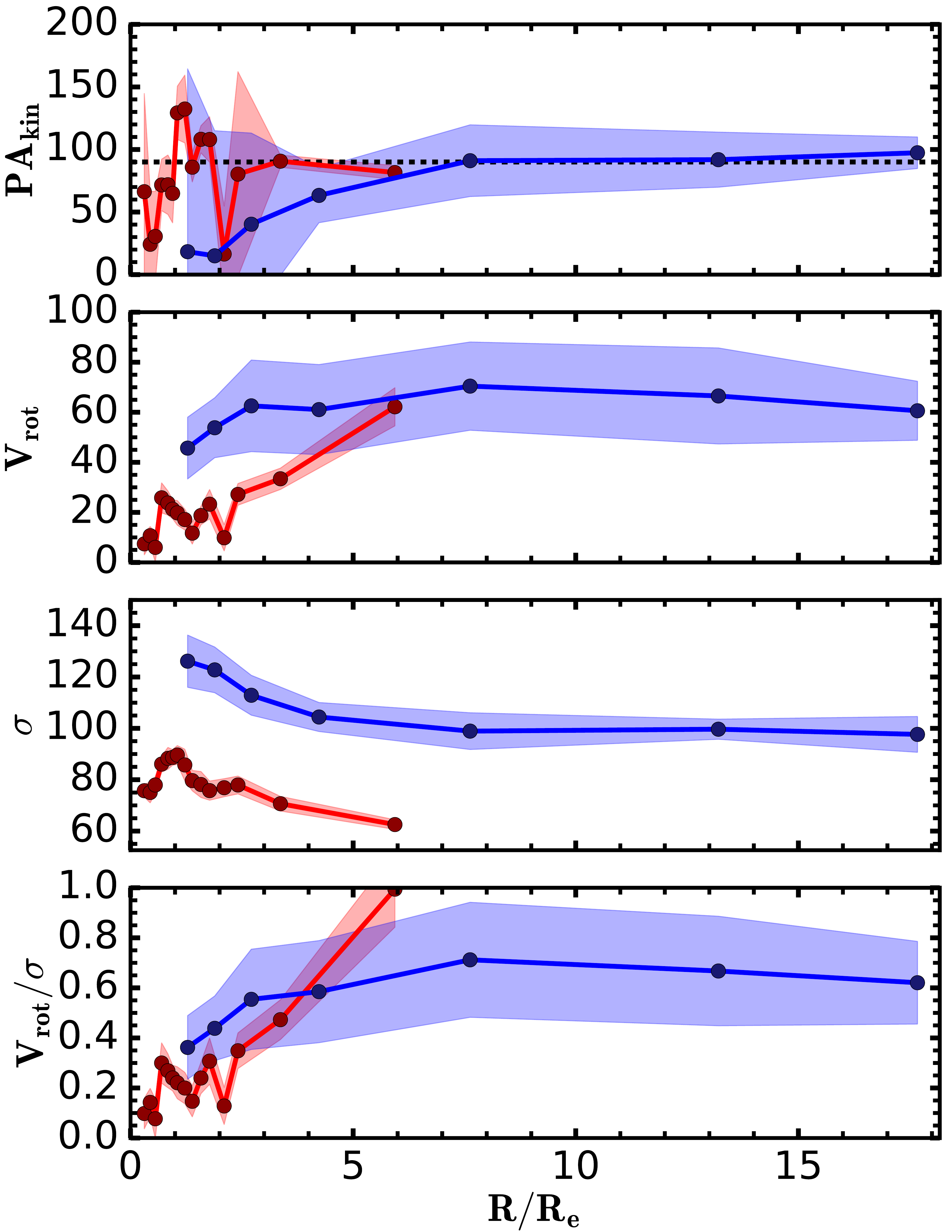}
\caption{GalaxyID = 9906061}
\label{fig:l}
\end{subfigure}
\begin{subfigure}[b]{0.24\textwidth}
\centering 
    \includegraphics[width=\textwidth]{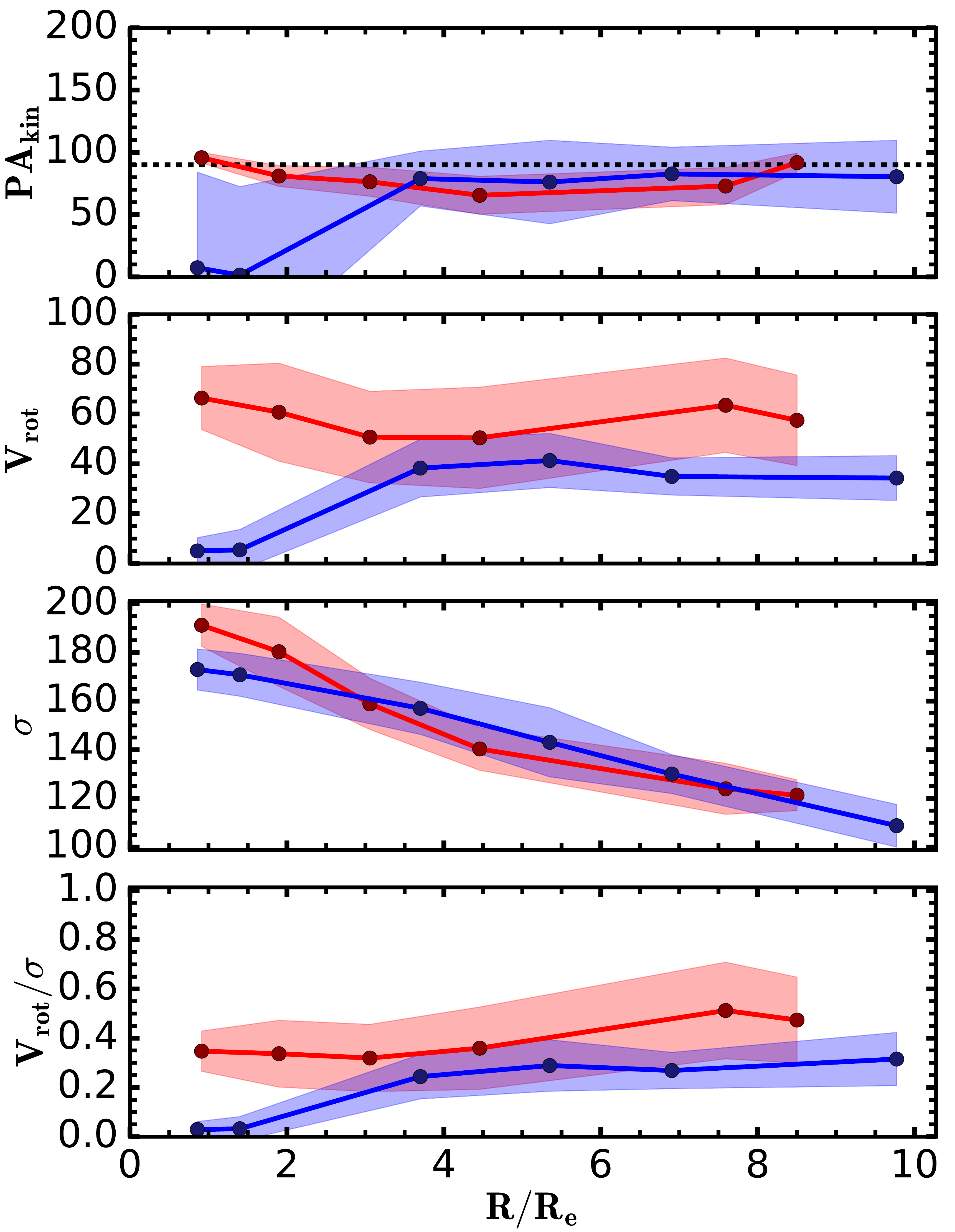}
\caption{GalaxyID = 12285754}
\label{fig:m}
\end{subfigure}
\begin{subfigure}[b]{0.24\textwidth}
\centering 
    \includegraphics[width=\textwidth]{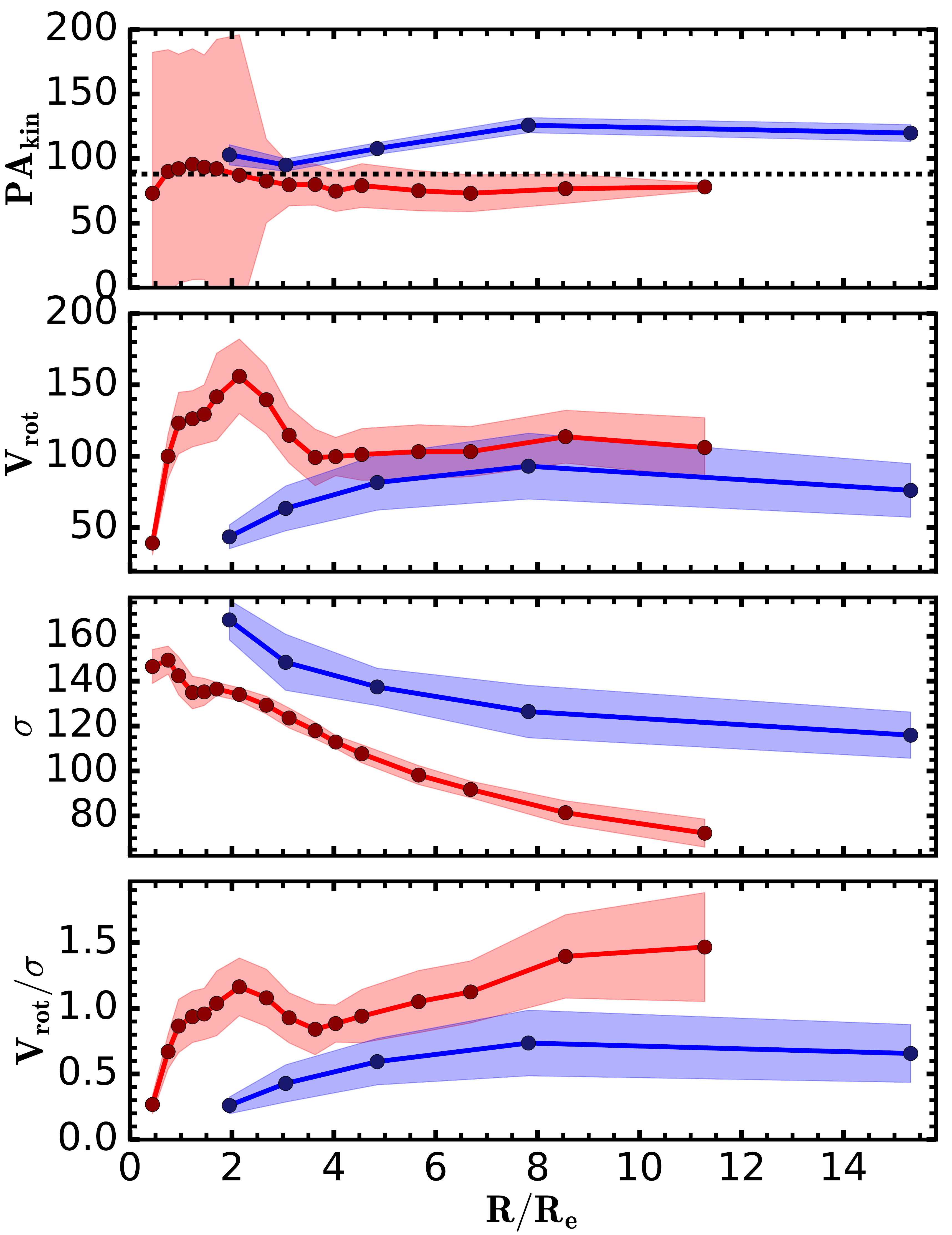}
\caption{GalaxyID = 18850581}
\label{fig:n}
\end{subfigure}
\caption{1D kinematic profiles of the metal-rich (red line) and metal-poor (blue line) GCs obtained by adopting a fixed metallicity split at $\mathrm{[Fe/H]}=-1$ for the $3$ \textit{aligned} S0 galaxies with \textit{increasing} $V_{\mathrm{rot}}/\sigma$ profiles characterized by a large enough number of GCs in each sub-population (i.e. more than $40$). The description is as in Fig. \ref{fig:GC_subpopulations_peaked}. In all three galaxies, the metal-rich and metal-poor GCs show consistent rotational velocity along the photometric major axis of the galaxy.}
\label{fig:GC_subpopulations_increasing}
\end{figure*}

\begin{figure*}
\centering
\huge
\textbf{{\it \textbf{Flat}} galaxies}\par\medskip
\begin{subfigure}[b]{0.24\textwidth}
\centering 
    \includegraphics[width=\textwidth]{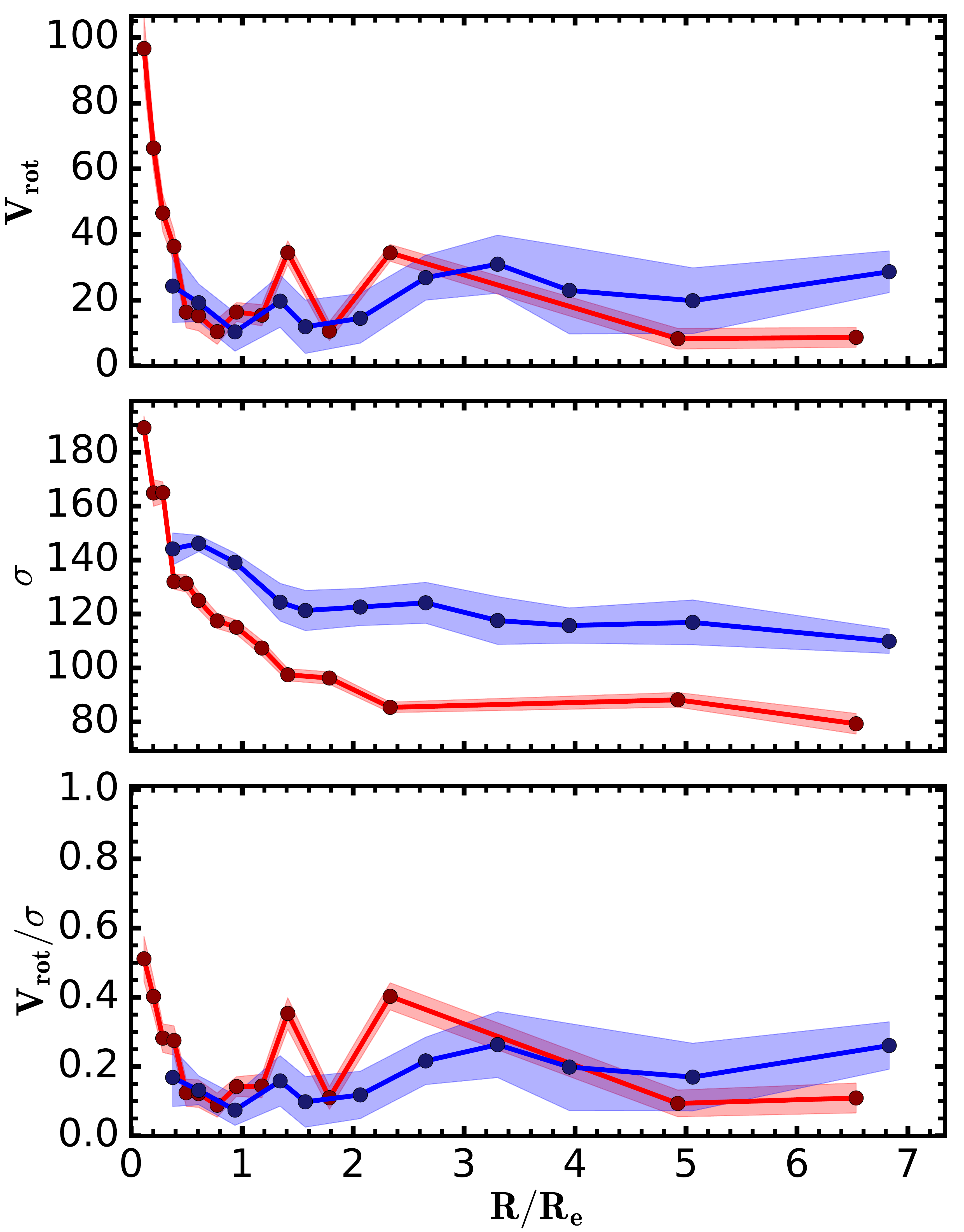}
\caption{GalaxyID = 2053416}
\label{fig:o}
\end{subfigure}
\begin{subfigure}[b]{0.24\textwidth}
\centering 
    \includegraphics[width=\textwidth]{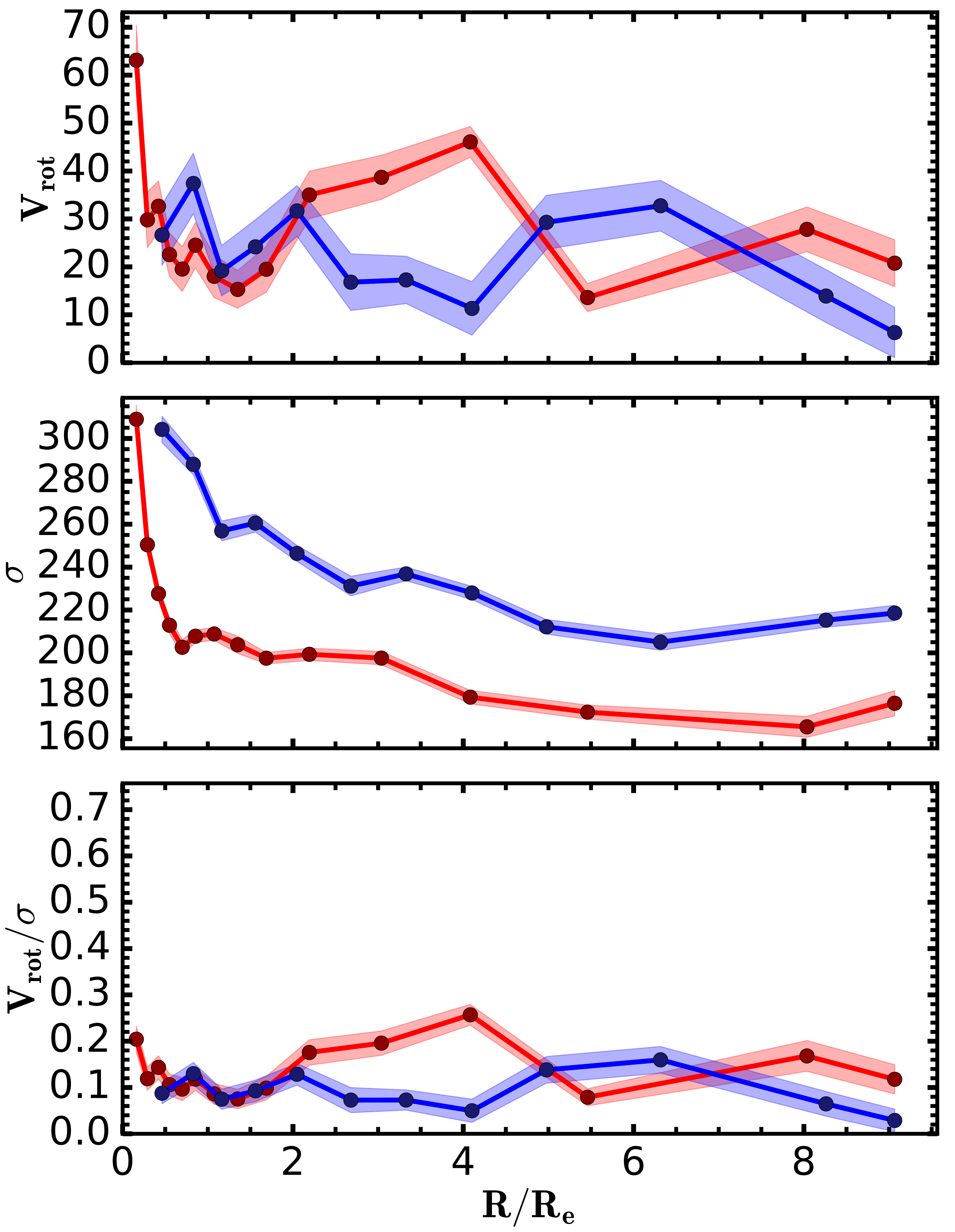}
\caption{GalaxyID = 7828321}
\label{fig:p}
\end{subfigure}
\begin{subfigure}[b]{0.24\textwidth}
\centering 
    \includegraphics[width=\textwidth]{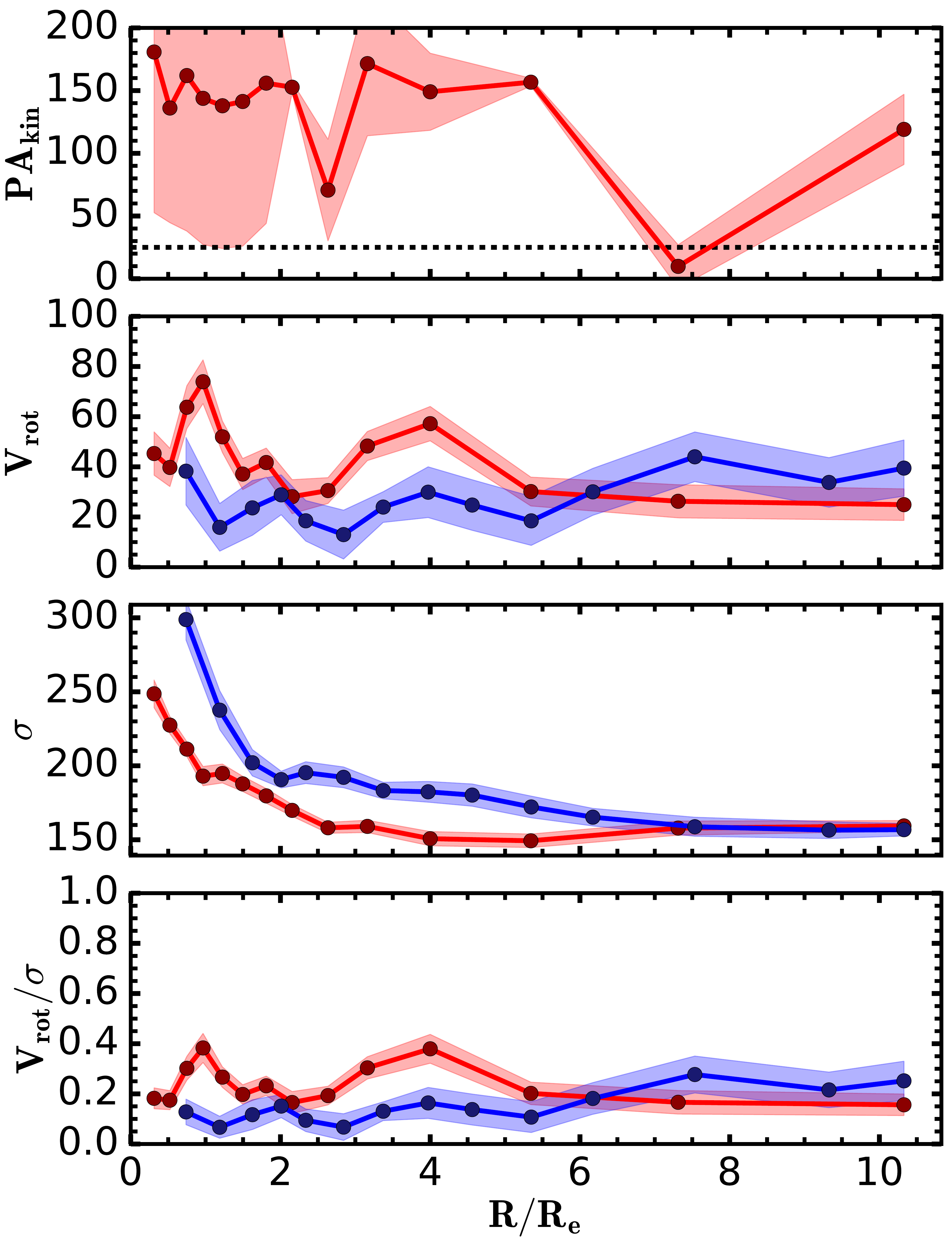}
\caption{GalaxyID = 12177862}
\label{fig:q}
\end{subfigure}
\begin{subfigure}[b]{0.24\textwidth}
\centering 
    \includegraphics[width=\textwidth]{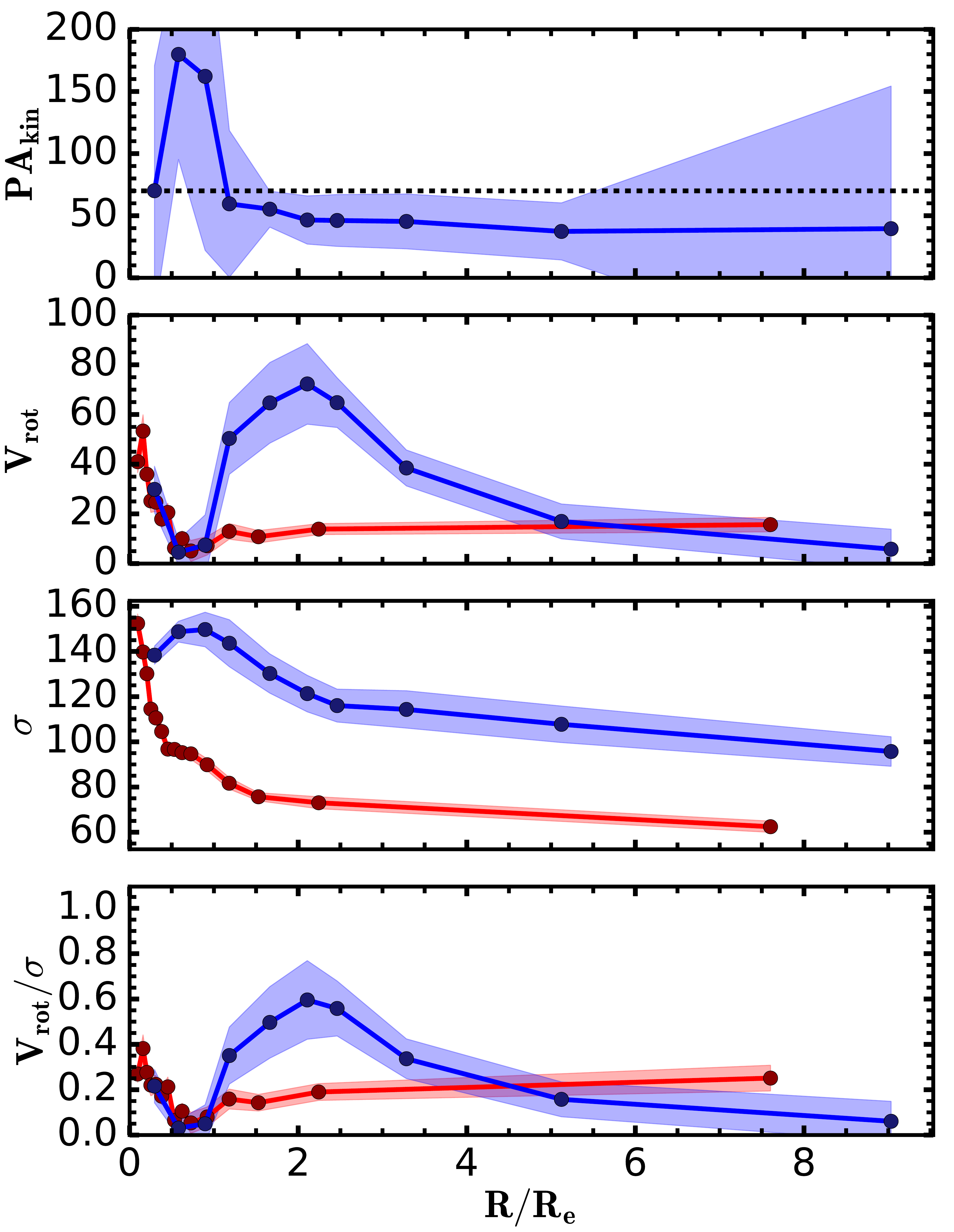}
\caption{GalaxyID = 6806325}
\label{fig:r}
\end{subfigure}
\caption{1D kinematic profiles of the metal-rich (red line) and metal-poor (blue line) GCs obtained by adopting a fixed metallicity split at $\mathrm{[Fe/H]}=-1$ for the $3$ \textit{aligned} S0 galaxies with \textit{flat} $V_{\mathrm{rot}}/\sigma$ profiles characterized by a large enough number of GCs in each sub-population (i.e. more than $40$). The description is as in Fig. \ref{fig:GC_subpopulations_peaked}. We see that the \textit{flat} galaxies are kinematically less well behaved than the \textit{peaked} and \textit{increasing} galaxies. The metal-rich and metal-poor GCs are characterized by overall lower rotational velocity than the \textit{peaked} and \textit{increasing} galaxies.}
\label{fig:GC_subpopulations_flat}
\end{figure*}

\begin{figure*}
\centering
\huge
\textbf{{\it \textbf{Misaligned}} galaxies}\par\medskip
\begin{subfigure}[b]{0.24\textwidth}
\centering 
    \includegraphics[width=\textwidth]{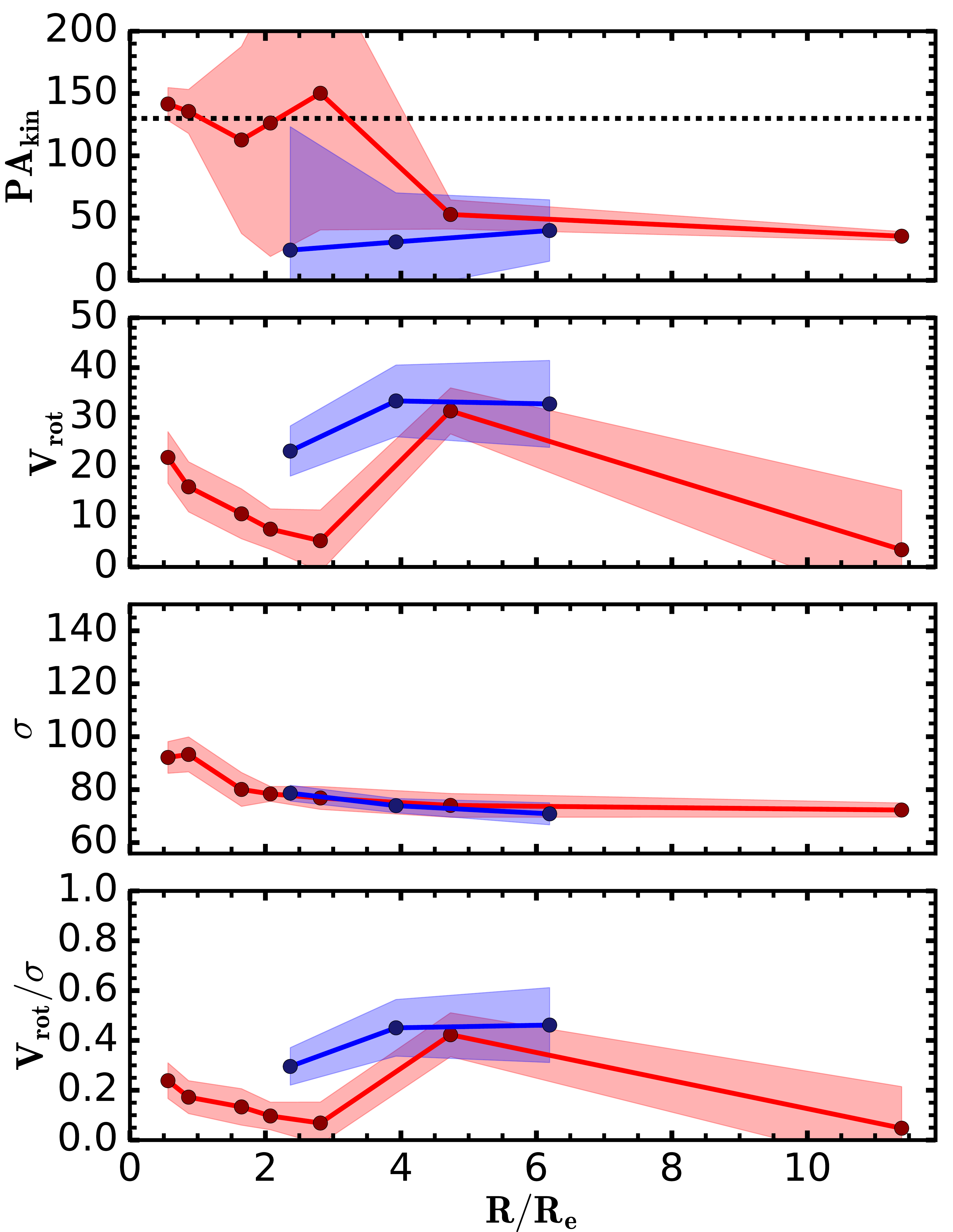}
\caption{GalaxyID = 12333262}
\label{fig:s}
\end{subfigure}
\begin{subfigure}[b]{0.24\textwidth}
\centering 
    \includegraphics[width=\textwidth]{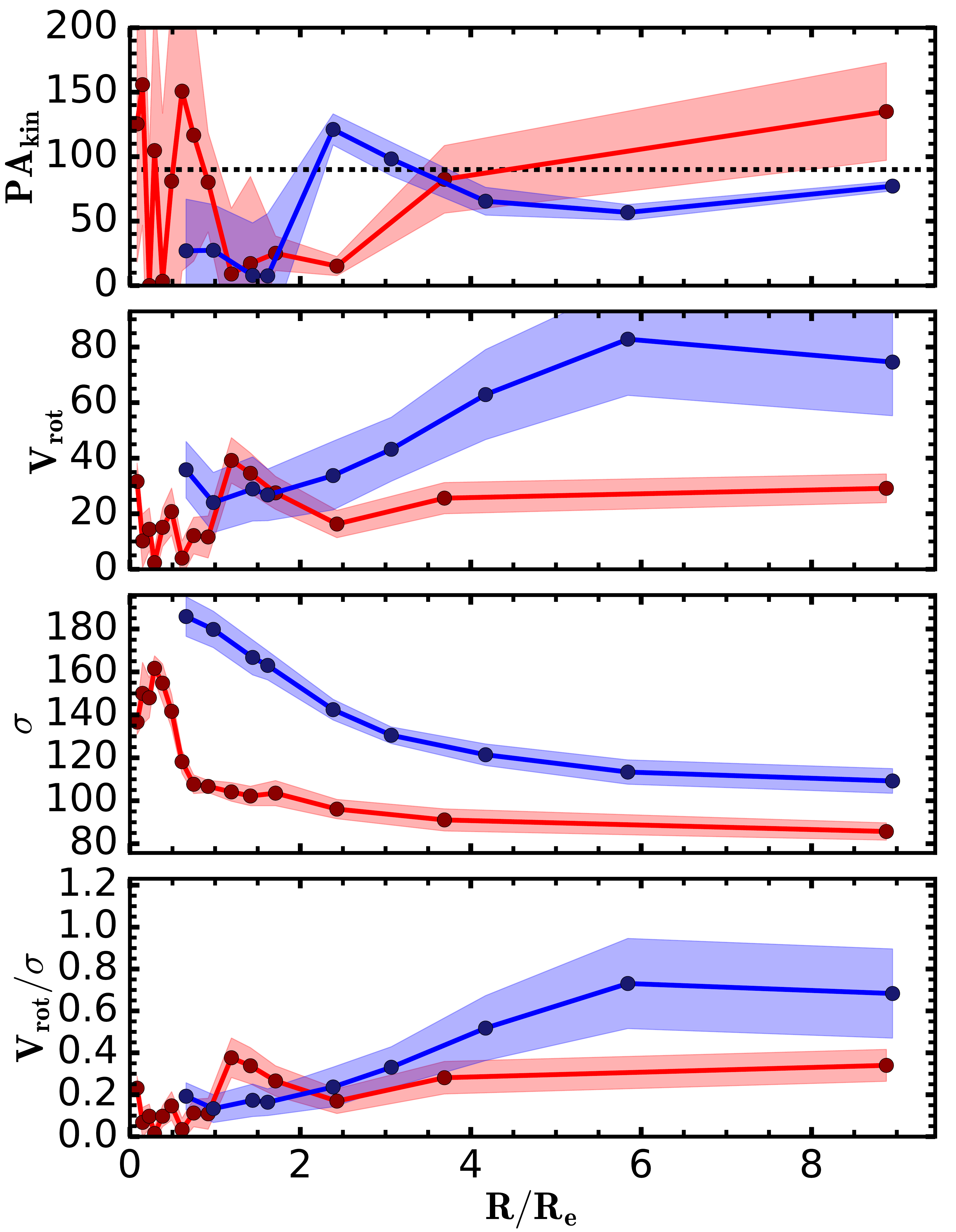}
\caption{GalaxyID = 16837060}
\label{fig:t}
\end{subfigure}
\begin{subfigure}[b]{0.24\textwidth}
\centering 
    \includegraphics[width=\textwidth]{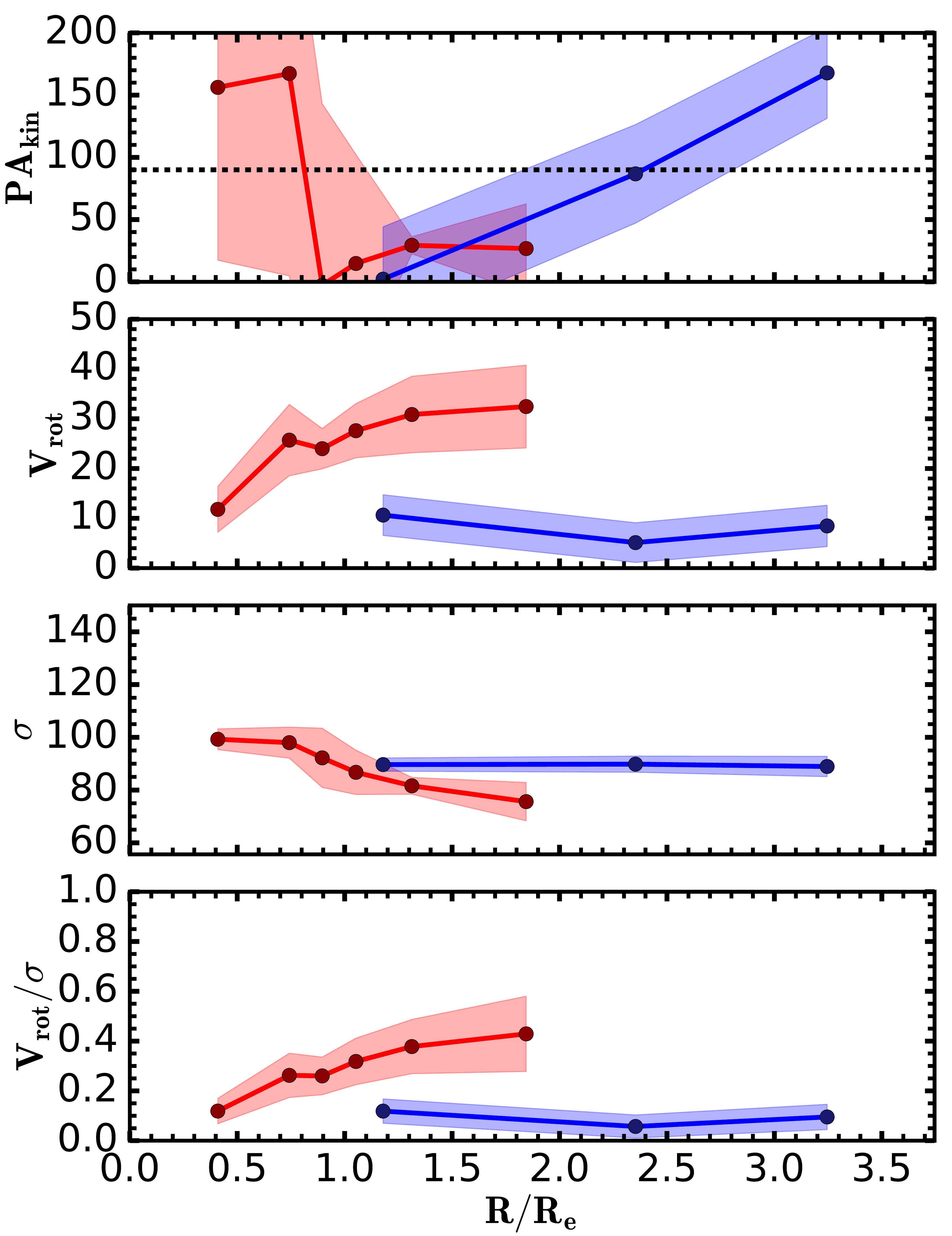}
\caption{GalaxyID = 12938857}
\label{fig:u}
\end{subfigure}
\begin{subfigure}[b]{0.24\textwidth}
\centering 
    \includegraphics[width=\textwidth]{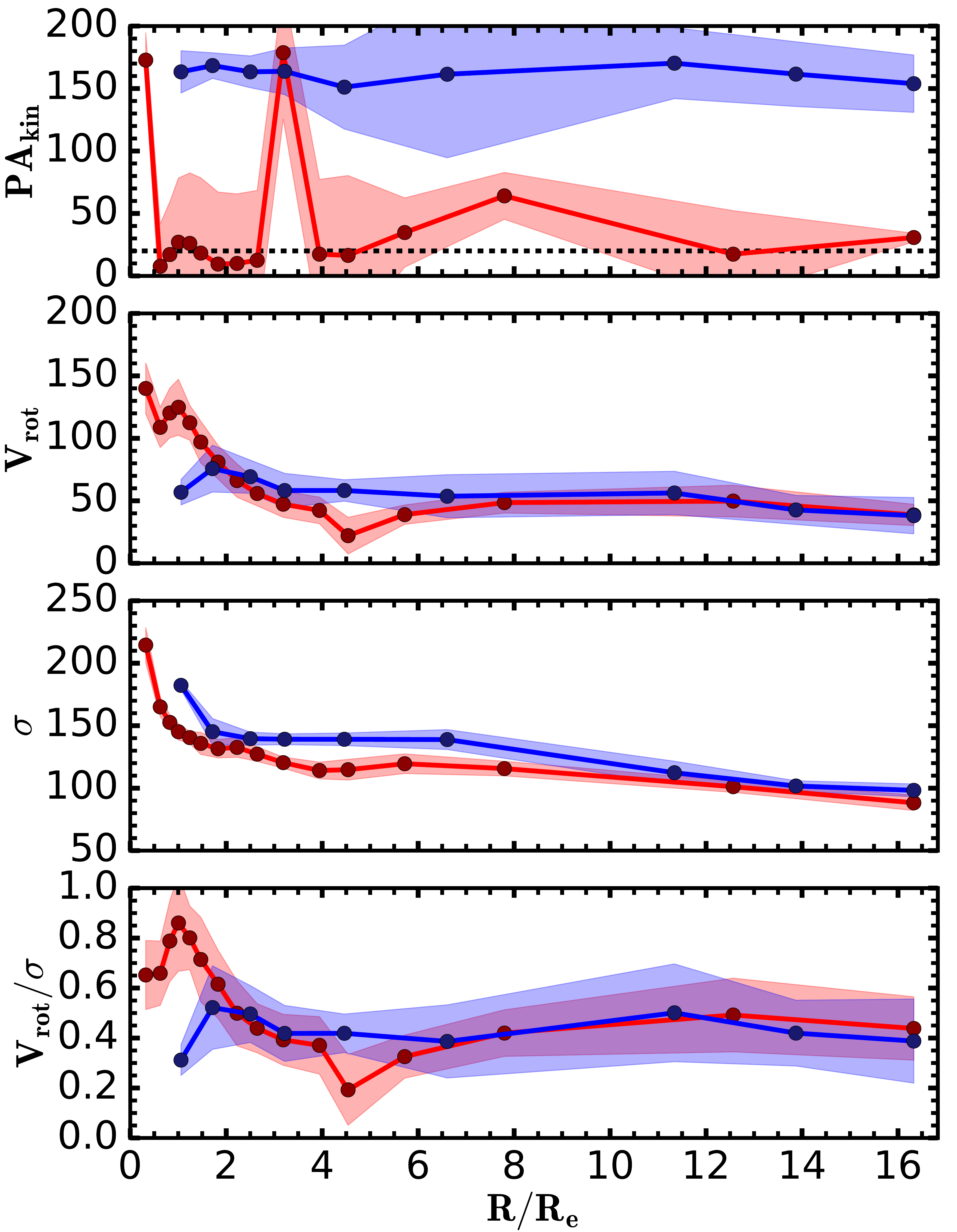}
\caption{GalaxyID = 1092793}
\label{fig:v}
\end{subfigure}
\begin{subfigure}[b]{0.24\textwidth}
\centering 
    \includegraphics[width=\textwidth]{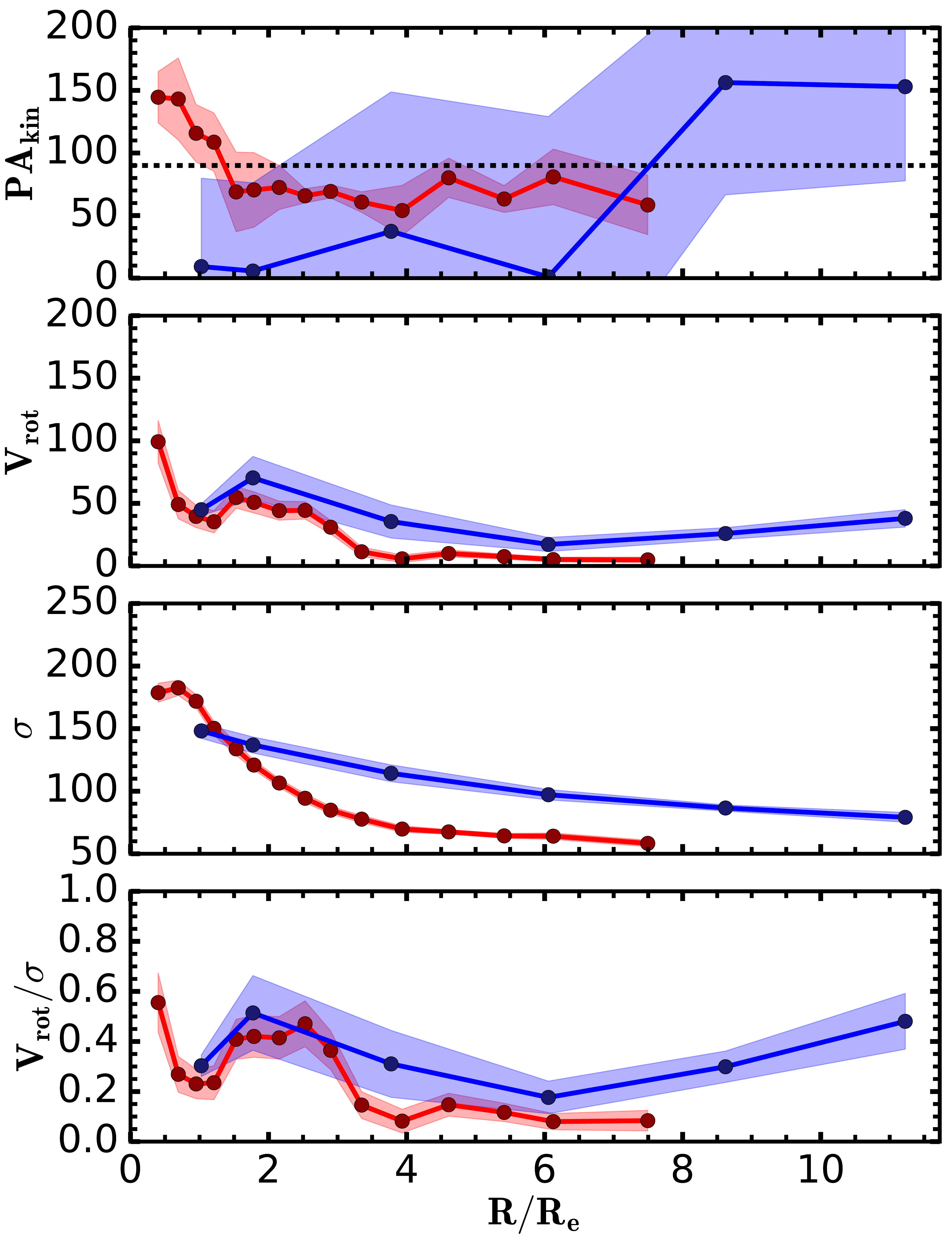}
\caption{GalaxyID = 59288}
\label{fig:w}
\end{subfigure}
\begin{subfigure}[b]{0.24\textwidth}
\centering 
    \includegraphics[width=\textwidth]{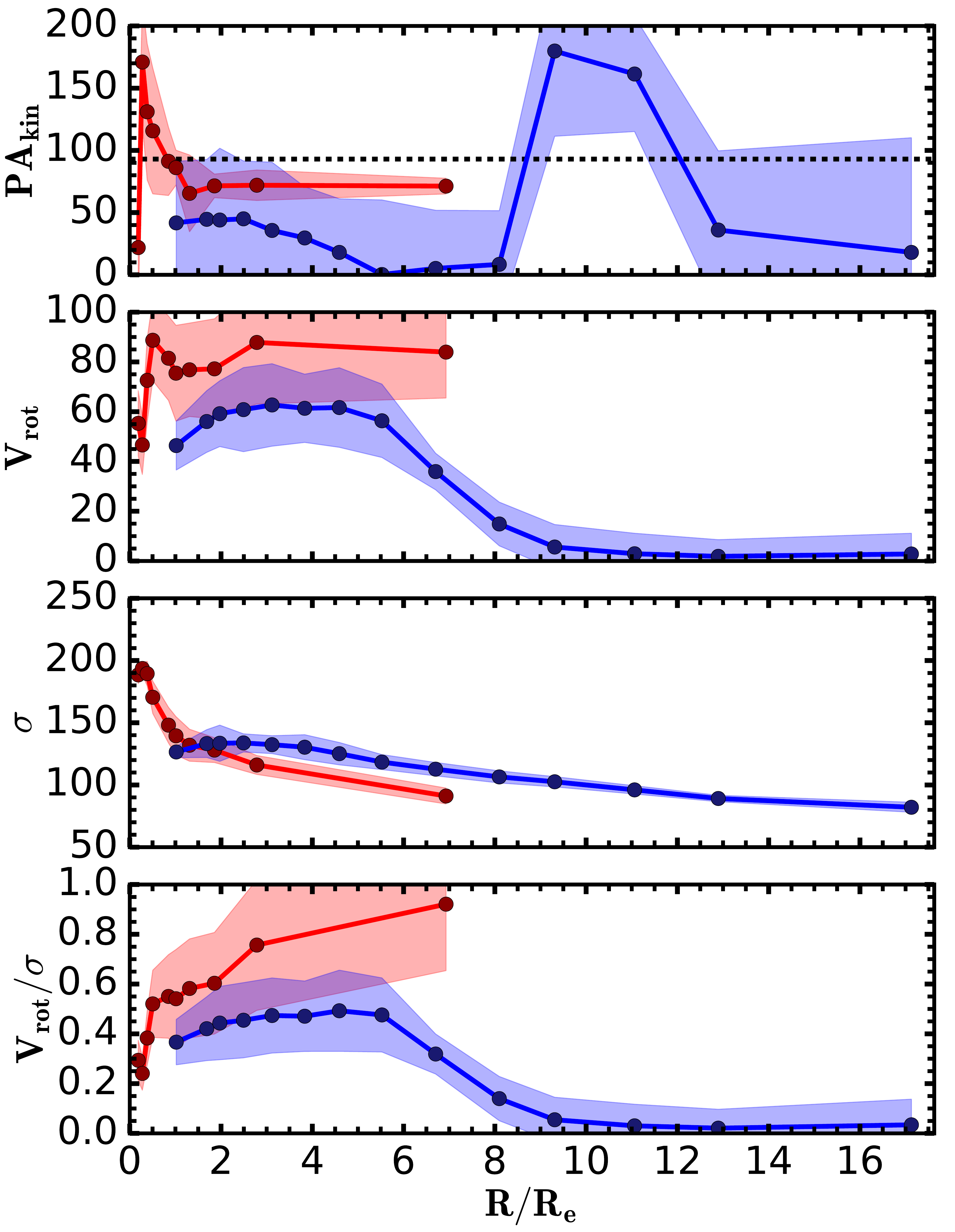}
\caption{GalaxyID = 6451189}
\label{fig:x}
\end{subfigure}
\begin{subfigure}[b]{0.24\textwidth}
\centering 
    \includegraphics[width=\textwidth]{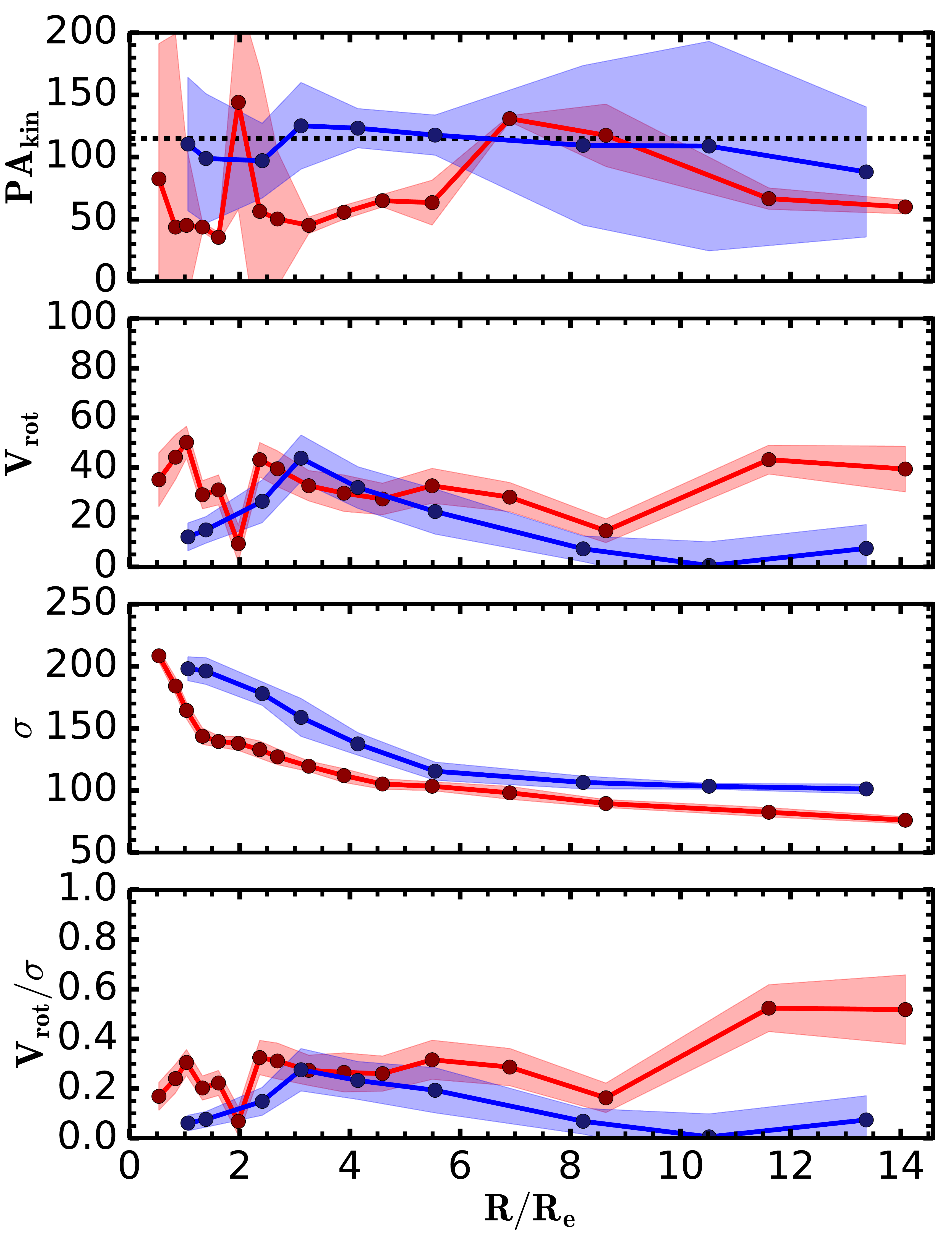}
\caption{GalaxyID = 1221119}
\label{fig:y}
\end{subfigure}
\begin{subfigure}[b]{0.24\textwidth}
\centering 
    \includegraphics[width=\textwidth]{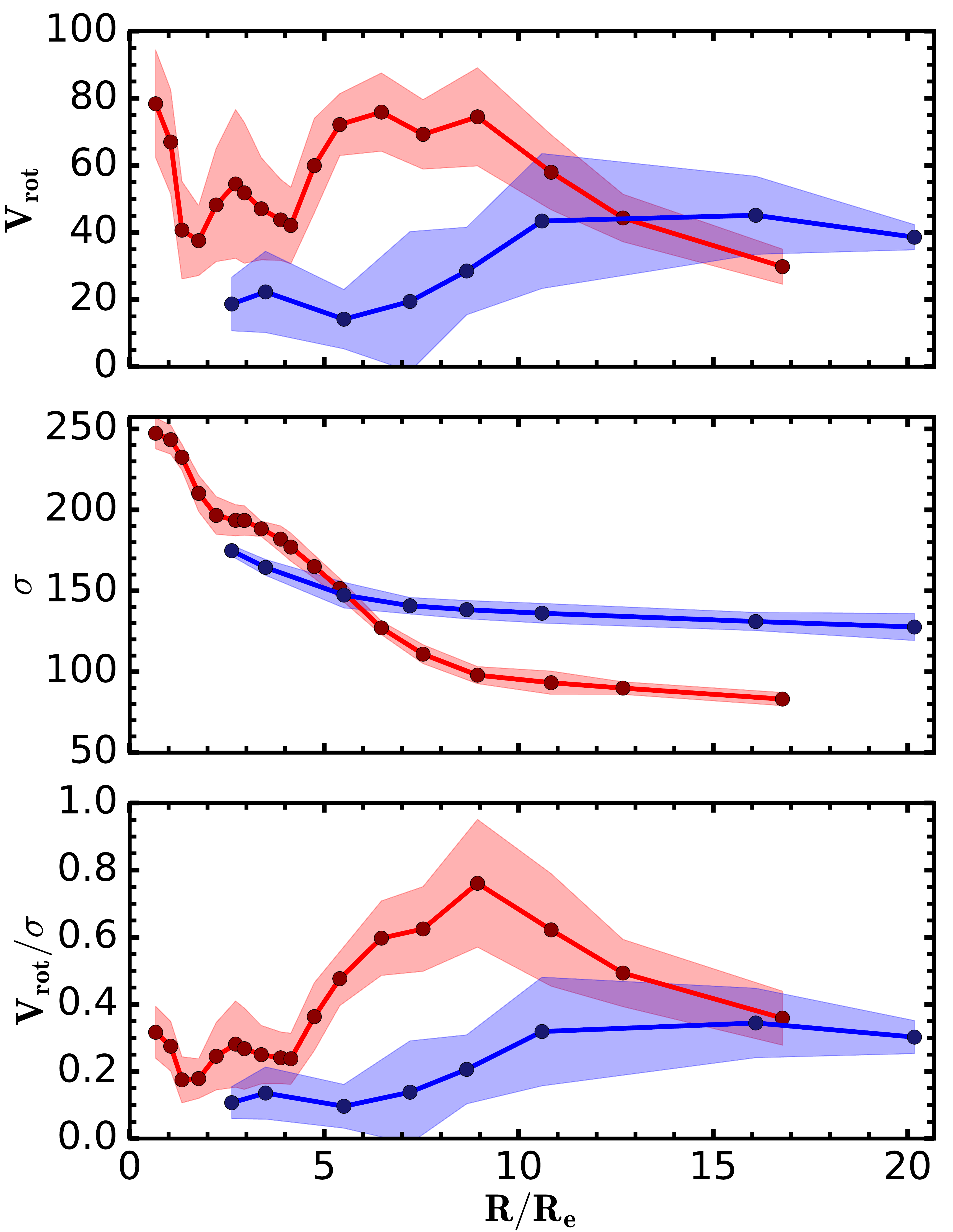}
\caption{GalaxyID = 4362302}
\label{fig:z}
\end{subfigure}
\caption{1D kinematic profiles of the metal-rich (red line) and metal-poor (blue line) GCs obtained by adopting a fixed metallicity split at $\mathrm{[Fe/H]}=-1$ for the $8$ \textit{misaligned} S0 galaxies characterized by a large enough number of GCs in each sub-population (i.e. more than $40$). The description is as in Fig. \ref{fig:GC_subpopulations_peaked}. We see that the \textit{misaligned} galaxies show a range of different kinematic behaviours, similarly to the \textit{flat} galaxies, with evidence of misaligned metal-poor as well as metal-rich GC kinematics.}
\label{fig:GC_subpopulations_misaligned}
\end{figure*}

Finally, across the entire stellar mass range covered by the simulated S0 galaxies (i.e. $10.0 < \log( \mathrm{M}_{*}/\mathrm{M}_{\odot} ) < 11.6$), we find that the \textit{peaked} $V_{\mathrm{rot}}/\sigma$ profile shape is still the dominant one among the \textit{aligned} galaxies with $\sim49\%$ fraction as compared to the \textit{flat} and \textit{increasing} galaxies that are found to occur with a $\sim24\%$ and $\sim27\%$ fraction, respectively.

\subsubsection{\textit{Misaligned} galaxies}
\label{sec:misaligned_galaxies}
Fig. \ref{fig:GC_kinematic_profiles} (bottom) shows the 1D kinematic profiles of the GCs of the $9$ \textit{misaligned} galaxies split into the low ($10 < \log( \mathrm{M}_{*}/\mathrm{M}_{\odot} ) < 10.5$; left column) and high ($10.5 < \log( \mathrm{M}_{*}/\mathrm{M}_{\odot} ) < 11.6$; right column) stellar mass bins, as for the \textit{aligned} galaxies.
Each sub-figure of Fig. \ref{fig:GC_kinematic_profiles} (bottom) also shows the 1D $V_{\mathrm{rot}}$, $\sigma$ and $V_{\mathrm{rot}}/\sigma$ profiles of GC systems as for the \textit{aligned} galaxies, but this time we are also plotting the absolute difference between the $\mathrm{PA}_{\mathrm{kin}}$ of the GCs and the $\mathrm{PA}_{\mathrm{phot}}$ of the S0 galaxy to show the degree of misalignment of the GCs from the $\mathrm{PA}_{\mathrm{phot}}$ of the galaxy.

As we see in Fig. \ref{fig:GC_kinematic_profiles} (bottom), the GC systems of the \textit{misaligned} galaxies are rotating along an axis different from the $\mathrm{PA}_{\mathrm{phot}}$ of the galaxy. 
Additionally, the $V_{\mathrm{rot}}/\sigma$ profile also show variations as a function of the galactocentric radius, with several galaxies being characterized by a second $V_{\mathrm{rot}}/\sigma$ peak at large radii.
This suggests that the \textit{misaligned} galaxies are characterized by a more complex assembly history that could likely be the result of multiple merger events occurring up to the most recent times in the formation history of the galaxies. Such an assembly history may have contributed to a significant fraction of the \textit{ex-situ} GCs that have still not acquired the overall underlying rotation of the galaxy.

In \citet{Dolfi2021}, we found that $3$ out of the $9$ S0 galaxies ($33\%$) from the SLUGGS survey showed kinematic misalignment and twists in the rotation of the GCs and PNe with respect to the underlying stars as a function of radius. 
Specifically, we found that one galaxy (i.e. NGC 4649) was characterized by a similar "double-peaked" $V_{\mathrm{rot}}/\sigma$ profile as we see for most of the \textit{misaligned} simulated S0s. Therefore, these "double-peaked" galaxies may be the results of a late major merger event, that spun up the rotation of the GCs at large radii \citep{Bekki2005}. 
On the other hand, the remaining two \textit{misaligned} galaxies from the SLUGGS survey could be the result of multiple, late minor and mini mergers that decreased the rotation of the GCs out to large radii.  

In Sec. \ref{sec:formation_history_S0s}, we aim at investigating the presence of differences between the past merger histories of the \textit{aligned} and \textit{misaligned} galaxies to understand what physical processes are driving the GC misalignment in our S0 galaxies. 

\subsection{Kinematics of the metal-rich and metal-poor GC sub-populations}
\label{sec:GC_subpopulations}
The colour-magnitude diagrams of the GC systems of our simulated S0 galaxies do not show evidence of any clear colour bimodality. For this reason, in Sec. \ref{sec:1D_kinematic_profiles_GC}, we did not adopt any colour cut to separate between the red and blue GC sub-populations. However, it is not clear whether this may be an issue of the simulations that are unable to reproduce the commonly observed colour-bimodality of the GC systems of observed galaxies.

In this section, we investigate whether the metal-rich and metal-poor GC sub-populations show any differences in their kinematic behaviours by adopting a fixed metallicity split at $\mathrm{[Fe/H]}=-1$, which is the same for all the GC systems of our simulated S0 galaxies. This metallicity value is consistent with that adopted for the ETGs from the SLUGGS survey to separate between the two GC sub-populations (e.g. \citealt{Usher2019}).
We note here that the issue of the overproduction of the metal-rich GCs with $-1.0 < \mathrm{[Fe/H]} < 0$ in the simulations could influence the comparison with the observations, specifically for the low-mass galaxies with $9.5 < \log( \mathrm{M}_{*}/\mathrm{M}_{\odot} ) < 10.5$, which are found to be the most affected (see Sec. \ref{sec:GC_metallicity_cut}).

In order to ensure that each simulated S0 galaxy has a large enough sample of metal-rich and metal-poor GCs for reliably deriving the corresponding 1D kinematic profiles, we only consider here the $26$ galaxies with a total number of metal-rich and metal-poor GCs greater than $40$. This threshold is selected based on the comparison with the GC systems of the observed S0 galaxies from the SLUGGS survey, whose GC sub-populations contain more than $40$ objects (see table 1 in \citealt{Dolfi2021}).

Following the classification in Sec. \ref{sec:1D_kinematic_profiles_GC}, we differentiate between \textit{aligned} and \textit{misaligned} galaxies. For the \textit{aligned} galaxies, we show the kinematics of the metal-rich and metal-poor GCs by separating between the galaxies previously classified as \textit{peaked}, \textit{flat} or \textit{increasing}.
Fig. \ref{fig:GC_subpopulations_peaked}-\ref{fig:GC_subpopulations_misaligned} show the 1D kinematic profiles of the metal-rich and metal-poor GCs for the $26$ S0 galaxies, as selected above. Each sub-figure shows the $\mathrm{PA}_{\mathrm{kin}}$ (where fitted), $V_{\mathrm{rot}}$, $\sigma$ and $V_{\mathrm{rot}}/\sigma$ profiles from the top to the bottom panel for the metal-rich (red line) and metal-poor (blue line) GCs of each galaxy.

From the 1D kinematic profiles, we see that the metal-rich GCs are rotating along the $\mathrm{PA}_{\mathrm{phot}}$ of the galaxy within the $1\sigma$ errors in $9$ out of the $11$ \textit{peaked} galaxies (see Fig. \ref{fig:GC_subpopulations_peaked}) as well as in all of the $3$ \textit{increasing} galaxies (see Fig. \ref{fig:GC_subpopulations_increasing}). Additionally, in the $3$ \textit{increasing} galaxies, the rotation of the metal-poor GCs is, overall, consistent with the metal-rich GCs along the $\mathrm{PA}_{\mathrm{phot}}$ of the galaxy. On the other hand, the metal-poor GCs show little rotation compared to the metal-rich GCs in $7$ of the \textit{peaked} galaxies (see Fig. \ref{fig:a}, \ref{fig:c}, \ref{fig:e}, \ref{fig:g}, \ref{fig:h}, \ref{fig:i}, \ref{fig:k}) and, in two of these galaxies, their rotation is not consistent with the $\mathrm{PA}_{\mathrm{phot}}$ of the galaxy (see Fig. \ref{fig:e}, \ref{fig:h}). In one of the \textit{peaked} galaxies, for which the $\mathrm{PA}_{\mathrm{kin}}$ is not well constrained, the 2D kinematic maps do not show evidence of clear rotation for either the metal-rich and metal-poor GCs (see Fig. \ref{fig:j}).

These results may suggest different origins for the metal-rich and metal-poor GCs. In fact, the $7$ \textit{peaked} galaxies with metal-poor GCs characterized by low rotation suggest that this may be the result of an accretion history involving multiple low-mass (i.e. minor and mini) mergers from random directions, which would be consistent with the expected formation scenario of the \textit{peaked} galaxies proposed by \citet{Schulze2020}. On the other hand, the $3$ \textit{peaked} (see Fig. \ref{fig:b}, \ref{fig:d}, \ref{fig:f}) and $3$ \textit{increasing} galaxies, with metal-rich and metal-poor GCs showing consistent rotation may have experienced fewer accretion events that preserved the overall rotation of the galaxy and of its GC sub-populations. 

In Fig. \ref{fig:GC_subpopulations_flat}, the \textit{flat} galaxies have metal-rich and metal-poor GCs showing typically low rotation at all radii, i.e. $V_{\mathrm{rot}}/\sigma\lesssim0.6$, as compared to the \textit{peaked} galaxies. One galaxy (see Fig. \ref{fig:q}) also shows misaligned metal-rich GC kinematics with respect to the $\mathrm{PA}_{\mathrm{phot}}$ of the galaxy. Overall, this would seem to suggest a more violent and complex accretion history, possibly from a late major merger, for these galaxies that enhanced the random motion of the GC sub-populations out to large radii.

Finally, in Fig. \ref{fig:GC_subpopulations_misaligned}, the \textit{misaligned} galaxies show mixed behaviours in the kinematics of their metal-rich and metal-poor GCs. In fact, in $4$ of the \textit{misaligned} galaxies, the rotation of the metal-rich GCs is, overall, aligned with the $\mathrm{PA}_{\mathrm{phot}}$ of the galaxy, while the metal-poor GCs are not (see Fig. \ref{fig:s}, \ref{fig:v}, \ref{fig:w}, \ref{fig:x}), similar to the $2$ \textit{peaked} galaxies described above. On the other hand, $2$ \textit{misaligned} galaxies are more similar to the \textit{flat} galaxies, as they show misaligned kinematics in both the metal-rich and metal-poor GC sub-populations that are characterized by less well behaved kinematic profiles with several transitions as a function of radius (see Fig. \ref{fig:t}, \ref{fig:y}).
Of the remaining $2$ \textit{misaligned} galaxies, one (see Fig. \ref{fig:z}) is characterized by low rotation in both the metal-rich and metal-poor GCs with unconstrained $\mathrm{PA}_{\mathrm{kin}}$, while the other (see Fig. \ref{fig:u}) shows low rotation in the metal-poor GCs.

\begin{figure}
\centering
    \includegraphics[width=0.5\textwidth]{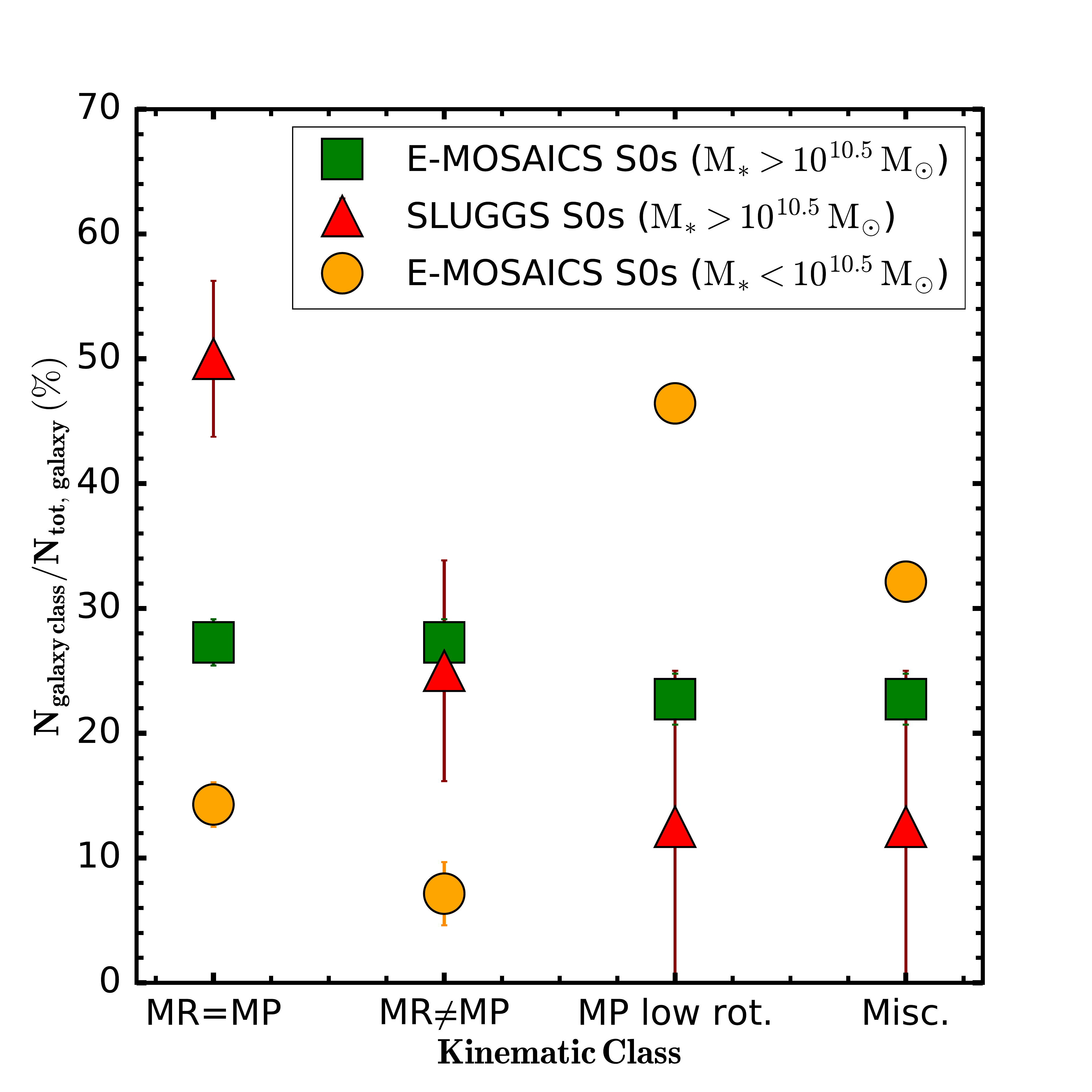}
\caption{Percentage of the high-mass ($\log( \mathrm{M}_{*}/\mathrm{M}_{\odot} ) > 10.5$) S0 galaxies from the SLUGGS survey (red triangles) and E-MOSAICS simulations (green squares) showing different kinematic behaviours in their metal-rich (red) and metal-poor (blue) GC kinematics. The distributions of the low-mass ($10 < \log( \mathrm{M}_{*}/\mathrm{M}_{\odot} ) < 10.5$) S0 galaxies from the E-MOSAICS simulations is shown with orange dots. The four kinematic classes, {\bf MR=MP}, {\bf MR$\neq$MP}, {\bf MP low rotation} and {\bf Miscellaneous}, are described in Sec. \ref{sec:GC_subpopulations}. We see that there is, overall, a good agreement between the kinematic behaviour of the metal-rich and metal-poor GCs in the simulated and observed S0 galaxies for $\log( \mathrm{M}_{*}/\mathrm{M}_{\odot} ) > 10.5$. For $\log( \mathrm{M}_{*}/\mathrm{M}_{\odot} ) < 10.5$, the majority of the simulated galaxies are characterized by metal-poor GCs with little rotation, suggesting few isotropic accretion events with low-mass galaxies.}
\label{fig:MR_vs_MP_kinematics_summary}
\end{figure}

In Fig. \ref{fig:GC_subpopulations_peaked}-\ref{fig:GC_subpopulations_misaligned}, we have only selected those galaxies with a large enough sample of metal-rich and metal-poor GCs to calculate the 1D kinematic profiles. Here, we investigate the kinematics of the metal-rich and metal-poor GCs of all $50$ S0 galaxies based on the 2D line-of-sight velocity maps (edge-on projection) to include in the analysis also those galaxies with less than $40$ GCs in one or both of the GC sub-populations. In Fig. \ref{fig:MR_vs_MP_kinematics_summary}, we summarize the fraction of the galaxies showing the following four different kinematic behaviours in their GC sub-populations as follows: \\
{\bf MR=MP:} the metal-rich (MR) and metal-poor (MP) GCs are rotating along a similar axis ($\mathrm{PA}_{\mathrm{kin}}$), which is consistent with the photometric major axis of the galaxy ($\mathrm{PA}_{\mathrm{phot}}$);\\ 
{\bf MR$\neq$MP:} the metal-rich GCs are rotating along an axis consistent with the photometric major axis of the galaxy ($\mathrm{PA}_{\mathrm{phot}}$), while the metal poor GCs are misaligned; \\
{\bf MP low rotation:} the metal-rich GCs are rotating along an axis consistent with the photometric major axis of the galaxy ($\mathrm{PA}_{\mathrm{phot}}$), while the metal-poor GCs have little or no rotation; \\
{\bf Miscellaneous:} the metal-rich and metal-poor GCs are both misaligned from the photometric major axis of the galaxy ($\mathrm{PA}_{\mathrm{phot}}$) or have little rotation.

For the comparison between the simulations and observations, in Fig. \ref{fig:MR_vs_MP_kinematics_summary}, we also show the distribution of the $8$ out of the $9$ S0 galaxies from the SLUGGS survey characterized by high-mass (i.e. $\log( \mathrm{M}_{*}/\mathrm{M}_{\odot} ) > 10.5$), which we studied in \citet{Dolfi2021}.

We find that the high-mass ($\log( \mathrm{M}_{*}/\mathrm{M}_{\odot} ) > 10.5$) simulated S0 galaxies are distributed with similar fractions in each of the four kinematic classes. On the other hand, the low-mass ($\log( \mathrm{M}_{*}/\mathrm{M}_{\odot} ) < 10.5$) simulated S0 galaxies are more frequently characterized by metal-poor GCs with low rotation velocity ({\bf MP low rotation} $\sim50\%$), suggesting that these galaxies have mainly experienced isotropic accretion events from low-mass dwarf galaxies that mostly contributed to few metal-poor GCs with low rotation.  

For the high-mass S0 galaxies, we find that the simulations predict a fraction of galaxies with misaligned metal-poor and metal-rich GC kinematics ({\bf MR$\neq$MP} $\sim30\%$) that is consistent with the observations.
On the other hand, the simulations underestimate the fraction of galaxies with aligned metal-rich and metal-poor GC kinematics ({\bf MR=MP} $\sim30\%$) with respect to the observations ({\bf MR=MP} $\sim50\%$). This discrepancy may be a result of sample selection effects as some of our simulated galaxies show signs of interaction from the Mpc-scale FoV images, while the observed galaxies from the SLUGGS survey are mainly undisturbed. Alternatively, it could be a result of viewing angle effects, as we are studying the edge-on view of the galaxies (see Sec. \ref{fig:GC_kinematic_profiles}).
Finally, we find an overall good agreement between the fraction of the simulated and observed S0 galaxies in the remaining two kinematic classes ({\bf MP low rotation} and {\bf Miscellaneous}) within the errors. Here, the discrepancy could also be the result of sample selection and viewing angle effects, as previously mentioned. Alternatively, it could also be due to differences in the splitting between the GC sub-populations that is colour-based in the observations and metallicity-based in the simulations.

\section{Phase-space diagrams of the GC systems}
\label{sec:phase_space_diagrams}
The aim of this section is to understand whether the location of the GC systems of the individual galaxies on the 3D phase-space diagrams correlates with their accretion times onto the galaxies. The results could subsequently be applied to the GC systems of observed galaxies for which we do not have explicit information about their \textit{in-situ} or \textit{ex-situ} origins as in the simulations.

In their study of simulated clusters, \citet{Rhee2017} separated the galaxies into three different accretion classes based on their accretion time onto the cluster: ancient infallers with $t_{\mathrm{inf}} (\mathrm{Gyr}) > 6.45$, intermediate infallers with $3.63 < t_{\mathrm{inf}} (\mathrm{Gyr}) < 6.45$, recent infallers with $t_{\mathrm{inf}} (\mathrm{Gyr}) < 3.63$ and first infallers that have just fallen onto the cluster. They found that the ancient and recent infallers were mainly located within one virial radius from the centre of the cluster (i.e. $\sim1\, R_{\mathrm{200}}$) and that the recent infallers had higher orbital velocities than the ancient infallers. On the other hand, the intermediate and first infallers were mainly located between $1$-$3\, R_{\mathrm{200}}$, with the intermediate infallers being characterized by lower orbital velocities than the first infallers.

\begin{figure*}
\centering
\includegraphics[width=0.25\textwidth]{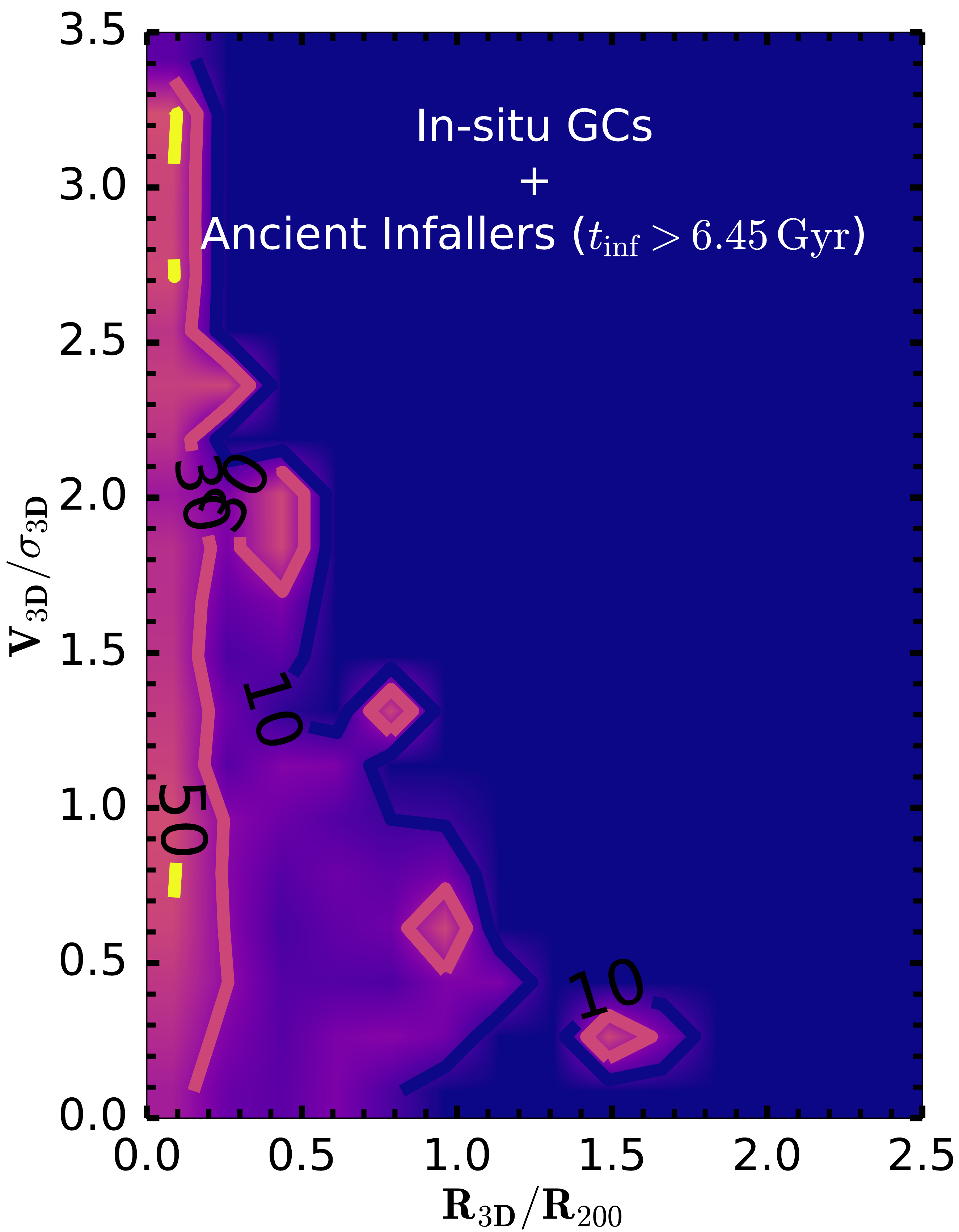}
\includegraphics[width=0.223\textwidth]{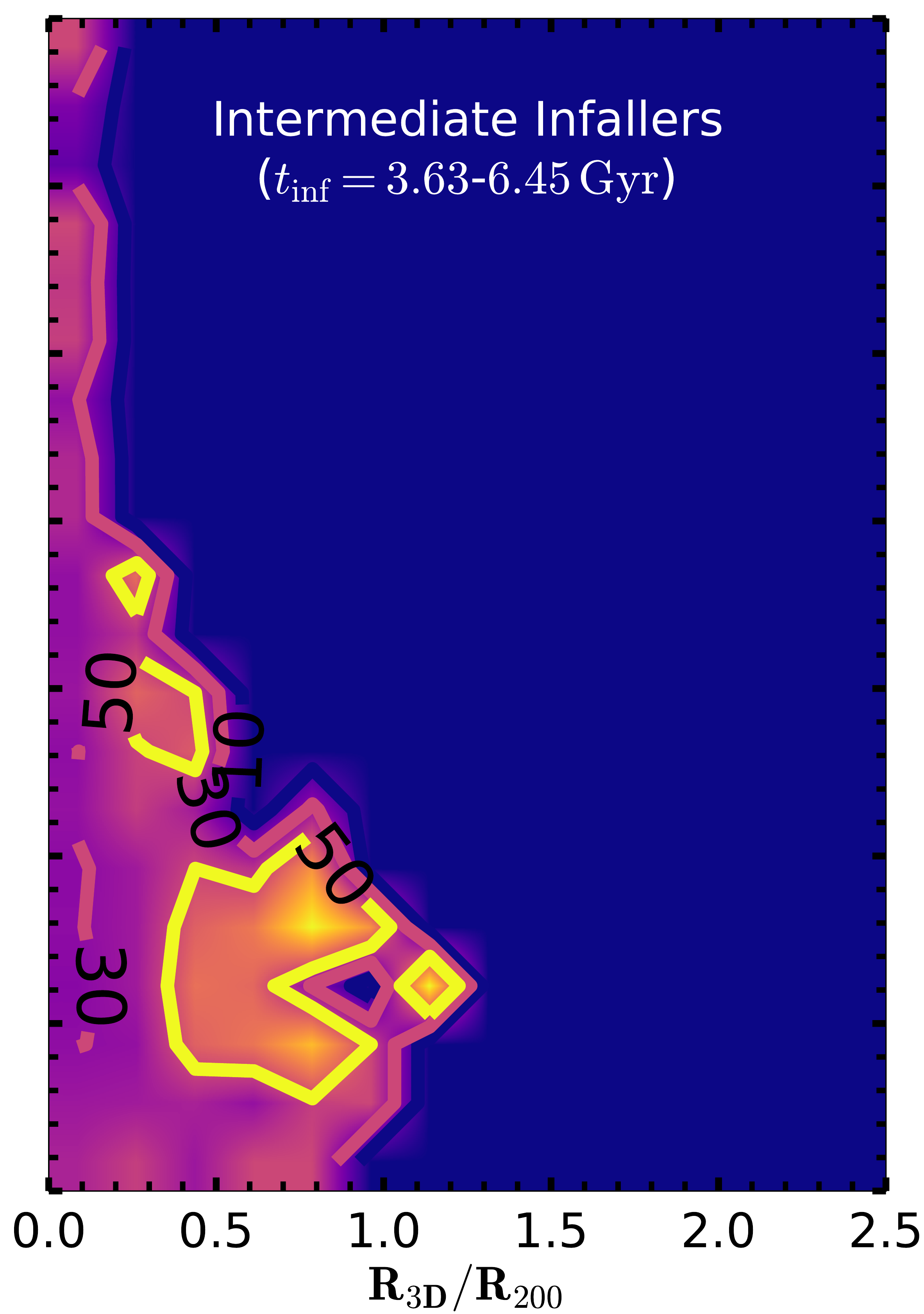}
\includegraphics[width=0.225\textwidth]{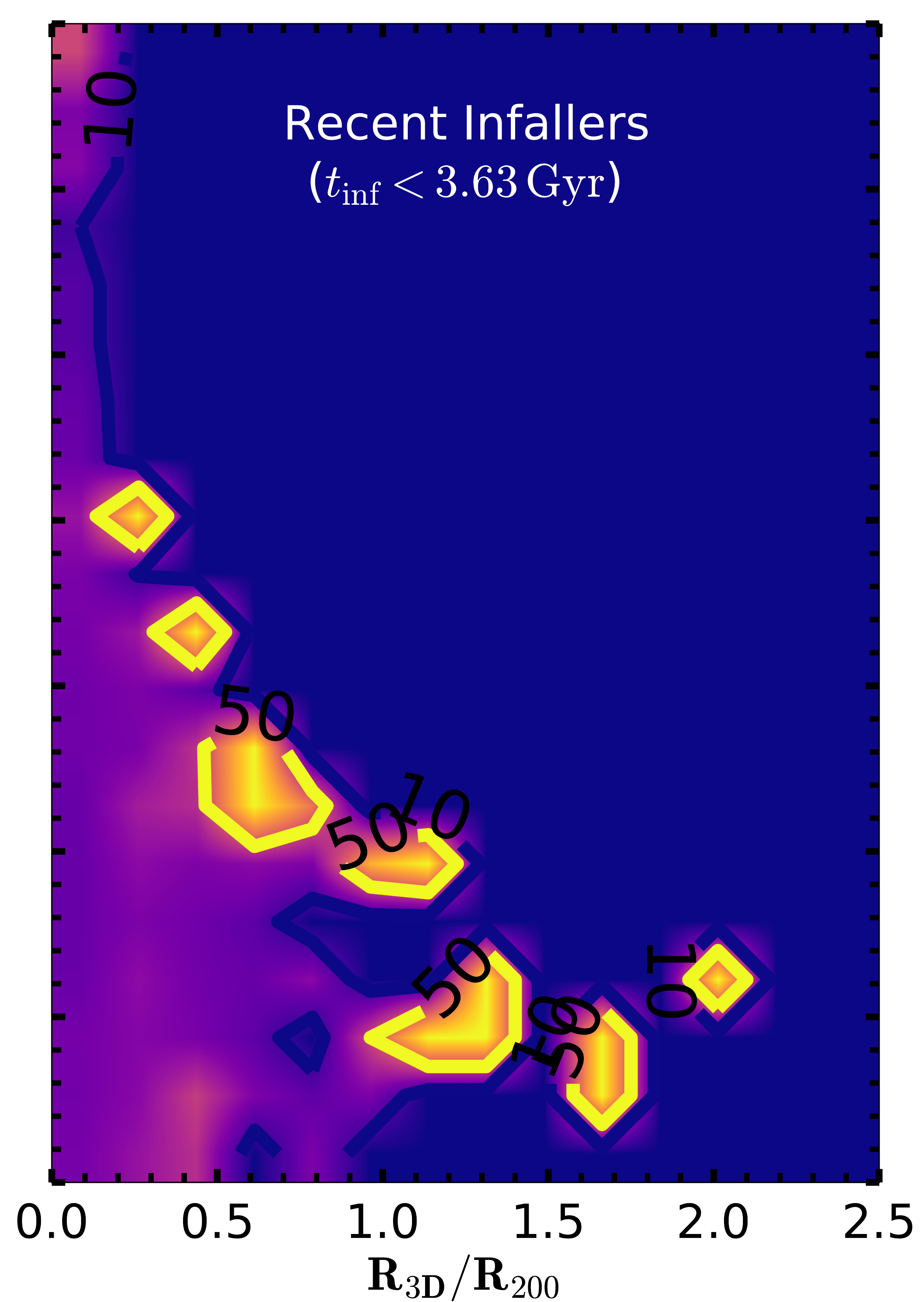}
\includegraphics[width=0.26\textwidth]{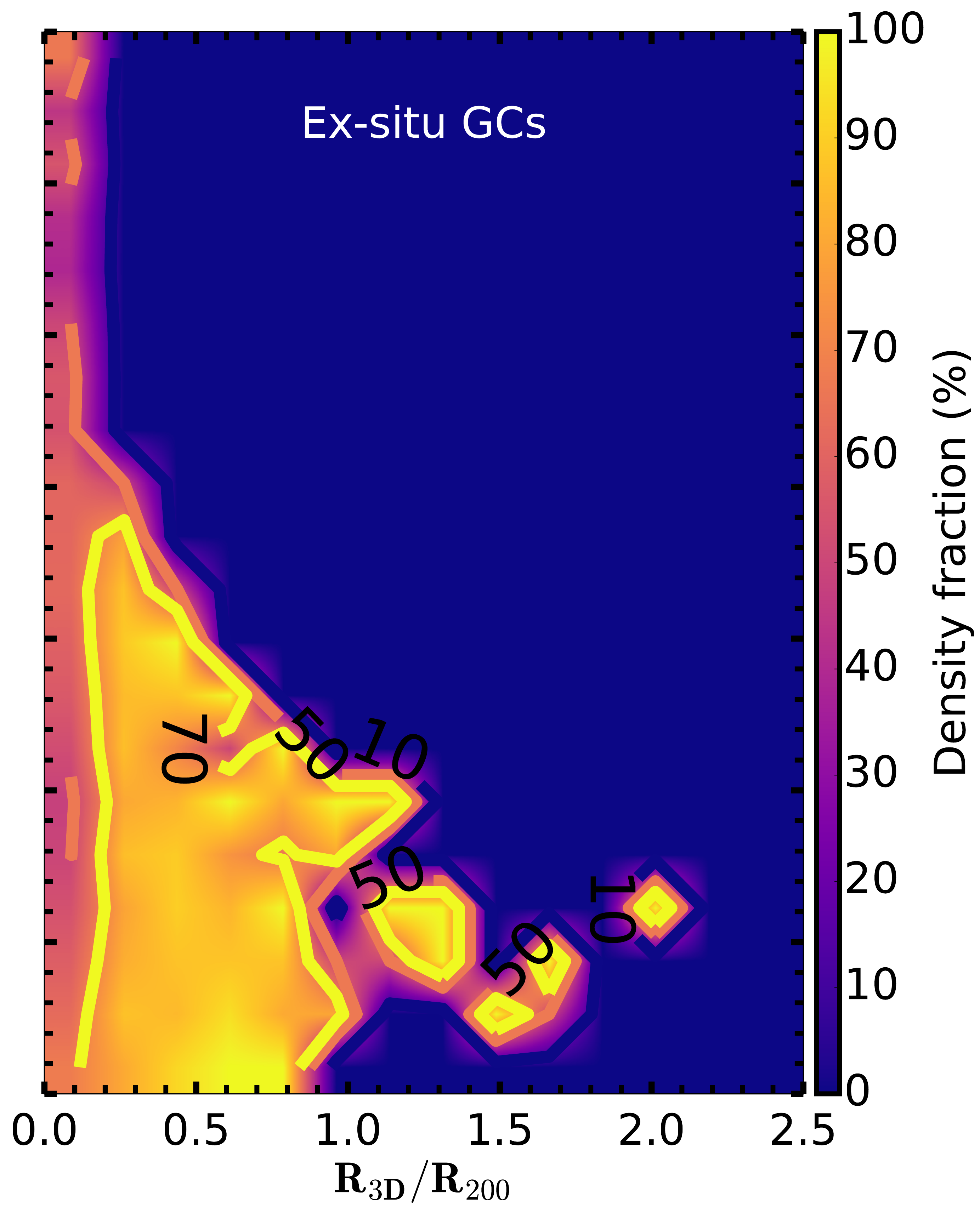}
\caption{3D phase-space diagrams of the \textit{in-situ} and ancient infaller GCs, intermediate infaller GCs, recent infaller GCs and \textit{ex-situ} GCs of all $50$ simulated S0 galaxies from the left to the right panel, respectively. The colorbar shows the distribution of the density fraction of the GCs in the 3D phase-space diagrams. We see that the ancient infaller (accreted more than $6.45\, \mathrm{Gyr}$ ago) and \textit{in-situ} GCs are concentrated within $0.2$ virial radii ($R_{200}$), while the GCs accreted more recently are mainly found at larger virial radii.}
\label{fig:infall_diagrams_3d}
\end{figure*}

Focusing on only the \textit{ex-situ} GC population of the sample of $50$ S0 galaxies, which are defined by the E-MOSAICS simulations, we do not apply any surface density cut here, unlike in Sec. \ref{sec:GC_density_cut}. Therefore, the spatial distributions of the GC systems will be more extended. Following \citet{Rhee2017}, we separate the \textit{ex-situ} GCs based on the time they were accreted onto the host galaxy as defined by the simulations: ancient infallers ($t_{\mathrm{inf}} (\mathrm{Gyr}) > 6.45$), intermediate infallers ($3.63 < t_{\mathrm{inf}} (\mathrm{Gyr}) < 6.45$) and recent infallers ($t_{\mathrm{inf}} (\mathrm{Gyr}) < 3.63$), and we produce the 3D phase-space diagrams of each one of these accreted GC classes.
The 3D phase-space diagrams are produced by plotting the 3D velocity ($V_{\mathrm{3D}}$) as a function of the 3D galactocentric radius ($R_{\mathrm{3D}}$) of each GC. We normalize the $V_{\mathrm{3D}}$ and $R_{\mathrm{3D}}$ of each GC by the velocity dispersion of the GC system, $\sigma_{\mathrm{3D}}$, and virial radius of the galaxy, $R_{\mathrm{200}}$, respectively, to which the GC belongs.
For the central galaxies, $R_{\mathrm{200}}$ is measured at $z=0$. However, for the satellite galaxies, $R_{\mathrm{200}}$ is not defined and can only be measured for the group as a whole. For this reason, for the satellite galaxies, we use the $R_{\mathrm{200}}$ measured when the galaxy was last a central of its own group prior to becoming a satellite. The velocity dispersion, $\sigma_{\mathrm{3D}}$, is calculated from the line-of-sight velocity dispersion along the $x$-, $y$- and $z$-axes estimated as the standard deviation of the corresponding GC line-of-sight velocities.

Fig. \ref{fig:infall_diagrams_3d} shows the 3D phase-space diagrams of the \textit{in-situ} GCs and \textit{ex-situ} GCs, split into the ancient, intermediate and recent infaller GCs, of all our $50$ S0 galaxies from the left to the right panel, respectively.
In the leftmost panel, we show together the \textit{in-situ} GCs and the ancient infaller GCs normalized by the total number of GCs and the total number of \textit{ex-situ} GCs of all our $50$ S0 galaxies, respectively. The rightmost panel shows together the \textit{ex-situ} GCs normalized by the total number of GCs of all our $50$ S0 galaxies. The intermediate and recent infaller GCs are normalized the total number of \textit{ex-situ} GCs of all our $50$ S0 galaxies, as for the ancient infaller GCs. 

First of all, we see that the \textit{in-situ} and \textit{ex-situ} GCs, which are split into ancient, intermediate and recent infallers GCs, show similar spatial distributions extending out to $\sim1$-$2\, R_{\mathrm{200}}$. 
However, the majority of the \textit{in-situ} and \textit{ex-situ} GCs are contained within $1\, R_{\mathrm{200}}$, with only a total of $10$ GC systems among all our $50$ S0 galaxies extending beyond $1\, R_{\mathrm{200}}$. 

Secondly, we see a trend that the GCs that were accreted more recently ($t_{\mathrm{inf}} (\mathrm{Gyr}) < 6.45$) are more likely found at larger radii (see also \citealt{Kruijssen2020,Pfeffer2020}). 
In fact, $\sim50\%$ of the \textit{ex-situ} GCs located within $0.2\, R_{\mathrm{200}}$ are ancient infallers ($t_{\mathrm{inf}} (\mathrm{Gyr}) > 6.45$). The \textit{in-situ} GCs are also concentrated within $0.2\, R_{\mathrm{200}}$, where they overlap with the ancient infaller GCs.
On the other hand, only $\sim30\%$ of the \textit{ex-situ} GCs within $0.2\, R_{\mathrm{200}}$ are intermediate infallers ($3.63 < t_{\mathrm{inf}} (\mathrm{Gyr}) < 6.45$) and $<30\%$ are recent infallers ($t_{\mathrm{inf}} (\mathrm{Gyr}) < 3.63$). The intermediate infallers are more numerous between $0.5$-$1\, R_{\mathrm{200}}$ for $V_{\mathrm{3D}}/\sigma_{3D}<1$ with $\geq50\%$ fraction, while $\geq50\%$ of the recent infallers are found beyond $1\, R_{\mathrm{200}}$ for $V_{\mathrm{3D}}/\sigma_{3D}<1$ or within $1\, R_{\mathrm{200}}$ for $V_{\mathrm{3D}}/\sigma_{3D}>1$.  

Overall, we see that the fraction of the \textit{in-situ} and ancient infaller GCs sharply drops beyond $0.2\, R_{\mathrm{200}}$ as opposed to the \textit{ex-situ} GCs, whose fraction increases to $\geq70\%$ out to $\sim1\, R_{\mathrm{200}}$.
Additionally, we note that the \textit{in-situ} GCs are also characterized by larger velocities on average than the \textit{ex-situ} GCs within $0.2\, \mathrm{R_{200}}$. This may be due to the nature of the \textit{in-situ} GCs, which are mainly metal-rich in a fast-rotating disk configuration.

The results from Fig. \ref{fig:infall_diagrams_3d} are overall consistent with those of the galaxies in individual galaxy clusters found by \citet{Rhee2017}, suggesting that similar physical processes do apply to both GCs and galaxies accreting onto an individual galaxy or cluster of galaxies, respectively. 
However, we note that the boundaries between the different accreted classes of the GCs in the galaxies are less sharp than those of the accreted galaxies in galaxy clusters. In fact, as we previously mentioned, the ancient, intermediate and recent GCs have overall similar distributions on the 3D phase-space diagrams. 
This may likely be a result of the degeneracy between the accretion redshift and the stellar mass of the accreted satellite galaxy. In fact, as shown in figure 2 of \citet{Pfeffer2020}, the apocentres of the accreted GCs decrease for increasing satellite stellar masses at fixed redshifts.

In summary, our results suggest that there exists a correlation between the location of the GCs of the individual galaxies on the 3D phase-space diagrams and the time they were accreted onto the galaxy. 
We also looked at the projected (2D) phase-space diagrams along the three line-of-sights, i.e. $x$, $y$ and $z$. We find that, overall, the boundaries delimiting the areas where the different accretion GC classes are mainly located are similar in the 3D and 2D phase-space diagrams. 
Therefore, the phase-space diagrams could be applied to the GC systems of the observed galaxies and we should expect that the GCs found within $0.2\, R_{\mathrm{200}}$ ($=20\, \mathrm{kpc}$ for $R_{\mathrm{200}}=100\, \mathrm{kpc}$) have likely an \textit{in-situ} origin or they were accreted at early times (more than $6.45\, \mathrm{Gyr}$ ago). On the other hand, the GCs found beyond $0.2\, R_{\mathrm{200}}$ have a higher probability of being \textit{ex-situ} and accreted more recently (less than $6.45\, \mathrm{Gyr}$ ago) onto the galaxy.   

\begin{figure*}
\centering
\includegraphics[width=0.46\textwidth]{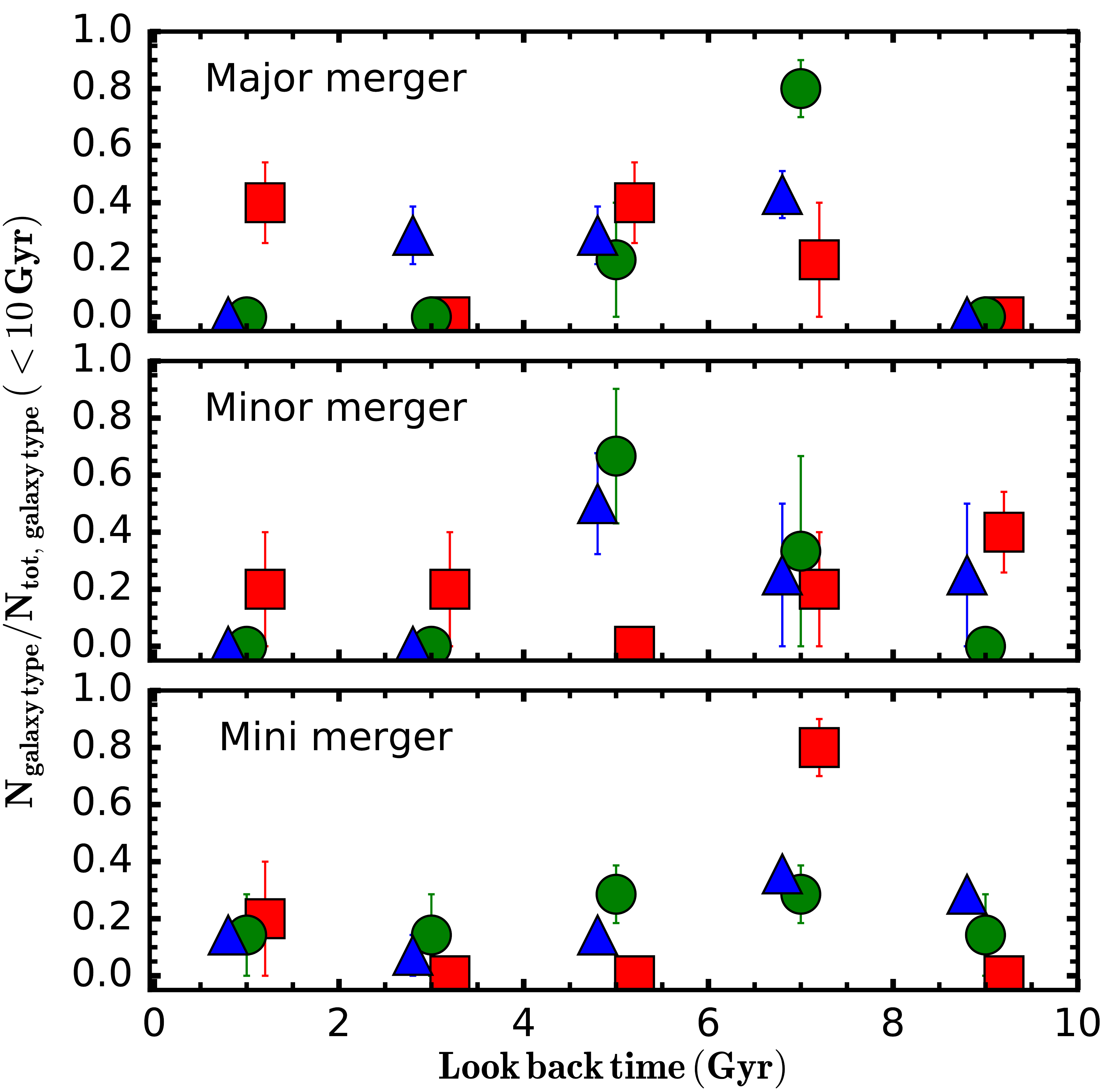}
\includegraphics[width=0.4\textwidth]{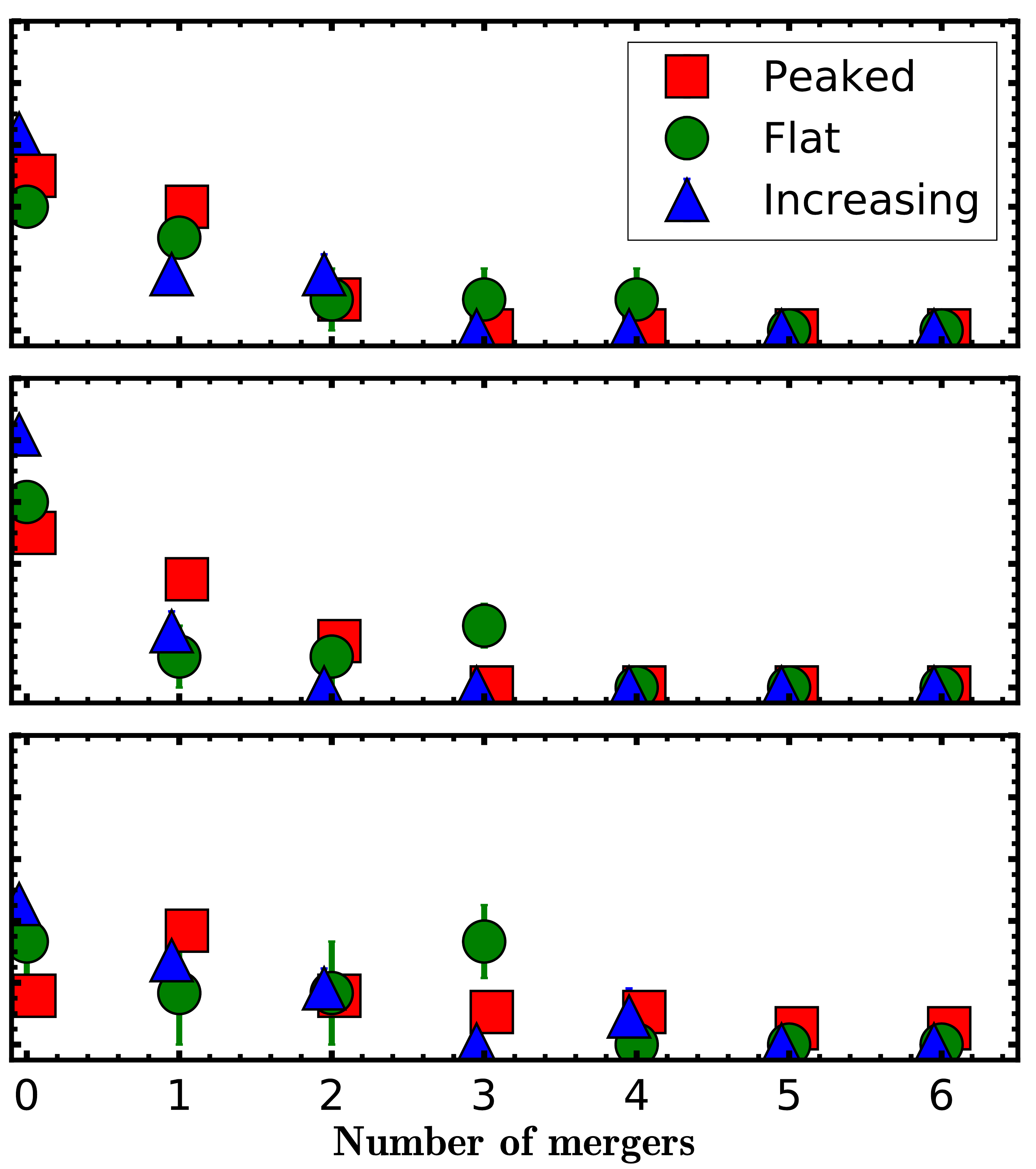}
\caption{\textit{Left:} the fraction of the \textit{peaked} (red), \textit{flat} (green) and \textit{increasing} (blue) galaxies that experienced their last major, minor and mini merger event within a time interval of $\pm1\, \mathrm{Gyr}$ for a given look back time in Fig. \ref{fig:fraction_galaxy_vs_merger} up to a maximum of $10\, \mathrm{Gyr}$ from the top to the bottom panels. Each panel is normalized by the total number of the galaxies of a given $V_{\mathrm{rot}}/\sigma$ class that experienced their last major, minor and mini merger within the last $10\, \mathrm{Gyr}$, such that the sum data points of a given type is equal to one in each panel. \textit{Right:} the fraction of the \textit{peaked}, \textit{flat} and \textit{increasing} galaxies that experienced a total number of major, minor and mini mergers within the last $10\, \mathrm{Gyr}$ from the top to the bottom panel. The data points are slightly shifted on the $x$-axis to avoid overlapping. We see that there is not a strong correlation between the present-day $V_{\mathrm{rot}}/\sigma$ profiles of the GCs and the past merger histories of the S0 galaxies.} 
\label{fig:fraction_galaxy_vs_merger}
\end{figure*}

\section{The assembly history of the simulated S0 galaxies}
\label{sec:formation_history_S0s}
One of the advantages of the simulations is that they contain information about the detailed merger and accretion histories of the galaxies. 
For this reason, we now investigate the presence of correlations between the distinct kinematic $V_{\mathrm{rot}}/\sigma$ profiles of the GCs, described in Sec. \ref{sec:1D_kinematic_profiles_GC}, and the corresponding assembly histories of the S0 galaxies. This adds to the previous work using E-MOSAICS, which has revealed correlations between the galaxy assembly history and GC age-metallicity distributions \citep{Kruijssen2019} and GC kinematics \citep{Gomez2021}.

Furthermore, in Sec. \ref{sec:GC_subpopulations}, we have seen that the metal-rich and metal-poor GC sub-populations can have misaligned kinematics. Therefore, we aim to understand whether the observed kinematic misalignment in the GC sub-populations of our simulated S0 galaxies is the result of specific merger events.

\subsection{The relation between the $V_{\mathrm{rot}}/\sigma$ profiles of the GCs and the accretion history of the S0 galaxies}
\label{sec:merger_history_and_kinematic_profiles}
From the E-MOSAICS simulations, we derive the merger trees of our simulated $50$ S0 galaxies. These merger trees include all the accretion events with different mass-ratio (i.e. $\mathrm{M_{*}^{\mathrm{sat}}}/\mathrm{M_{*}}$, where $\mathrm{M_{*}^{\mathrm{sat}}}$ is the satellite galaxy mass and $\mathrm{M_{*}}$ the target galaxy stellar mass) experienced by our target S0 galaxy as a function of redshift.

Following \citet{Schulze2020}, we classify the merger events into four categories depending on the mass-ratio of the merger: major mergers (mass-ratio$>$1:4), minor mergers (1:10$<$mass-ratio$<$1:4), mini mergers (1:100$<$mass-ratio$<$1:10) and smooth accretion (mass-ratio$<$1:100). We focus here on the mergers that occurred within the last $10\, \mathrm{Gyr}$ for the comparison with \citet{Schulze2020}. 

On the left-hand side of Fig. \ref{fig:fraction_galaxy_vs_merger}, we show the fraction of the \textit{peaked}, \textit{flat} and \textit{increasing} galaxies that experienced their last major, minor and mini merger event within a time interval of $\pm1\, \mathrm{Gyr}$ for a given look back time in Fig. \ref{fig:fraction_galaxy_vs_merger} up to a maximum of $10\, \mathrm{Gyr}$, similarly to figure 10 of \citet{Schulze2020}. Each panel is normalized by the total number of the galaxies of a given $V_{\mathrm{rot}}/\sigma$ class that experienced their last major, minor and mini merger within the last $10\, \mathrm{Gyr}$, such that the sum data points of a given type is equal to one in each panel.
However, overall, we do not see any clear trend between the merger histories of the different galaxy types as a function of redshift, as it was previously found by \citet{Schulze2020}. In fact, \citet{Schulze2020} found that $\sim60\%$ of the \textit{peaked} galaxies had no major mergers and that, if a major merger occurred, it was most probably more than $5\, \mathrm{Gyr}$ ago. On the other hand, they found that only $40\%$ and $30\%$ of the \textit{flat} and \textit{increasing} galaxies experienced no major mergers, respectively, while the remaining ones more likely experienced at least a major merger between $3$-$7\, \mathrm{Gyr}$ ago. Finally, \citet{Schulze2020} found that all galaxies experienced typically between $1$-$4$ mini mergers with a probability increasing towards low redshifts.

On the right-hand side of Fig. \ref{fig:fraction_galaxy_vs_merger}, we show the fraction of the \textit{peaked}, \textit{flat} and \textit{increasing} galaxies that experienced a total number of major, minor and mini mergers in the last $10\, \mathrm{Gyr}$ from the top to the bottom panel, respectively.
We see that the majority of the galaxies are likely to have experienced either zero or one major merger. A small fraction of the galaxies ($\sim10$-$20\%$) may have also experienced two major mergers, however the \textit{peaked} and \textit{increasing} galaxies are unlikely to have experienced more than two. On the other hand, a small fraction of the \textit{flat} galaxies may have experienced up to four major mergers.
Almost all of the \textit{increasing} galaxies ($80\%$) did not experience any minor merger, while roughly half of the \textit{peaked} and \textit{flat} galaxies may have experienced up to two or three minor mergers, respectively. 
Approximately $40\%$ of the \textit{peaked} galaxies have had at least one mini merger and a similar fraction of the \textit{flat} galaxies have had at least three. On the other hand, half of the \textit{increasing} galaxies have had no mini mergers, while the remaining half is likely to have experienced between one and two mini mergers. 
These results show that the \textit{peaked}, \textit{flat} and \textit{increasing} galaxies are overall similar in terms of the number of the different types of mergers that they experienced in the last $10\, \mathrm{Gyr}$. An exception is represented by the frequency of minor mergers that are more typical for the \textit{peaked} and \textit{flat} than the \textit{increasing} ones.

The discrepancy with the results found by \citet{Schulze2020} may be due to several possible reasons. It could be a result of low sample statistics as we only have $50$ galaxies compared to the $\sim500$ of \citet{Schulze2020}. Additionally, the fact that we are also only selecting S0 galaxies in this work could add a certain bias. Secondly, it could be due to the fact that we are studying the $V_{\mathrm{rot}}/\sigma$ profiles of the GCs, while \citet{Schulze2020} looked at the stellar $V_{\mathrm{rot}}/\sigma$ profiles. Additionally, our $V_{\mathrm{rot}}/\sigma$ profiles extend beyond $5\, R_{\mathrm{e}}$, while those from \citet{Schulze2020} were limited to $5\, R_{\mathrm{e}}$.
Finally, it could be a result of differences between the EAGLE and \texttt{Magneticum} simulations.

To test whether our results may partly depend on the radial extension of the $V_{\mathrm{rot}}/\sigma$ profiles, we perform an alternative classification of the $V_{\mathrm{rot}}/\sigma$ profiles restricted to $5\, R_{\mathrm{e}}$. However, the results in Fig. \ref{fig:fraction_galaxy_vs_merger} do not change significantly and we still do not see the trends found by \citet{Schulze2020} for the major and mini mergers specifically.

The results in Fig. \ref{fig:fraction_galaxy_vs_merger} could possibly suggest that the distinct present-day kinematic profiles are not the result of the specific different merger histories of the S0 galaxies. Other factors, such as the orbital configuration of the merging galaxies, may be playing a more dominant role. A possible follow-up project could involve investigating the kinematic properties of the GC systems back in time to understand whether and how the different physical processes (e.g. mergers) influence the GC kinematics in our galaxies during their evolution and increase the sample size beyond $50$ galaxies. 

\subsection{The origin of the GC misalignment in S0 galaxies}
\label{sec:origin_kinematic_misalignment}
In Sec. \ref{sec:1D_kinematic_profiles_GC}, we classified $9$ out of the $50$ simulated S0 galaxies as \textit{misaligned}, since their GCs were rotating along a different $\mathrm{PA}_{\mathrm{kin}}$ compared to the $\mathrm{PA}_{\mathrm{phot}}$ of the galaxy.

In this section, we aim to investigate what is the cause of the misalignment in these galaxies and, specifically, whether it is a result of the specific merger histories of the \textit{misaligned} galaxies with respect to the other \textit{aligned} galaxies.
For each one of the $9$ \textit{misaligned} galaxies, we identify those accretion events that each contributed to a significant fraction of the present-day GC population (i.e. $\geq10\%$) and we study the kinematic properties of these accreted GCs, with the main focus being their 2D velocity map and 1D $\mathrm{PA}_{\mathrm{kin}}$ and $V_{\mathrm{rot}}$ profiles.

\begin{figure*}
\centering
\Large
\textbf{Galaxy with a significant accretion event}\par\medskip
\Large GalaxyID = 4362302\par\medskip
\includegraphics[width=0.6\textwidth]{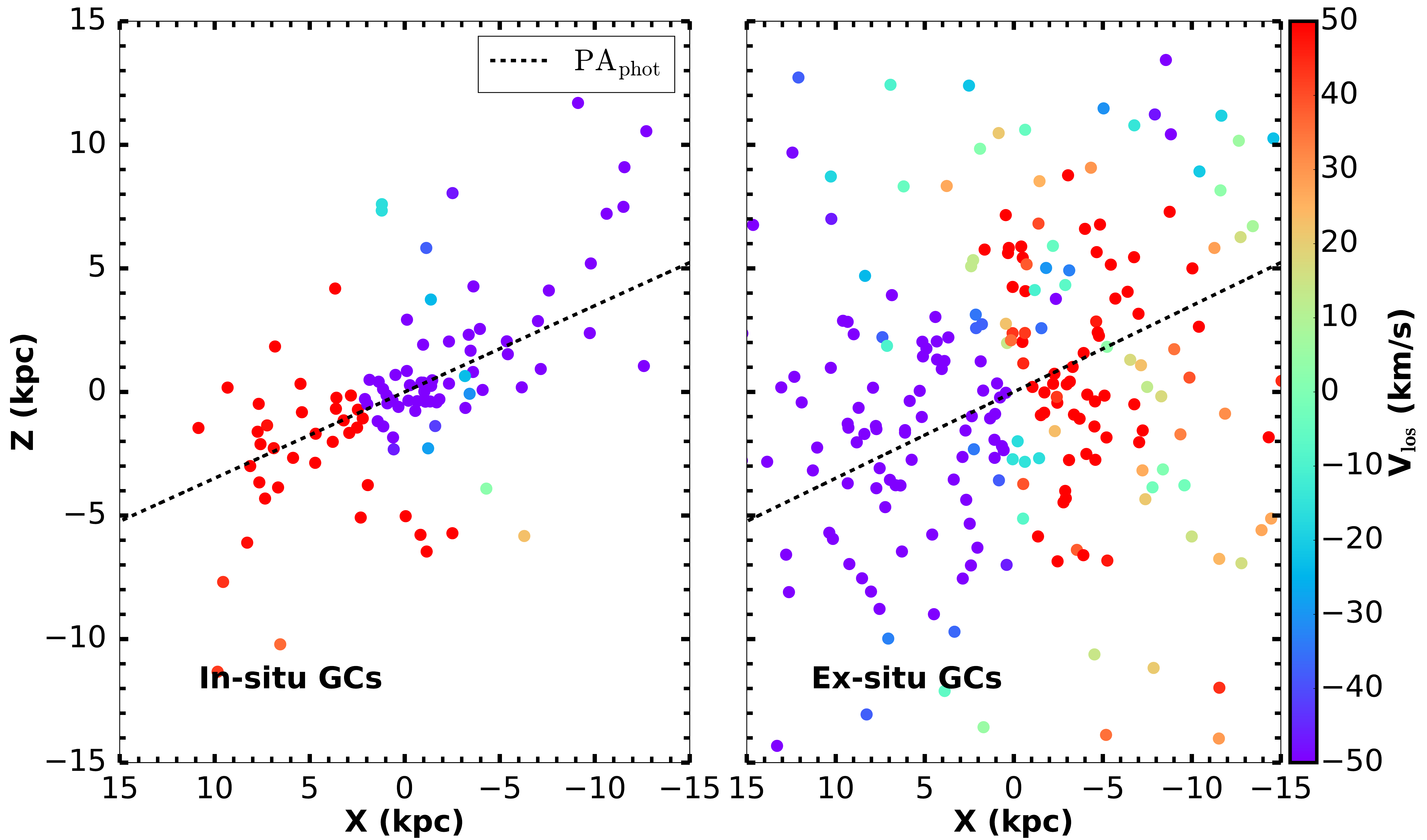}
\includegraphics[width=0.31\textwidth]{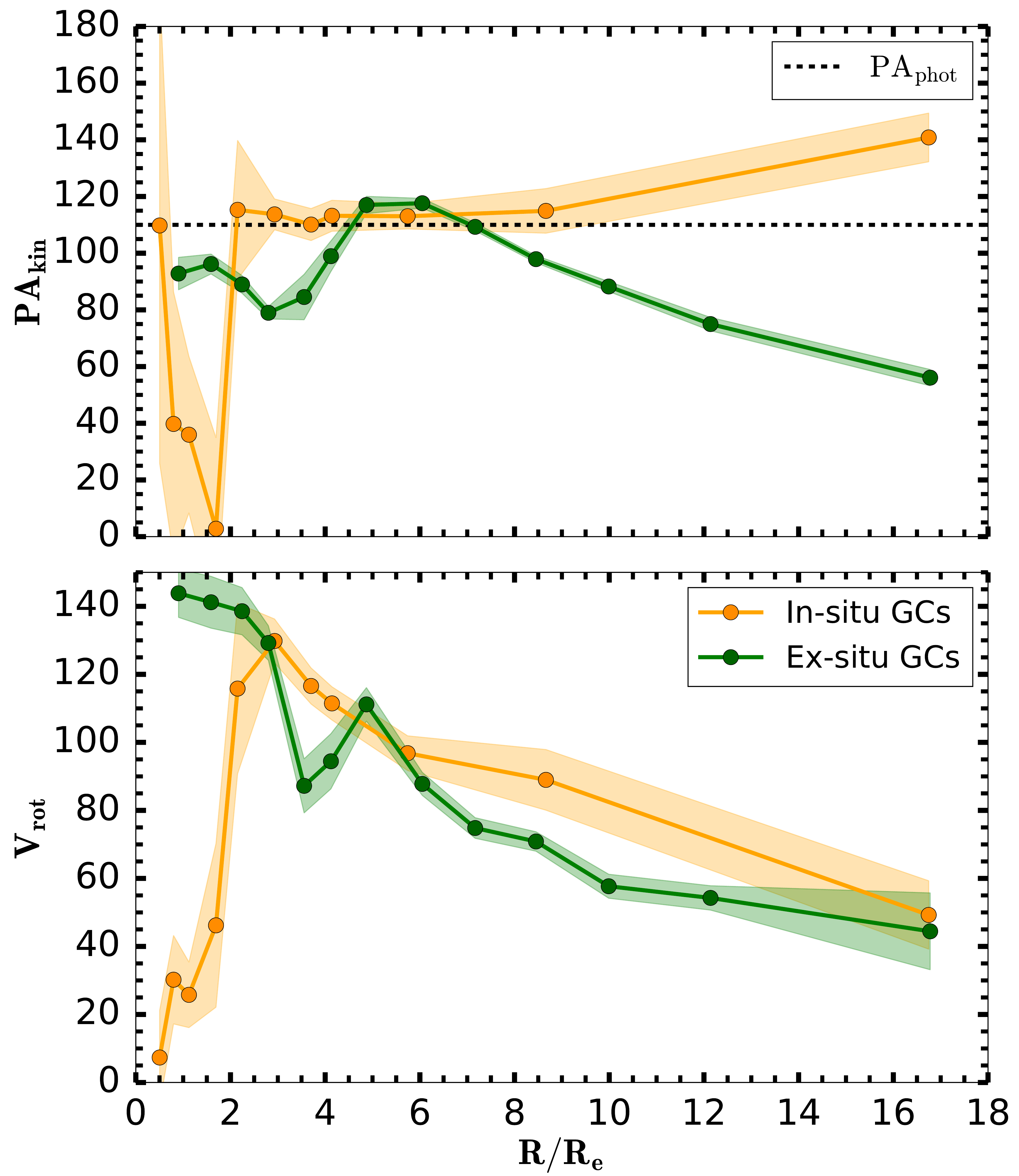}
\caption{\textit{Continued}}
\end{figure*}
\begin{figure*}
\ContinuedFloat
\centering
\Large
\textbf{Galaxy with no significant accretion event}\par\medskip
\Large GalaxyID = 6451189\par\medskip
\includegraphics[width=0.6\textwidth]{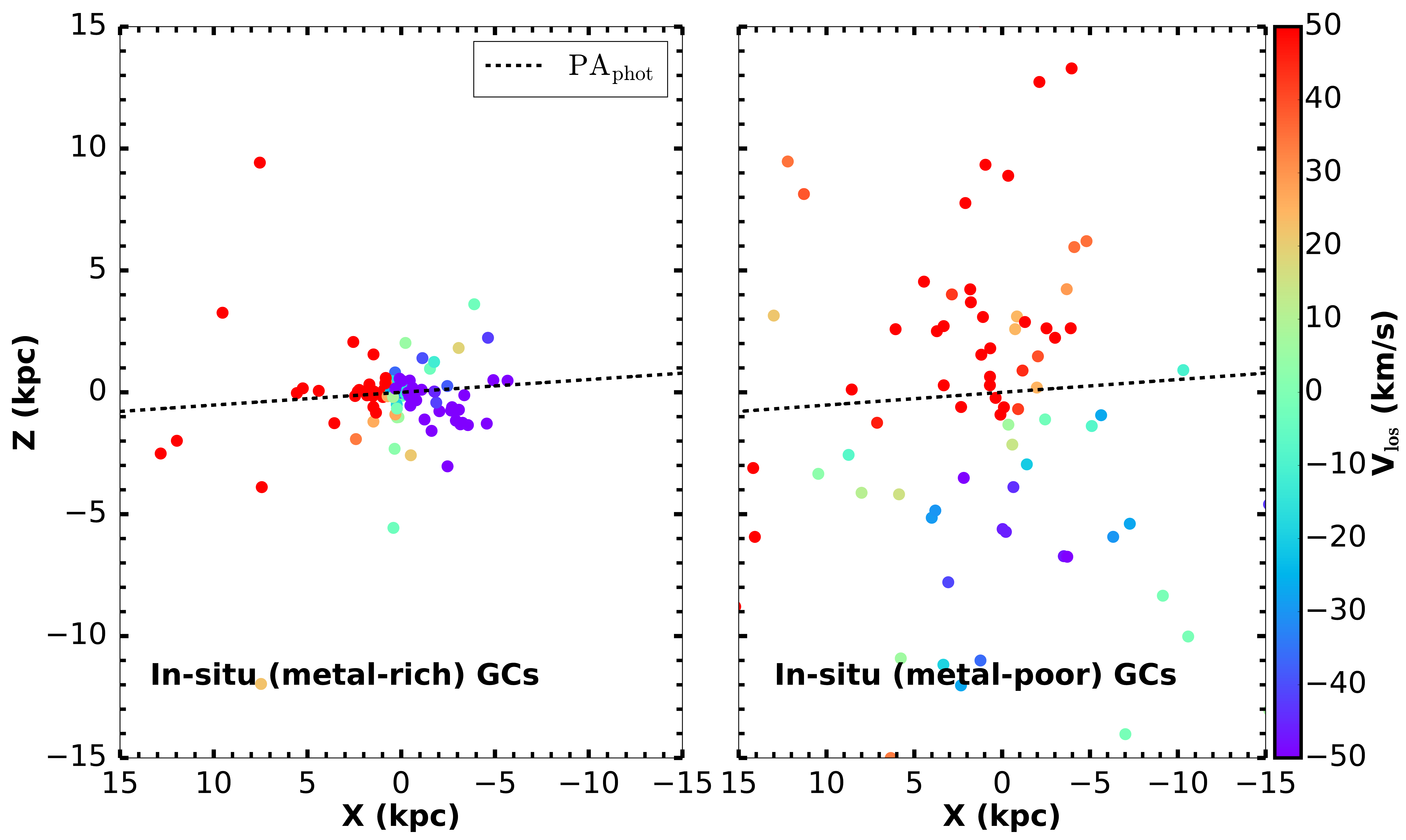}
\includegraphics[width=0.31\textwidth]{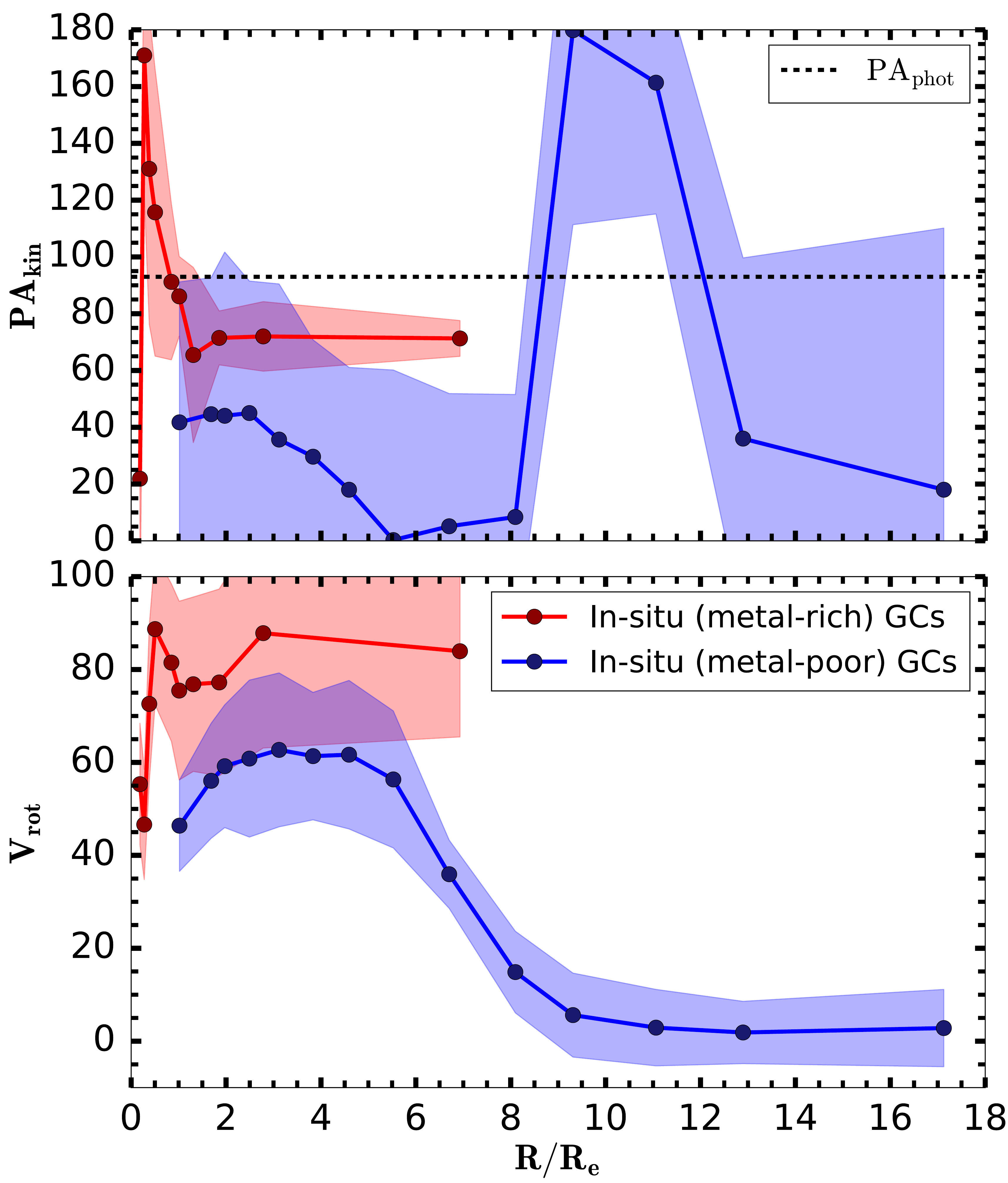}
\caption{\textit{Top panel:} 2D line-of-sight velocity maps and 1D $\mathrm{PA}_{\mathrm{kin}}$ and $V_{\mathrm{rot}}$ profiles of the \textit{in-situ} (yellow line) and \textit{ex-situ} (green line) GCs of an example galaxy that experienced a major merger event that contributed to more than $30\%$ of the total present-day GC population. \textit{Bottom panel:} 2D line-of-sight velocity maps and 1D $\mathrm{PA}_{\mathrm{kin}}$ and $V_{\mathrm{rot}}$ profiles of the \textit{in-situ} metal-rich (red line) and \textit{in-situ} metal-poor (blue line) GCs of an example galaxy that did not experience any significant accretion events and, as a result, has low accreted GC fraction (i.e. $\sim15\%$). In all figures, the black dashed line represents the $\textit{PA}_{\mathrm{phot}}$ of the galaxy and the 1D kinematic profiles are calculated as described in Sec. \ref{sec:1D_kinematic_profiles_GC}. We see that the GC misalignment can be either driven by the different kinematics of the \textit{in-situ} and \textit{ex-situ} GCs or by the different kinematics of the \textit{in-situ} metal-rich and \textit{in-situ} metal-poor GCs in our galaxies.}
\label{fig:misalignment}
\end{figure*}

We find that $5$ of the $9$ \textit{misaligned} galaxies ($56\%$) have experienced a major merger event within the last $10\, \mathrm{Gyr}$ that contributed between $30$-$80\%$ of the total present-day GC population. Three of these galaxies have also experienced a minor merger event between $4$-$7\, \mathrm{Gyr}$ ago that contributed between $10\%$-$20\%$ of the GC population.
The \textit{ex-situ} GCs are largely metal-rich (i.e. $\gtrsim40\%$) in these $5$ galaxies and they show different rotation properties with respect to the \textit{in-situ} GCs. In fact, while the \textit{in-situ} GCs show typically strong rotation, which is overall well aligned with the $\mathrm{PA}_{\mathrm{phot}}$ of the galaxy, the \textit{ex-situ} GCs show rotation which can be misaligned by angles varying between $0\degr < \Delta \mathrm{PA} < 180\degr$ (where $0\degr$ is co-rotating and $180\degr$ is counter-rotating) with respect to that of the \textit{in-situ} GCs. 
In Fig. \ref{fig:misalignment}, we show the edge-on projection of one of these $5$ galaxies (i.e. GalaxyID $=4362302$) as an example, which experienced a major merger $\sim9\, \mathrm{Gyr}$ ago that contributed to $\sim70\%$ of the total present-day GC population with $60\%$ of these \textit{ex-situ} GCs being metal-rich. We see that the \textit{ex-situ} GCs are counter-rotating with respect to the \textit{in-situ} GCs in this galaxy.

We also find that few of the \textit{aligned} galaxies in Sec. \ref{sec:1D_kinematic_profiles_GC} (e.g. GalaxyID $=13394165$) have accreted a significant fraction of the \textit{ex-situ} metal-rich GCs from a major merger (i.e. $\sim40\%$), but these \textit{ex-situ} metal-rich GCs are co-rotating with the \textit{in-situ} metal-rich GCs. These \textit{aligned} galaxies have experienced the major merger event between $6$-$9\, \mathrm{Gyr}$ ago, similarly to the $5$ \textit{misaligned} galaxies. Therefore, we suggest that the misalignment between the \textit{in-situ} and \textit{ex-situ} metal-rich GCs in these galaxies is not necessarily connected to the time of occurrence of the major merger, but it, possibly, depends on other properties of the mergers, such as the orbital configuration of the merging galaxies. In this scenario, the misalignment between the \textit{in-situ} and \textit{ex-situ} GCs could then be the result of the change of the axis of accretion from the cosmic web during the galaxy formation process, such that the spin axis of the accreted GCs was aligned differently with respect to the spin axis of the host galaxy.

The remaining $4$ of the $9$ \textit{misaligned} galaxies ($44\%$) did not experience any significant accretion events and, thus, they have low \textit{ex-situ} GC fractions (i.e. $\lesssim30\%$ of the total present-day GC population). The \textit{ex-situ} GCs, which are mostly metal-poor in these galaxies, do not show evidence of any clear rotation. For this reason, we investigate the 2D velocity maps and the 1D $\mathrm{PA}_{\mathrm{kin}}$ and $V_{\mathrm{rot}}$ profiles of the \textit{in-situ} metal-rich and \textit{in-situ} metal-poor GCs in these galaxies. 
In two of these galaxies, we find that the rotation of the \textit{in-situ} metal-rich GCs is, overall, aligned with the $\mathrm{PA}_{\mathrm{phot}}$ of the galaxy, while the \textit{in-situ} metal-poor GCs are rotating along a kinematic axis offset by $\Delta \mathrm{PA} \simeq 90\degr$ with respect to the $\mathrm{PA}_{\mathrm{phot}}$ of the galaxy. In other two galaxies, the \textit{in-situ} metal-poor GCs do not show evidence of any net rotation, while the \textit{in-situ} metal-rich GCs show rotation that oscillates around the $\mathrm{PA}_{\mathrm{phot}}$ of the galaxy, but it is overall consistent within the large $1\sigma$ errors.
In Fig. \ref{fig:misalignment}, we show one of these $4$ galaxies (i.e. GalaxyID $=6451189$) as an example, whose \textit{in-situ} metal-poor GCs are misaligned by $\sim90\degr$ with respect to the \textit{in-situ} metal-rich GCs.

From the 2D kinematic maps of all $50$ S0 galaxies, we find that the majority of the galaxies have the \textit{in-situ} GCs co-rotating with the stars about the photometric major-axis of the galaxy (i.e. $70\%$). On the other hand, a small fraction of the galaxies have \textit{in-situ} GCs that are not co-rotating with the stars (i.e. $16\%$) or that do not show evidence of any net rotation (i.e. $14\%$). Additionally, we find that $20\%$ of the S0 galaxies have the \textit{in-situ} metal-poor GCs that are rotating about an axis different from that of the \textit{in-situ} metal-rich GCs, which is overall consistent with the photometric major axis of the galaxy (e.g. GalaxyID $=13394165$ and $17827697$ in Fig. \ref{fig:GC_subpopulations_peaked}). 

We do not find any clear differences in the merger histories of the galaxies with misaligned \textit{in-situ} metal-rich and \textit{in-situ} metal-poor GC kinematics. Therefore, these results suggest that the misalignment of the \textit{in-situ} GC sub-populations is not related to the specific merger histories of the galaxies. 

The \textit{in-situ} metal-poor GCs likely formed at early times, i.e. $z>2$, in the turbulent and high-pressure disks of low-mass gas-rich galaxies. Prior to their formation, the \textit{in-situ} metal-poor GCs suffered kinematic perturbations from merger events (more frequent at high redshifts) that redistributed these GCs in the host galaxy halo where they survived until the present-day \citep{Kruijssen2015,Keller2020}. The \textit{in-situ} metal-rich GCs likely formed more recently after the host galaxy built up more of its stellar mass. For this reason, the \textit{in-situ} metal-rich GCs are less likely to have suffered kinematic perturbations from merger events (less frequent and with smaller masses) and are generally co-rotating with the host galaxy stars.

We find that the \textit{in-situ} metal-rich GCs are typically characterized by a more centrally concentrated and flattened spatial distribution than the \textit{in-situ} metal-poor GCs, and they are rotating consistent with the stars about an axis consistent with the photometric major axis of the galaxy in a disk-like configuration (i.e. $70\%$), as commonly found in the observations (e.g. \citealt{Brodie2006,Forbes1997,Pota2013,Dolfi2020,Dolfi2021,Arnold2011}).
On the other hand, the \textit{in-situ} metal-poor GCs typically have a more extended spatial distribution.

This \textit{in-situ} GC formation scenario could also explain the misalignment of the \textit{in-situ} metal-rich GCs seen in $16\%$ of the galaxies if the merger events had a large enough mass-ratio to perturb the orbits of these GCs and the subsequent star formation produced a stellar disk that was aligned differently from these previously formed \textit{in-situ} metal-rich GCs.

Additionally, previous observational works have also found evidence of ETGs showing "kinematically distinct cores" (KDC). These types of galaxies are generally not very common (i.e. $\sim7\%$ identified in ATLAS$^{\mathrm{3D}}$ by \citealt{Krajnovic2011}), but some previous works have shown that the stars and GC sub-populations in these galaxies may have distinct and misaligned rotation (e.g. \citealt{Blom2012}), suggesting different origins. Therefore, the distinct rotational properties of the \textit{in-situ} metal-rich and \textit{in-situ} metal-poor GCs could also have their origins in a similar decoupling event that leads to KDCs. 

\section{Summary and conclusions}
\label{sec:summary}
In this work, we have studied the kinematic profiles of the GCs in a selected sample of $50$ S0 galaxies from the E-MOSAICS simulations \citep{Pfeffer2018,Kruijssen2019}, with the aim of finding the link between the present-day kinematic properties and the past merger histories of the S0 galaxies in low-density environments (i.e. field and small galaxy groups).

From the 2D kinematic maps and 1D kinematic profiles of the GCs (see Sec. \ref{sec:1D_kinematic_profiles_GC}), we find that $82\%$ of the galaxies have GCs that are rotating about an axis that is consistent with the photometric major axis of the galaxy (\textit{aligned}), while the remaining $18\%$ of the galaxies do not (\textit{misaligned}), in general agreement with the observations from the SLUGGS survey \citep{Dolfi2021} 

Among the \textit{aligned} galaxies, we find that $49\%$, $24\%$ and $27\%$ show a \textit{peaked}, \textit{flat} and \textit{increasing} $V_{\mathrm{rot}}/\sigma$ profile, respectively (see Sec. \ref{sec:1D_kinematic_profiles_GC}). 
From the comparison of these three distinct $V_{\mathrm{rot}}/\sigma$ profiles with the past merger histories of the S0 galaxies derived from the E-MOSAICS simulations (see Sec. \ref{sec:merger_history_and_kinematic_profiles}), we do not find evidence of any clear correlation between the present-day $V_{\mathrm{rot}}/\sigma$ profile shape and the past merger histories of the galaxies, unlike in the \texttt{Magneticum} simulations of the stellar profiles by \citet{Schulze2020}. 
A future work could investigate the kinematic properties of the GC systems back in time to understand whether and how the different physical processes can influence the GC kinematics in our galaxies during their evolution and increase the sample size beyond $50$ galaxies.

Among the \textit{misaligned} galaxies, we find that (see Sec. \ref{sec:origin_kinematic_misalignment}):
\begin{itemize}

    \item $56\%$ of the galaxies experienced at least one major merger event that contributed between $30\%$-$80\%$ of the total present-day GC population. In these galaxies, a large fraction of the \textit{ex-situ} GCs are metal-rich and they are rotating about an axis that is different from that of the \textit{in-situ} metal-rich GCs (which is consistent with the photometric major axis of the galaxy). The misalignment angle between the \textit{ex-situ} and \textit{in-situ} metal-rich GCs can vary between $0\degr$ (co-rotation) to $180\degr$ (counter-rotation), while the metal-poor GCs do not show evidence of any clear rotation in these galaxies (see Fig. \ref{fig:misalignment}; top panel).

    \item $44\%$ of the galaxies did not experience any major merger event and have low \textit{ex-situ} GC fraction (i.e. $\lesssim30\%$). In these galaxies, the \textit{ex-situ} GCs are predominantly metal-poor with no evidence of any clear rotation. In two of these galaxies, the \textit{in-situ} metal-poor GCs are rotating about an axis offset by $90\degr$ from that of the \textit{in-situ} metal-rich GCs (which is consistent with the photometric major axis of the galaxy).

\end{itemize}

For the total sample, from the 2D kinematic maps, we find that the majority of the galaxies (i.e. $70\%$) have the \textit{in-situ} GCs that are co-rotating with the stars about an axis that is consistent with the photometric major axis of the galaxy, as commonly found in the observations (e.g. \citealt{Forbes1997,Brodie2006,Arnold2011,Pota2013,Dolfi2020,Dolfi2021}). A minority of the galaxies (i.e. $16\%$) have the \textit{in-situ} GCs that are rotating about an axis that is different from the photometric major axis of the galaxy, while the remaining $14\%$ of the galaxies have the \textit{in-situ} GCs that do not show evidence of any net rotation. Finally, for the total sample, we find that $20\%$ of the galaxies have the \textit{in-situ} metal-poor GCs that are rotating about an axis different from that of the \textit{in-situ} metal-rich GCs.
Therefore, we suggest that there are likely two possible origins for the GC misalignment in our S0 galaxies:
\begin{enumerate}

    \item \textit{in-situ} GC formation during the evolution of the host galaxy. The \textit{in-situ} metal-poor GCs formed early ($z>2$) and suffered kinematic perturbations from the frequent merger events, which redistributed these GCs in the host galaxy halo \citep{Kruijssen2015,Keller2020}. The \textit{in-situ} metal-rich GCs are less likely to have suffered kinematic perturbations since they formed more recently when mergers became less frequent and the galaxy built up more of its stellar mass.

    \item a major merger event that contributed to a large fraction of the \textit{ex-situ} metal-rich GCs that can be misaligned by angles varying between $0\degr$ (co-rotation) to $180\degr$ (counter-rotation) with respect to the \textit{in-situ} metal-rich GCs (which are rotating consistent with the photometric major axis of the galaxy). The degree of misalignment will depend on the orbital configuration of the merging galaxies. 

\end{enumerate}

Finally, we have studied the distribution of the GC systems of our S0 galaxies in 3D phase-space diagrams (see Sec. \ref{sec:phase_space_diagrams}). We find that there is a correlation between the accretion time of the GCs onto the galaxy with its location on the 3D and 2D phase-space diagrams, with the \textit{in-situ} and ancient infaller GCs (accreted more than $6.45\, \mathrm{Gyr}$ ago) being predominantly located within $0.2\, \mathrm{R}_{\mathrm{200}}$. On the other hand, the GCs that were accreted more recently (less than $6.45\, \mathrm{Gyr}$ ago) are most likely found at larger virial radii, i.e. beyond $0.2\, \mathrm{R}_{200}$ (see Fig. \ref{fig:infall_diagrams_3d}). 
Therefore, in future work, we can use the 2D phase-space diagrams of observed GC systems to assign a probability of the GCs of being \textit{in-situ} or ancient/recent infallers.

\section*{Acknowledgements}
We thank the anonymous referee for their very constructive comments and suggestions that helped improving this paper. We thank the E-MOSAICS team for providing access to the data in the simulations used in this paper. DF, WC, KB and AD acknowledge support from the Australian Research Council under Discovery Project 170102344. JP acknowledges support from the Australian Research Council under Discovery Project 200102574. AJR was supported by National Science Foundation grant AST-1616710 and as a Research Corporation for Science Advancement Cottrell Scholar. JMDK gratefully acknowledges funding from the Deutsche Forschungsgemeinschaft (DFG) in the form of an Emmy Noether Research Group (grant number KR4801/1-1), as well as from the European Research Council (ERC) under the European Union’s Horizon 2020 research and innovation programme via the ERC Starting Grant MUSTANG (grant agreement number 714907).

This work used the DiRAC Data Centric system at Durham University, operated by the Institute for Computational Cosmology on behalf of the STFC DiRAC HPC Facility (\url{www.dirac.ac.uk}). This equipment was funded by BIS National E-infrastructure capital grant ST/K00042X/1, STFC capital grants ST/H008519/1 and ST/K00087X/1, STFC DiRAC Operations grant ST/K003267/1 and Durham University. DiRAC is part of the National E-Infrastructure. The work also made use of high performance computing facilities at Liverpool John Moores University, partly funded by the Royal Society and LJMU's Faculty of Engineering and Technology.

\section*{Data Availability}
The data are available upon request from the E-MOSAICS project (\url{https://www.astro.ljmu.ac.uk/~astjpfef/e-mosaics/}).



\bibliographystyle{mnras}
\bibliography{biblio} 





\bsp	
\label{lastpage}
\end{document}